\documentclass[12pt]{article}
\usepackage{amsmath}
\usepackage{graphicx}
\usepackage{natbib}
\usepackage{url} 
\usepackage{xcolor}
\usepackage{adjustbox} 
\usepackage{tocloft}
\renewcommand{\cftsecfont}{\bfseries}
\usepackage{titletoc}

\newcommand{\blind}{0}
\usepackage{amsmath}
\DeclareMathOperator{\arcosh}{arcosh}
\usepackage{natbib}

\newcommand{\bi}{\begin{itemize}}
\newcommand{\be}{\begin{equation}\begin{array}{lllllllllllll} \displaystyle}
\newcommand{\beno}{\begin{equation}\begin{array}{lllllllllllll}\nonumber \displaystyle}
\newcommand{\ee}{\end{array}\end{equation}}

\newcommand{\ei}{\end{itemize}}
\newcommand{\mE}{\mathscr{E}}
\newcommand{\hide}[1]{}
\newcommand{\dsum}{\displaystyle\sum\limits}

\newcommand{\dint}{\displaystyle\int\limits}

\newcommand{\q}[1]{``#1''}
\newcommand{\pd}{Poincar\'e disk}
\newcommand{\prodL}[1]{\left\langle #1 \right\rangle_{\mathscr{L}}}
\newcommand{\s}{\vspace{0.25cm}}
\usepackage{verbatim}
\usepackage{hyperref,bbm}
\usepackage{textcomp}
\usepackage{amsmath,mathtools,latexsym,mathtools}
\usepackage{bm}
\usepackage{amsfonts,amsmath}
\usepackage[mathscr]{eucal}
\usepackage{amssymb}
\usepackage{algorithm2e}
\newtheorem{theorem}{Theorem}
\usepackage{graphicx}

\def\bse{\begin{eqnarray*}}
\def\ese{\end{eqnarray*}}
\def\bc{\[\begin{array}{ccccccc}}
\def\ec{\end{array}\]}
\def\bsq{\begin{equation*}}
\def\esq{\end{equation*}}
\def\bq{\begin{equation}}
\def\eq{\end{equation}}
\newcommand{\Fnorm}[1]{{|\!|\!|#1|\!|\!|_F}}

\newcommand{\vecnormsqrt}[1]{{|\!|#1|\!|_2^2}}
\newcommand{\vecnorm}[1]{{|\!|#1|\!|_2}}

\newtheorem{prop}{Proposition}
\usepackage{bm}
\usepackage{amsmath}

\def\pos{\bm{\theta}}
\def\Pos{\bm{\Theta}}

\usepackage{amsmath}

\newcommand{\alert}[1]{\textcolor{red}{\bf{#1}}}

\newcommand{\alertC}[1]{\textcolor{red}{\bf{CF: #1}}}

\renewcommand{\eqref}[1]{Equation~(\ref{#1})}

\usepackage{subcaption}
\usepackage{tikz}
\usetikzlibrary{decorations.pathmorphing, patterns,shapes}
\DeclareRobustCommand\sampleline[1]{%
  \tikz\draw[#1] (0,0) (0,\the\dimexpr\fontdimen22\textfont2\relax)
  -- (0.99em,\the\dimexpr\fontdimen22\textfont2\relax);%
}

		\newcounter{assumption}
		\newenvironment{assumption}[1][]{\refstepcounter{assumption}\par\noindent%
			\textbf{Condition~\theassumption #1}. \rmfamily}

\makeatletter
\renewcommand\p@subfigure{\thefigure~}         
\makeatother

\newcommand{\msim}{\mathop{\rm \sim}}
\newcommand{\ind}{\msim\limits^{\mbox{\tiny ind}}}

\usepackage{multibib}
\newcites{supp}{Supplementary Material References}

\date{}

\begin{document}

\def\spacingset#1{\renewcommand{\baselinestretch}%
{#1}\small\normalsize} \spacingset{1}

\spacingset{1.8} 


\if0\blind
{
  \title{\bf Scalable Sample-to-Population Estimation of Hyperbolic Space Models for Hypergraphs}
  \author{Cornelius Fritz\thanks{The first two authors are joint first authors and are listed in alphabetical order.}\hspace{1cm} Yubai Yuan\footnotemark[1]\\
   Michael Schweinberger}
  \maketitle
} \fi

\if1\blind
{
  \mbox{}
  \begin{center}
  { 
  \LARGE\bf Scalable Sample-to-Population Estimation of\s 
  \\
  Hyperbolic Space Models for Hypergraphs
  }
  \end{center}
  \medskip
} \fi

\begin{abstract}
Hypergraphs are useful mathematical representations of overlapping and nested subsets of interacting units, including groups of genes or brain regions, economic cartels, political or military coalitions, and groups of products that are purchased together. Despite the vast range of applications, the statistical analysis of hypergraphs is challenging: There are many hyperedges of small and large sizes, and hyperedges can overlap or be nested. Existing approaches to hypergraphs are either not scalable or achieve scalability at the expense of model realism. We develop a statistical framework that enables scalable estimation, simulation, and model assessment of hypergraph models, which is supported by non-asymptotic and asymptotic theoretical guarantees. First, we introduce a novel model of hypergraphs capturing core-periphery structure in addition to proximity, by embedding units in an unobserved hyperbolic space. Second, we achieve scalability by developing manifold optimization algorithms for learning hyperbolic space models based on samples from a population hypergraph. Third, we provide non-asymptotic and asymptotic theoretical guarantees for learning hyperbolic space models based on samples from a population hypergraph. We use the proposed statistical framework to detect core-periphery structure along with proximity among U.S.\ politicians based on historical media reports.
\end{abstract}

\noindent%

\section{Introduction}
\label{sec:intro}

Hypergraphs are useful mathematical representations of overlapping and nested subsets of interacting units, 
including groups of genes or brain regions performing vital biological functions;
economic cartels consisting of companies that collude with an eye to reducing competition and inflating the prices of goods and services;
political or military coalitions of state or non-state actors seeking to achieve shared political or military goals;
and groups of products purchased together.
A hypergraph with $N \geq 2$ units consists of $\binom{N}{2} + \binom{N}{3} + \ldots + \binom{N}{N}$ possible hyperedges of sizes $2$,
$3$,
$\ldots$,
$N$.
In other words,
hypergraphs consist of many possible hyperedges of small and large sizes that can overlap or be nested,
whereas graphs consist of edges of size two.
An example of a hypergraph is shown in Figure \ref{fig:hyperedge}.

Despite the vast range of applications highlighted above,
the statistical analysis of hypergraphs is challenging,
for at least three reasons.
First,
a hypergraph with $N$ units consists of $\sum_{k=2}^N {N \choose k} = 2^N - N - 1$ possible hyperedges.
In other words,
the number of possible hyperedges is exponential in $N$,
whereas the number of possible edges in a graph is quadratic in $N$.
As a result,
the statistical analysis of hypergraphs is more challenging than the statistical analysis of graphs.
These challenges are exacerbated by recent advances in collecting digital data,
which have enabled the collection of large hypergraphs \citep{kim2023, maleki2022}.
Second, 
hyperedges can overlap,
and the probabilities of overlapping hyperedges should be related:
e.g.,
the probabilities of publications involving researchers 1, 2, and 3 (hyperedge $\{1, 2, 3\}$) and researchers 1, 2, and 4 (hyperedge $\{1, 2, 4\}$) should be related,
because both publications involve researchers 1 and 2.
Third,
small hyperedges are nested in larger ones,
and the probabilities of nested hyperedges should be related:
e.g.,
if researchers 1 and 2 publish with researcher 3 (hyperedge $\{1, 2, 3\}$) and researcher 4 (hyperedge $\{1, 2, 4\}$), 
then 1 and 2 may very well publish with both 3 and 4 (hyperedge $\{1, 2, 3, 4\}$),
suggesting that the probabilities of hyperedges $\{1, 2, 3\}$,
$\{1, 2, 4\}$,
and $\{1, 2, 3, 4\}$ should be related.

\begin{figure}[t!]
  \centering
    \includegraphics[width=0.22\textwidth,trim=4cm 2.5cm 3cm 2.5cm]{figures/watergate_network_alpha.pdf}  
    \caption{
    \hide{
    A hypergraph with five U.S.~politicians, 
    where hyperedges represent subsets of politicians mentioned together in U.S.\ Newswire articles between 1900 and 1977. 
    }
    A hypergraph with five U.S.~politicians,
    which is a subgraph of a hypergraph with 678 U.S.\ politicians described in Section \ref{sec:application}.
    There are ten hyperedges of size two,
    five hyperedges of size three,
    and one hyperedge of size four,
    represented by colored contours.
    \hide{\alertC{There are 16 hyperedges (which are too many to be written out, they are very layered in the picture, I added a version of the hypergraph with higher opaqueness to see all hyperedges, which might be an option):  "Fulbright" "Nixon"    
[[2]] "Kennedy" "Nixon"  
[[3]] "Fulbright" "Kennedy"   "Nixon"    
[[4]] "Fulbright" "Kennedy"  
[[5]] "Fulbright" "Ford"     
[[6]]  "Kennedy" "Ford"   
[[7]] "Nixon" "Ford" 
[[8]] "Nixon"     "Kissinger"
[[9]]"Ford"      "Kissinger"
[[10]]"Kennedy"   "Nixon"     "Kissinger"
[[11]] "Fulbright" "Kissinger"
[[12]]"Fulbright" "Nixon"     "Kissinger"
[[13]] "Nixon"     "Ford"      "Kissinger"
[[14]]"Kennedy"   "Nixon"     "Ford"      "Kissinger"
[[15]] "Kennedy"   "Kissinger"
[[16]] "Kennedy"   "Ford"      "Kissinger"}}
}
\label{fig:hyperedge}
\end{figure}

Existing approaches to hypergraphs include hyperedge incidence models \citep{YuZh25},
random geometric hypergraphs with positions in Euclidean space \citep{turnbull2023latent},
stochastic block models \citep{ng2022model,yuan2022testing,brusa2024},
tensor-based models \citep{lyu2023latent,ghoshdastidar2014consistency,yuan2023high},
and hypergraph neural networks \citep{kim2024},
among other models \citep[e.g.,][]{wu2025, nandy2024}.
These approaches are either not scalable (i.e., these approaches can be applied to small hypergraphs but not large ones) or achieve scalability at the expense of model realism.
For example,
some approaches restrict attention to hyperedges of a fixed size \citep[e.g.,][]{lyu2023latent,ghoshdastidar2014consistency},
which neglects hyperedges of all other sizes and ignores the fact that small hyperedges are nested in larger ones.
Other approaches assume that realized hyperedges of all sizes are independent and identically distributed draws from a common probability law \citep[e.g.,][]{YuZh25},
which specifies a model for realized hyperedges but ignores all unrealized hyperedges.
Unrealized hyperedges can be both interesting and informative:
A failure to create a hyperedge can be as telling as a success.
Still other approaches reduce hypergraphs to bipartite graphs \citep{ng2022model,wu2025},
but such approaches condition on the observed number of hyperedges.

\paragraph*{Contributions}
We develop a statistical framework that enables scalable estimation, simulation, and model assessment of hypergraph models and is supported by non-asymptotic and asymptotic theoretical guarantees:
\begin{enumerate}
\item We introduce a novel class of hyperbolic space models for hypergraphs in Section \ref{sec:model}.\break
In contrast to existing approaches to hypergraphs,
we capture core-periphery structure in addition to proximity in hypergraphs.
Many hypergraphs exhibit core-periphery structure in the sense that some units are more central than others.
As such, capturing core-periphery structure is an important task in hypergraph modeling:
e.g.,
in collaboration networks some scientists lead labs while others are members of labs,
hence some scientists can be expected to be more central to the hypergraph than others.
While hyperbolic space models have been used to capture core-periphery structure in graphs \citep{krioukov_hyperbolic_2010,SmAsCa19,lubold2023identifying},
none of the cited publications tackles hypergraphs with $2^N - N - 1$ possible hyperedges of small and large sizes that can overlap or be nested.
\item We achieve scalability by estimating hypergraph models based on samples of hyperedges from a population hypergraph without placing unnecessary restrictions on the model,
i.e.,
we consider both realized and unrealized hyperedges of varying sizes
and we do not condition on the observed number of hyperedges.
Sample-to-population inference is an important problem in statistical network analysis \citep{Ko17} and is indispensable to the statistical analysis of large hypergraphs,
because the number of possible hyperedges $2^N - N - 1$ is exponential in $N$.
The large number of possible hyperedges has two implications.
First,
it may not be possible to observe the entire population hypergraph due to data collection constraints.
Second,
even when the entire population hypergraph is observed,
using observations of all $2^N - N - 1$ possible hyperedges may be infeasible due to computational constraints (including memory and computing time).
As a consequence,
regardless of whether the entire population hypergraph is observed,
scalable statistical inference requires sampling hyperedges from the population hypergraph of interest and inferring the mechanism that generated the population hypergraph from sampled hyperedges.
To prepare the ground for sample-to-population inference,
we first discuss sampling hyperedges from a population hypergraph in Section \ref{sec:sampling},
and then introduce scalable manifold optimization algorithms for learning hyperbolic space models from sampled hyperedges in Section \ref{sec:optimization}.
We complement scalable learning methods by scalable simulation methods in Section \ref{sec:simulation} to facilitate simulation studies and simulation-based model assessment.
\item We show in Section \ref{sec:identifiability} that the positions of units in hyperbolic space are identifiable up to rotations and establish non-asymptotic theoretical guarantees based on fixed $N$ along with asymptotic theoretical guarantees based on $N \to \infty$ in Section \ref{sec:theory}.
The theoretical results in Section \ref{sec:theory} are supported by simulation results in Section \ref{sec:simulation_study}.
\item We use hyperbolic space models in Section \ref{sec:application} to detect core-periphery structure along with proximity among U.S.\ politicians based on historical media reports.
\end{enumerate}
To pave the way for hyperbolic space models for hypergraphs,
we first provide background on hyperbolic space in Section \ref{sec:back}. 

\paragraph*{Notation}
Let $\mathscr{V} \coloneqq \{1, \ldots, N\}$ be a population consisting of $N \geq 3$ units.
We consider hyperedges $e \subseteq \mathscr{V}$ of sizes $|e| \in \mathscr{K} \coloneqq \{2, \ldots, K\}$,
where $K \in \{3, \ldots, N\}$ is chosen by the investigator independent of $N$:
For instance, 
we choose $\mathscr{K} = \{2, 3, 4\}$ in the application to U.S.\ politicians in Section \ref{sec:application},
because 89\% of the observed hyperedges among U.S.\ politicians is of size 2, 3, or 4 and therefore the observed hypergraph does not contain much information about hyperedges of sizes $5, \ldots, N$.
Throughout the remainder of the manuscript,
we assume that $\mathscr{K} \coloneqq \{2, \ldots, K\}$ with $K \in \{3, \ldots, N\}$,
although one could specify $\mathscr{K}$ as a proper subset of $\{2, \ldots, K\}$,
so that $\mathscr{K} \subset \{2, \ldots, K\}$.
The set of all possible hyperedges of size $k$ is denoted by $\mathscr{E}_k \coloneqq \{e:\; e \subseteq \mathscr{V},\; |e| = k\}$ ($k = 2, \ldots, K$) and the set of all hyperedges of all sizes is denoted by $\mathscr{E} \coloneqq \bigcup_{k=2}^K \mathscr{E}_k$.
Let $Z_{e} \in \{0, 1\}$ be an indicator of an hyperedge,
where $Z_e \coloneqq 1$ if hyperedge $e$ exists and $Z_e \coloneqq 0$ otherwise.
The collection of indicators $\bm{Z}_{\mathscr{E}} = (Z_e)_{e \in \mE}$ represents a hypergraph, 
while $\bm{Z}_{\mathscr{E}_k} = (Z_e)_{e \in \mE_k}$ refers to all hyperedges of size $k \in \{2, \ldots, K\}$.
For any vector $\bm{x} \coloneqq (x_1, \ldots, x_p) \in \mathbb{R}^p$ of dimension $p \geq 1$, 
the Euclidean norm is defined by $\vecnorm{\bm{x}} \coloneqq (\sum_{i=1}^p x_i^2)^{1/2}$.
For any matrix $\bm{A} \in \mathbb{R}^{p \times p}$,
the Frobenius norm is denoted by $\Fnorm{\bm{A}} \coloneqq (\sum_{i=1}^p \sum_{j=1}^p A_{i,j}^2)^{1/2}$. 
We write $a(n) = O(b(n))$ if there exists a finite constant $C > 0$ such that $|a(n)\, /\, b(n)| \leq C$ holds for all large enough $n$,
and $a(n) = O_p(b(n))$ if $|a(n)\, /\, b(n)| \leq C$ holds with a probability tending to $1$.
We write $a(n) = o(b(n))$ if,
for all $\epsilon > 0$,
$|a(n)\, /\, b(n)|\, \,<\, \epsilon$ holds for all large enough $n$,
and $f(n) \asymp g(n)$ if there exist finite constants $C_1 > 0$ and $C_2 > 0$ such that $C_1 \leq f(n)/g(n) \leq C_2 $ holds for all large enough $n$.

\section{Hyperbolic Space}
\label{sec:back}

\begin{figure}[t]
\centering
  \begin{subfigure}[c]{0.375\textwidth}   
    \adjustbox{valign=t}{\includegraphics[width=\textwidth]{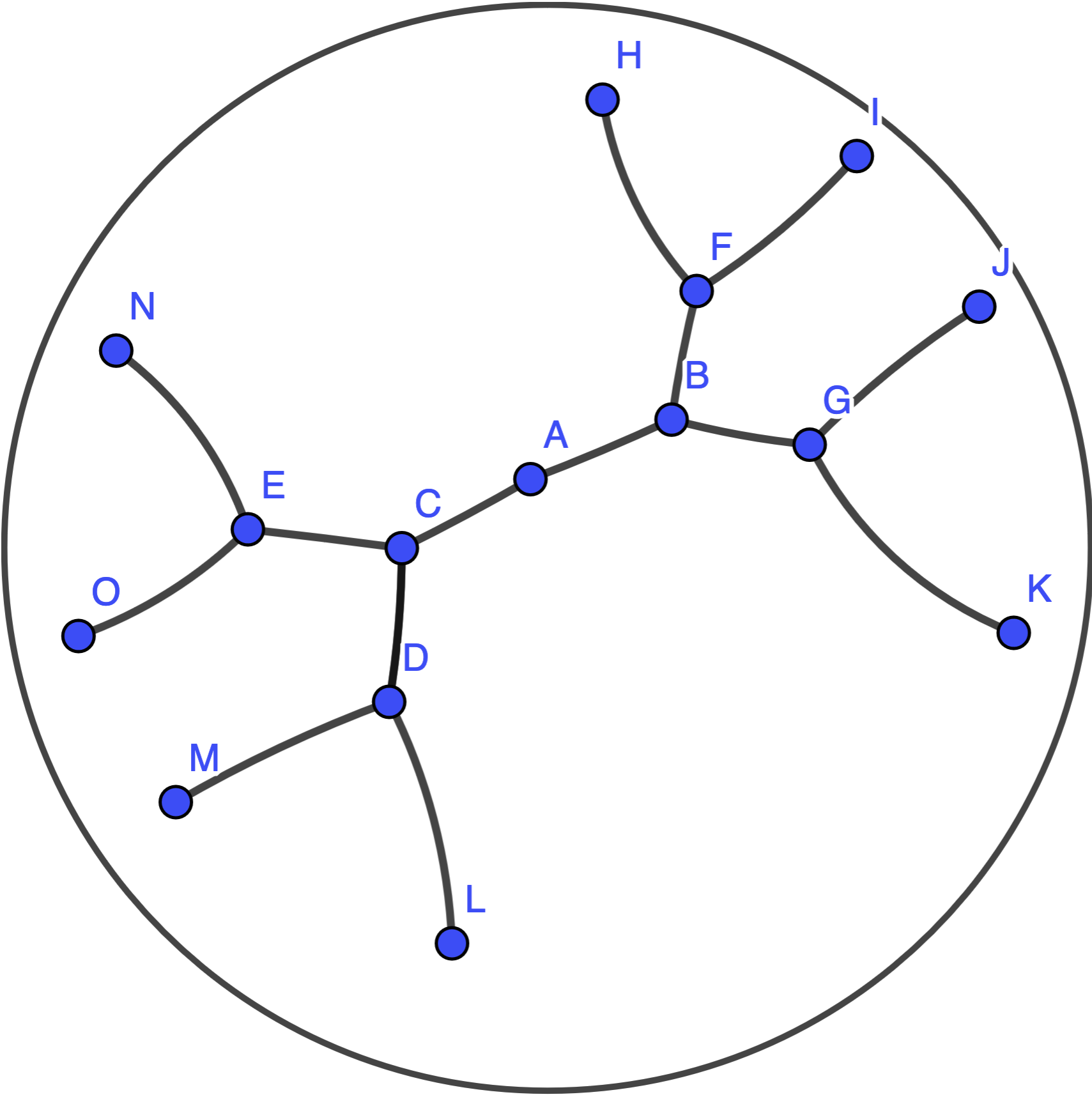}}
    \caption{\hangindent=1.7em \pd{ }with tree \\ ~}
    \label{fig:subfig2}
  \end{subfigure}\hspace{20mm}
   \begin{subfigure}[c]{0.375\textwidth}   
    \adjustbox{valign=t}{\includegraphics[width=\textwidth]{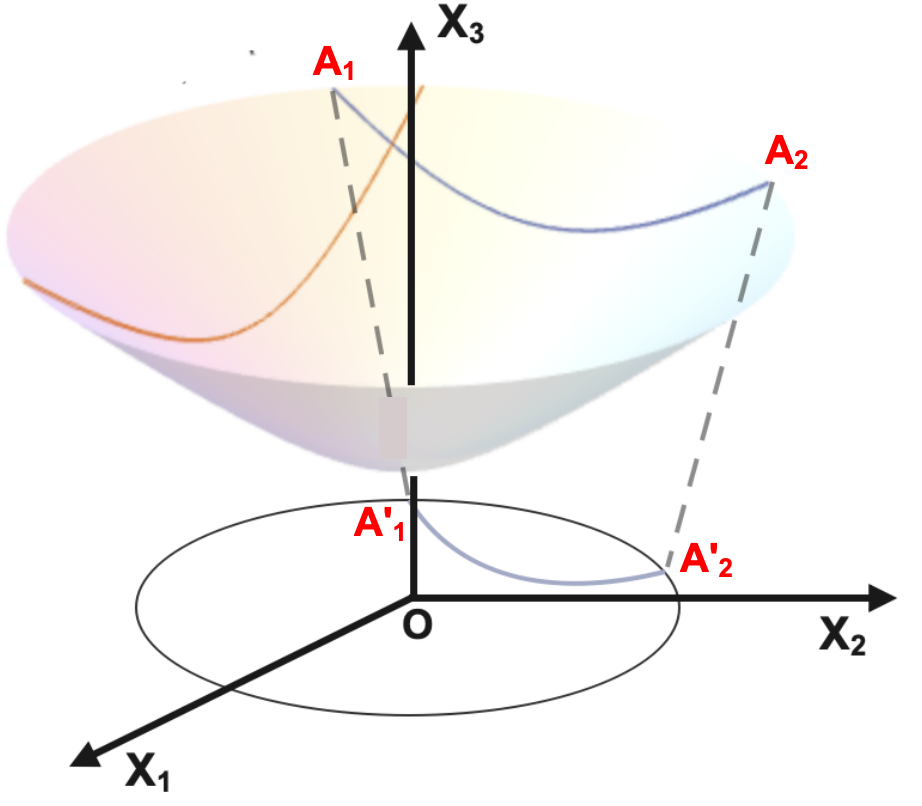}} 
    \vspace{17pt}
     \caption{\hangindent=1.7em \hspace*{-5pt}\; Equivalent representations of hyperbolic space}
    \label{fig:example}
  \end{subfigure}
  \caption{Hyperbolic space:
  (a) \pd, including an embedded tree with root $A$ close to the center of the \pd{ }and  leaves $H$, $I$, $J$, $K$, $L$, $M$, $N$, $O$ close to the boundary of the \pd.
  (b){ } Equivalent representations of hyperbolic space: \pd{ } and Lorentz model.
  The segment $A'_1$--$A'_2$ is the projection of the segment $A_1$--$A_2$ in the Lorentz model onto the \pd.
  }
  \label{fig:main_fig}
\end{figure}

To pave the way for hyperbolic space models for hypergraphs,
we first provide background on hyperbolic space in Section \ref{sec:back} and then introduce hyperbolic space models for hypergraphs in Section \ref{sec:model}.
To describe the properties of hyperbolic space,
we focus on two-dimensional hyperbolic space,
although we will consider $r$-dimensional hyperbolic space ($r \geq 2$) throughout the remainder of the manuscript.
Hyperbolic space has multiple equivalent representations,
including the \pd,
which facilitates interpretation and visualization;
and the Lorentz model,
which facilitates computing.
These two representations of hyperbolic space are visualized in Figure \ref{fig:main_fig}.

\paragraph{\pd} 
The \pd{ }with radius $1$ is defined by
\beno
\mathscr{P} 
&\coloneqq& \{\bm{x}\, \in\, \mathbb{R}^2:\, |\!|\bm{x}|\!|_2 < 1\}.
\ee
The distance between two points $\bm{x}_1 \in \mathscr{P}$ and $\bm{x}_2 \in \mathscr{P}$ on the \pd{ }is
\beno
d_{\mathscr{P}}(\bm{x}_1,\, \bm{x}_2)
\coloneqq \arcosh\left(1 + \dfrac{2 \, \vecnormsqrt{\bm{x}_1 - \bm{x}_2}}{\left(1-\vecnormsqrt{\bm{x}_1}\right) \,\left(1-\vecnormsqrt{\bm{x}_2}\right)}\right),
\ee
where $\mathop{\arcosh}(a) \coloneqq \log(a + \sqrt{a^2 -1})$ ($a \geq  1$).
A useful property of hyperbolic space is its curvature.
The curvature quantifies how much the space deviates from a flat surface  \citep{boumal2023introduction}:
While the curvature of Euclidean space is $0$, 
the curvature of hyperbolic space is negative.
The negative curvature of hyperbolic space implies that hyperbolic distance grows as one approaches the boundary of the \pd.
As a result,
hyperbolic space can embed trees into a low-dimensional space as demonstrated in Figure \ref{fig:main_fig} (a),
helping capture core-periphery structure along with local structure in hypergraphs. 

\paragraph{Lorentz model}
The Lorentz model offers an alternative representation of hyperbolic space and facilitates computing, 
by providing closed-form representations of geodesics and tangent space \citep{nickel2018learning}. 
The Lorentz model of hyperbolic space is given by 
\beno
\mathscr{L}
&\coloneqq& \left\{\pos\, \in\, \mathbb{R}^{r+1}:\, \left\langle\pos,\, \pos\right\rangle_{\mathscr{L}}=-1,\;\, \theta_{1}>0\right\},
\ee
where $\left\langle\pos,\, \pos\right\rangle_{\mathscr{L}} \coloneqq -\theta_{1}^2 + \sum_{i=2}^{r+1}\, \theta_i^2$.
The associated distance function on $\mathscr{L}$ is 
\beno
d_{\mathscr{L}}(\pos_1,\, \pos_2)
&\coloneqq& \arcosh\left(-\prodL{\pos_1,\, \pos_2}\right). 
\ee

\section{Hyperbolic Space Models for Hypergraphs}
\label{sec:model}

We introduce hyperbolic space models for hypergraphs,
assuming that the units $1, \ldots, N$ of a hypergraph have positions $\pos_1, \ldots, \pos_N$ in $r$-dimensional hyperbolic space $\mathscr{L}$.
In line with other latent space models \citep[e.g.,][]{HpRaHm01,krioukov_hyperbolic_2010},
we assume that the dimension $r \geq 2$ is a known constant independent of $N$,
chosen by the investigator based on domain knowledge or ease of visualization:
e.g.,
in the application to U.S.\ politicians in Section \ref{sec:application},
we choose $r = 2$ to facilitate the visualization of results in a two-dimensional hyperbolic space.
Recently,
\citet{LoCh25} introduced a principled Bayesian approach for selecting the dimension $r$ of a generalized linear network eigenmodel,
which is a latent space model for graphs rather than hypergraphs and is not based on hyperbolic geometry.
Having said that,
a data-driven approach to selecting the dimension $r$ of other latent space models,
for either graphs or hypergraphs,
is an open problem.

We assume that the indicators $Z_e$ of hyperedges $e$ are independent Bernoulli$(\pi(\alpha_{|e|},\, \Pos_{e}))$ random variables conditional on sparsity parameters $\alpha_{|e|}$ and positions $\Pos_{e} \coloneqq (\pos_i)_{i \in e}$ of units involved in hyperedge $e$:
\beno
\label{eq:model}
Z_e 
\mid \alpha_{|e|},\, \Pos_{e} 
&\ind& \text{Bernoulli}(\pi(\alpha_{|e|},\, \Pos_{e})),
\ee
where $\pi(\alpha_{|e|}, \Pos_{e})$ is the probability of observing hyperedge $e$.
The probability $\pi(\alpha_{|e|}, \Pos_{e})$ of observing hyperedge $e$ is of the form
\beno
\label{eq_2}
\pi(\alpha_{|e|},\, \Pos_{e})
&\coloneqq& \alpha_{|e|}\, \sigma\left(- g(\Pos_{e})\right),
\ee
where 
\beno
\label{sigma.def}
\sigma(x) 
&\coloneqq& \dfrac{2\, \exp(x)}{1+\exp(x)},
& x \in \mathbb{R}.
\ee
The parameter $\alpha_{|e|} \in(0,\, 1]$ determines the sparsity of hyperedges $e$ of size $|e|$,
while the function $g: \Pos_{e} \mapsto [0,\, \infty)$ specified in Section \ref{sec:model.specification1} captures core-periphery structure along with local structure in hypergraphs.
The factor $2$ ensures that $\pi(\alpha_{|e|}, \Pos_{e}) \in (0,\, 1)$.

\subsection{Model Specification}
\label{sec:model.specification1}

To specify the function $g(\Pos_{e})$,
let
\beno
\label{eq:dd}
d^{(e)}_{i}(\Pos_e)
&\coloneqq& \dsum_{j \in e \setminus \{i\}}\, d_{\mathscr{L}}(\pos_i,\, \pos_j)
\ee
be the sum of distances of unit $i \in e$ to other units $j \in e \setminus \{i\}$ involved in hyperedge $e$.\break
We introduce a flexible family of hyperbolic distance models by specifying $g(\Pos_{e})$ as the H\"older mean of $d^{(e)}_i(\Pos_e)$ ($i \in e$):
\be
\label{eq:conc}
g(\Pos_{e}) 
\;\coloneqq\; \left\{\dfrac{1}{|e|}\, \dsum_{i \in e} \left[d^{(e)}_i(\Pos_e)\right]^p \right\}^{1/p},
\ee 
where $p \in \mathbb{R} \setminus \{0\}$;
note that $g(\Pos_{e})$ is undefined when $p = 0$ because $1/0$ is undefined.
We assume that $p$ is a constant independent of $N$.
Depending on the choice of $p$,\,
a wide range of distance-based functions $g(\Pos_{e})$ can be specified:
e.g.,
$g(\Pos_{e})$ approaches $\min_{i \in  e} \{d^{(e)}_i(\Pos_{e})\}$ as $p \rightarrow - \infty$ and $\max_{i \in  e} \{d^{(e)}_i(\Pos_{e})\}$ as $p \rightarrow  +\infty$,
while $p=-1$ gives the harmonic mean,
$p\rightarrow 0$ the geometric mean,
and $p=+1$ the arithmetic mean.
While any choice of $p \in \mathbb{R} \setminus \{0\}$ is legitimate,
we focus on $p \ll 0$ motivated by two considerations:
\bi
\item {\bf Core-periphery structure}
Specifications of $g(\Pos_{e})$ with $p \ll 0$ allow hyperedges to emerge from a hierarchical core-periphery structure resembling a tree,
whose root is a central unit close to the center of the \pd{ }and whose leaves are peripheral units close to the boundary of the \pd.
The ability of $g(\Pos_{e})$ with $p \ll 0$ to capture hierarchical core-periphery structure in the form of a tree is demonstrated in Figure \ref{fig:main_fig} (a).
Specifications of $g(\Pos_{e})$ with $p \ll 0$ are better suited to capturing hierarchical core-periphery structure than specifications with $p \gg 0$:
While specifications with $p \gg 0$ require each unit in a hyperedge to be close to all other units involved in the hyperedge,
specifications with $p \ll 0$ require each unit in a hyperedge to be close to one unit,
but not necessarily close to all other units involved in the hyperedge.
\item {\bf Pooling strength across overlapping and nested hyperedges}
The function $g(\Pos_{e})$ helps pool strength across overlapping and nested hyperedges of different sizes:
e.g.,
if $e = \{i, j\}$ and $e' = \{i, j, k\}$,
the distance $d_{\mathscr{L}}(\pos_i,\, \pos_j)$ between units $i$ and $j$ is contained in both $g(\Pos_{e})$ and $g(\Pos_{e'})$.
As a result,
the probabilities of hyperedges $e$ and $e'$ are related,
and the proposed model is able to leverage information from overlapping and nested hyperedges,
helping estimate the distance between $i$ and $j$.
\ei
In the simulations and application in Sections \ref{sec:simulation_study} and \ref{sec:application},
we choose $p = -20$,
for two reasons.
First, 
the function $g(\Pos_{e})$ with $p = -20$ provides a reasonable approximation of $\min_{i \in  e} \{d^{(e)}_i(\Pos_{e})\}$ and is differentiable,
in contrast to $\min_{i \in  e} \{d^{(e)}_i(\Pos_{e})\}$.
Second, 
the choice $p = -20$ avoids numerical instability (e.g., overflow) in evaluations of $g(\Pos_{e})$,
which can occur when $p$ is too large in absolute value.

\subsection{Identifiability}
\label{sec:identifiability}

The hyperbolic distances among units are determined by the Gram matrix $\bm{D} \coloneqq \bm{\Theta}\, \bm{J}\, \bm{\Theta}^\top$, 
where $\bm{J}~\coloneqq~\text{diag}(-1,\, \bm{1}_r) \in \mathbb{R}^{(r+1)\times(r+1)}$ and $\bm{1}_r ~\coloneqq~(1, \ldots, 1)\in \mathbb{R}^r$, 
whereas $\bm{\Theta} \coloneqq (\bm{\theta}_1^\top, \ldots, \bm{\theta}_N^\top)^\top$ stores the positions $\bm\theta_1, \ldots, \bm\theta_N \in \mathscr{L}$ of units $1, \ldots, N$.
The following proposition establishes identifiability of the Gram matrix $\bm{D}$ and sparsity parameters $\alpha_2, \ldots, \alpha_K$,
assuming that $K \geq 3$,\,
$p \in \mathbb{R} \setminus \{0\}$,\,
and $r \geq 2$ are constants independent of $N$;
recall that $g(\Pos_{e})$ is undefined when $p = 0$. 

\begin{prop}{\bf Identifiability}
\label{prop1}
The Gram matrix $\bm{D} \coloneqq \bm{\Theta}\, \bm{J}\, \bm{\Theta}^\top$ and the sparsity parameters $\alpha_2, \ldots, \alpha_K$ are identifiable provided that $N>r+2$.
\end{prop}
A proof of Proposition \ref{prop1} can be found in Supplement \ref{sec:proof_prop1}. 
Proposition \ref{prop1} assumes that the sparsity parameters $\alpha_2, \ldots, \alpha_K$ satisfy $\alpha_k \in (0,\, 1]$ ($k = 2, \ldots, K$) and holds when constraints on sparsity parameters are imposed,
such as $\alpha_2 > \ldots > \alpha_K$.

Since $\bm{D}$ is identifiable, 
the positions of units can be identified up to a hyperbolic rotation matrix $\bm{R}$ such that $\bm{R}\, \bm{J}\, \bm{R}^\top = \bm{J}$. 
In other words, 
we can identify an equivalence class of positions,
$\{\bm{\Theta}\, \bm{R} \mid \bm{R}\, \bm{J}\, \bm{R}^\top = \bm{J}\}$. The rotation matrix $\bm{R}$ is the hyperbolic counterpart of the orthogonal transformation on Euclidean space. 
For example, 
when $r=2$, 
the hyperbolic rotation $\bm{R}$ is
\beno 
\mathbf{R} = \begin{bmatrix} 
  1 & 0 & 0\\
   0 & \cos x&  -\sin x \\
  0  &\sin x & \cos x  \\
\end{bmatrix}
\begin{bmatrix}
  \cosh x  &0  &\sinh x \\
  0 & 1 & 0 \\
  \sinh &x  &0  \cosh x
\end{bmatrix}
\begin{bmatrix}
  \cosh x&  \sinh x&  0 \\
  \sinh x & \cosh x &  0 \\
   0 & 0&  1 
\end{bmatrix}.
\ee
To address the rotation invariance,
we perform an eigenvalue decomposition:
$\bm{D} = \bm{U} \bm{S}\, \bm{U}^\top$.
Since $\bm{D}$ has one negative and $r$ positive eigenvalues \citep{tabaghi2020hyperbolic}, $\bm{S} = \text{diag}(\lambda_{1}, \ldots, \lambda_{r+1}) \in \mathbb{R}^{(r+1)\times (r+1)}$ is a diagonal matrix with all nonzero eigenvalues of $\bm{D}$ on the diagonal and $\bm{U} \in \mathbb{R}^{N \times (r+1)}$ is a matrix of corresponding eigenvectors. 
The eigenvalues are assumed to have the following order: $\lambda_1$ is the negative eigenvalue and $\lambda_2, \ldots, \lambda_{r+1}$ are all positive eigenvalues in decreasing order. 
Within this equivalence class, 
we fix positions $\widetilde{\bm{\Theta}} := \bm{U}|\,\bm{S}|^{1/2}\bm{J} \in \{\bm{\Theta}\, \bm{R} \mid \bm{R}\, \bm{J}\, \bm{R}^\top = \bm{J}\}$, where $|\bm{S}|^{1/2}$ denotes the diagonal matrix whose entries are the square root of the absolute values of the eigenvalues in $\bm{S}$. 
Thus, 
the transformed latent positions $\widetilde{\bm{\Theta}}$ are identifiable.

\section{Sample-to-Population Estimation}
\label{sec:comp}

Sample-to-population estimation is indispensable,
because the number of possible hyperedges $2^N - N - 1$ is exponential in $N$,
which implies that collecting data on all possible hyperedges and using all of them for the purpose of statistical learning may not be feasible.
While reducing the statistical analysis to hyperedges $e$ of size $|e| \leq K$ alleviates the problem,
it does not solve it:
e.g.,
in the application in Section \ref{sec:application} with $N = 678$ units,
there are more than 1 trillion possible hyperedges of size $|e| \leq 5$ and 8 billion possible hyperedges of size $|e| \leq 4$.
A body of data of such size is too large to be analyzed in its entirety,
necessitating sample-to-population estimation.

To prepare the ground for sample-to-population estimation,
we first describe methods for sampling from hypergraphs in Section \ref{sec:sampling}.
We then introduce scalable statistical methods for learning hyperbolic space models based on samples in Section \ref{sec:optimization},
complemented by scalable simulation methods in Section \ref{sec:simulation}.

\subsection{Sampling from Hypergraphs}
\label{sec:sampling}

To sample from a hypergraph,
one can adapt sampling designs for graphs \citep[e.g.,][Section 5.2]{Ko17,ScKrBu17} 
to hypergraphs.
At least two broad classes of sampling designs can be distinguished:
\bi 
\item {\bf Indirect sampling} generates a probability sample of hyperedges $e \in \mathscr{E}$ by first generating a probability sample of units $i \in \mathscr{V}$ and then collecting data on hyperedges of sampled units and those connected to them.
Examples are ego-centric sampling and link-tracing adapted to hypergraphs.
\item {\bf Direct sampling} generates a probability sample of hyperedges $e \in \mathscr{E}$,
where the sample inclusion probabilities of hyperedges $e$ may depend on the size $|e|$, 
status $Z_e \in \{0,\, 1\}$, 
and observed attributes of hyperedges $e$.
\ei 
While either direct or indirect sampling can be used,
indirect sampling is less convenient than direct sampling for the purpose of statistical learning,
for two reasons.
First, 
sample loss functions based on Horvitz-Thompson estimators require the sample inclusion probabilities of hyperedges,
which are more involved when sampling is indirect rather than direct.
Second, 
indirect sampling is less developed for hyperedges than graphs \citep{luo2024}.

In the simulations and application in Sections \ref{sec:simulation_study} and \ref{sec:application},
we use direct sampling,
by sampling hyperedges $e$ with sample inclusion probabilities
\bi
\item $\mu_e = n\, |\mE_k^{(1)}|/|\mE_k^{(0)}|$ for all $e \in \mathscr{E}_k^{(0)}$ with $1 \leq n \leq |\mE_k^{(0)}|/|\mE_k^{(1)}|$,
\item $\mu_e = 1$ for all $e \in \mathscr{E}_k^{(1)}$,
\ei
where $\mE_k^{(0)} \coloneqq \{e \in \mE_k:\; Z_e = 0\}$ is the set of unrealized hyperedges of size $k$ and $\mE_k^{(1)} \coloneqq \{e \in \mE_k:\; Z_e = 1\}$ is the set of realized hyperedges of size $k$ ($k = 2, \ldots, K$).

The described sampling design is motivated by the observation that many real-world networks are sparse.
As a result, 
the number of realized hyperedges is small relative to the number of unrealized hyperedges.
We therefore sample all realized hyperedges and,
for each realized hyperedge,
we sample $n$ unrealized hyperedges,
called controls.

\subsection{Learning From Samples}
\label{sec:optimization}

The target of statistical inference is the positions $\bm\Theta$ of units in hyperbolic space along with the sparsity parameters $\alpha_2, \ldots, \alpha_K$,
which we denote by $\bm\Lambda \coloneqq (\bm\Theta, \alpha_2, \ldots, \alpha_K)$.
To learn $\bm\Lambda$ from data,
we first introduce a population loss and then introduce a sample loss.

\paragraph*{Population loss}
A natural loss function is the negative loglikelihood function.
If $\mathscr{E}_k^{(0)}$ and $\mathscr{E}_k^{(1)}$ are observed ($k = 2, \ldots, K$),
the negative loglikelihood function is
\beno
\label{eq:like}
\ell(\bm\Lambda)
&\coloneqq& {\ell_0(\bm\Lambda)} \,+\, {\ell_1(\bm\Lambda)},
\ee
where 
\beno 
\ell_0(\bm\Lambda) 
&\coloneqq& -\dsum_{k = 2}^K\; \dsum_{e \in \mE_k^{(0)}} \log(1 - \pi(\alpha_{|e|}, \Pos_{e}))
\\
\ell_1(\bm\Lambda) 
&\coloneqq& -\dsum_{k = 2}^K\; \dsum_{e \in \mE_k^{(1)}}\, \log \pi(\alpha_{|e|}, \Pos_{e}).
\ee
Since the number of possible hyperedges is $\binom{N}{k} = O(N^k)$,
observing $\mathscr{E}_k^{(0)}$ and $\mathscr{E}_k^{(1)}$ ($k = 2, \ldots, K$) and minimizing the population loss $\ell(\cdot)$ may be infeasible when $N$ or $K$ are large.
We hence approximate the population loss $\ell(\cdot)$ by a sample loss $\widehat\ell(\cdot)$ defined below.

\paragraph*{Sample loss}
If samples $\mathscr{S}_k^{(0)}$ and $\mathscr{S}_k^{(1)}$ from $\mathscr{E}_k^{(0)}$ and $\mathscr{E}_k^{(1)}$ ($k = 2, \ldots, K$) are available,
one can approximate the population loss $\ell(\cdot)$ by the sample loss $\widehat\ell(\cdot)$ defined by
\beno
\label{eq:approx_llh}
\widehat\ell(\bm\Lambda)
&\coloneqq& \widehat\ell_0(\bm\Lambda) \,+\, \widehat\ell_1(\bm\Lambda),
\ee
where $\widehat\ell_0(\cdot)$ and $\widehat\ell_1(\cdot)$ are Horvitz-Thompson estimators of the corresponding population quantities $\ell_0(\cdot)$ and $\ell_1(\cdot)$:
\beno 
\widehat\ell_0(\bm\Lambda)  
~\coloneqq~ - \dsum_{k = 2}^K\; \dsum_{e \in \mathscr{S}_k^{(0)}}\,
\dfrac1{\mu_e}\, \log (1 - \pi(\alpha_{|e|}, \Pos_{e}))
\\
\widehat\ell_1(\bm\Lambda)  
 ~\coloneqq~ - \dsum_{k = 2}^K\; \dsum_{e \in \mathscr{S}_k^{(1)}}\,
 \dfrac1{\mu_e} \log\, \pi(\alpha_{|e|}, \Pos_{e}).
\ee
The weights $\mu_e \in [0,\, 1]$ are the sample inclusion probabilities of hyperedges $e \in \mathscr{E}$.
We consider them known and allow them to depend on 
the size $|e|$, 
status $Z_e \in \{0,\, 1\}$, 
and observed attributes of hyperedges $e$.

We minimize the sample loss $\widehat\ell(\cdot)$ by blockwise optimization,
cycling through Riemannian gradient descent updates of $\bm\theta_1, \ldots, \bm\theta_N$ and Quasi-Newton updates of $\alpha_2, \ldots, \alpha_K$.
The minimization of $\widehat\ell(\cdot)$ is based on the Lorentz model of hyperbolic geometry,
because it provides closed-form representations of geodesics and tangent space \citep{nickel2018learning}.
While the minimization of $\widehat\ell(\cdot)$ could be based on the \pd,
the \pd{} is less convenient than the Lorentz model,
because it lacks closed-form representations of geodesics and tangent space and hence would necessitate additional approximations.

\paragraph*{Riemannian gradient descent updates of $\bm\theta_1, \ldots, \bm\theta_N$}
We update the positions\break 
$\pos_1, \ldots, \pos_N \in \mathscr{L}$ of units $1, \ldots, N$ by minimizing $\widehat{\ell}(\cdot)$ using Riemannian gradient descent \citep[Chapter 4]{absil2008}.
At iteration $t+1$,
we update the position $\pos_i^{(t)}$ of unit $i$ to
\beno
\label{eq:updated_pos}
\pos_i^{(t+1)}
&=& \exp_{\pos_{i}^{(t)}}\left(-\eta_i^{(t+1)}\, \text{proj}_{\pos_i^{(t)}}(\bm \vartheta_i)\right),
\ee
where $\bm\vartheta_i \coloneqq \bm{J}\,  \nabla_{\pos_i}\, \widehat\ell(\bm\Lambda)|_{\pos_i = \pos_i^{(t)}}$.
The gradient $\nabla_{\pos_i}\,\widehat\ell(\bm\Lambda)$ is derived in  Supplement \ref{sec:gradient}.
The functions $\exp_{\pos}(\bm{x})$ and $\text{proj}_{\pos}(\bm{x})$
refer to the exponential map and the orthogonal projection onto the tangent space of $\pos$ with respect to the Lorentz model:
\beno 
\exp_{\pos}(\bm{x}) &\coloneqq & \cosh(\prodL{\boldsymbol{x},\, \boldsymbol{x}})\, \pos + \dfrac{\sinh(\prodL{\boldsymbol{x},\, \boldsymbol{x}})\,\boldsymbol{x}}{\prodL{\boldsymbol{x},\, \boldsymbol{x}}}\s
\\
\text{proj}_{\pos}(\bm{x}) &\coloneqq &\bm{x} + \prodL{\pos,\, \bm{x}}\,\pos.
\ee
The learning rate $\eta_i^{(t+1)} \in [0,\, \infty)$ is computed using the algorithm of \citet{brentAlgorithmGuaranteedConvergence1971}.

\paragraph*{Quasi-Newton updates of $\alpha_2, \ldots, \alpha_K$}
We update sparsity parameters $\alpha_2, \ldots, \alpha_K$ by minimizing $\widehat\ell(\cdot)$.
Setting the partial derivative of $\widehat\ell(\cdot)$ with respect to $\alpha_k$ to $0$ gives
\beno
\label{fun}
\dsum_{e\, \in \mathscr{S}_k^{(0)} \cup\, \mathscr{S}_k^{(1)}}\,  \dfrac1{\mu_e} \left( \dfrac{z_{e}}{\alpha_k} - \dfrac{\sigma\left(- g(\Pos_{e})\right) (1-z_e)}{1 - \alpha_k\, \sigma\left(- g(\Pos_{e} )\right)}\right)
&=& 0,
\ee
which can be solved using the Quasi-Newton method of \citet{byrd_limited_1995} subject to the constraint $\alpha_k \in (0,\, 1]$ ($k = 2, \ldots, K$).

\paragraph*{Remark 1 Convergence and uniqueness of minimizers}
The described updates of $\pos_1, \ldots, \pos_N$ and $\alpha_2, \ldots, \alpha_K$ decrease the sample loss $\widehat{\ell}(\cdot)$.
We declare convergence when $|\widehat\ell({\bm\Lambda}_t) - \widehat\ell(\bm\Lambda_{t+1})|\, /\, |\widehat\ell(\bm\Lambda_{t+1})| < 10^{-5}$, 
where $\bm\Lambda_t$ and ${\bm\Lambda}_{t+1}$ are the values of $\bm\Lambda$ at iterations $t$ and $t+1$,
respectively.
The sample loss $\widehat{\ell}(\cdot)$ is non-convex,
so minimizers of $\widehat{\ell}(\cdot)$ may not be unique.
In the simulations and applications in Sections \ref{sec:simulation_study} and \ref{sec:application},
we therefore use multiple starting values chosen at random as described in Supplement \ref{sec:initialization},
and report the results based on the smallest sample loss.

\paragraph*{Remark 2 Computational complexity}
The computational complexity of algorithms minimizing the sample loss $\widehat{\ell}(\cdot)$ is dominated by the number of sampled hyperedges, 
denoted by $\mathscr{S}_{\max} \coloneqq \max_{2 \leq k \leq K}\, |\mathscr{S}_{k}^{(0)} \cup\, \mathscr{S}_{k}^{(1)}|$.
Approximating $\ell(\cdot)$ by $\widehat{\ell}(\cdot)$ comes at a computational cost of $O(\mathscr{S}_{\max})$ operations. 
Updating positions requires $O(N\, \mathscr{S}_{\max})$ operations, 
whereas updating sparsity parameters requires $O(\mathscr{S}_{\max})$ operations. 

\hide{
\paragraph*{Convergence criterion}
We declare convergence when $|\widehat\ell({\bm\Lambda}_t) - \widehat\ell(\bm\Lambda_{t+1})|\, /\, |\widehat\ell(\bm\Lambda_{t+1})| < 10^{-5}$, 
where $\bm\Lambda_t$ and ${\bm\Lambda}_{t+1}$ are the values of $\bm\Lambda$ at iterations $t$ and $t+1$.

\paragraph*{Non-convexity of $\ell(\cdot)$ by $\widehat{\ell}(\cdot)$}
The loss functions $\ell(\cdot)$ and $\widehat{\ell}(\cdot)$ are not convex,
so that the minimizers of $\ell(\cdot)$ and $\widehat{\ell}(\cdot)$ are not guaranteed to be unique. On the other hand, given the strong convexity of negative log-likelihood and identifiability of model parameters from $\pi(\alpha_{|e|},\, \Pos_{e})$ established in Proposition 1, both $\ell(\cdot)$ and $\widehat{\ell}(\cdot)$ are locally convex around local minimizer. In applications,
we therefore use multiple starting values chosen at random and report the results with the smallest loss.

\paragraph*{Computational complexity}
\alert{\bf Why were the last three paragraphs outcommented? Why is the text of "Computational complexity" gone?}
}

\subsection{Simulating Hypergraphs}
\label{sec:simulation}

To complement the scalable learning methods in Section \ref{sec:optimization},
we develop scalable simulation methods for generating hypergraphs,
with an eye to facilitating large-scale simulation studies and simulation-based model assessment \citep{HuGoHa08}.

Direct simulation of hypergraphs requires enumerating all $N \choose k$ possible hyperedges $e$ of size $k \in \{2, \ldots, K\}$,
which may be infeasible when $N$ or $K$ are large.
We propose a scalable method for generating hyperedges of size $k \in \{2, \ldots, K\}$ in two steps:
\bi
\item[] {\bf Step 1} Generate $U_k$,
the number of hyperedges of size $k$.
\item[] {\bf Step 2} Conditional on $U_k = u_k$,
generate $u_k$ hyperedges of size $k$.
\ei
We describe these steps in turn.

\paragraph*{Step 1 Generate the number of hyperedges $U_k$}
The number of hyperedges $U_k$ of size $k$ is distributed as
\beno
\label{eq:exact}
U_k
&\sim& \text{Poisson-Binomial}(\pi(\alpha_{k}, \Pos_{e}),\; e \in \mathscr{E}_k).
\ee
To generate $U_k$ from $\text{Poisson-Binomial}(\pi(\alpha_{k}, \Pos_{e}),\; e \in \mathscr{E}_k)$,
one needs to evaluate the probabilities $\pi(\alpha_{k}, \Pos_{e})$ of all $N \choose k$ hyperedges $e$ of size $k$,
which may be infeasible when $N$ or $k$ are large.
A scalable alternative is suggested by combining an inequality based on Stein's method \citep[Theorem 1,][]{barbour1984} with 
$\pi(\alpha_{k}, \Pos_{e}) \leq \alpha_k$,
which shows that the total variation distance $|\!|\mathbb{P} - \mathbb{Q}|\!|_{\mbox{\tiny TV}}$ between the Poisson-Binomial distribution $\mathbb{P}$ and the Poisson distribution $\mathbb{Q}$ with mean $\lambda_k \coloneqq \sum_{e\in \mathcal{E}_k} \pi(\alpha_{k}, \Pos_{e})$ vanishes as long as the hypergraph is sparse in the sense that $\alpha_k = o(1)$:
\beno 
\label{eq:lecam}
|\!|\mathbb{P} - \mathbb{Q}|\!|_{\mbox{\tiny TV}}
&\leq& \dfrac{\sum_{e\in \mathcal{E}_k} \pi(\alpha_{k}, \Pos_{e})^2}{\sum_{e\in \mathcal{E}_k} \pi(\alpha_{k}, \Pos_{e})}
\hide{
&\leq& \alpha_{k}^2\, \displaystyle{N \choose k}
}
&\leq& \alpha_{k}
&=& o(1).
\ee
As a result,
$U_k$ can be generated from $\text{Poisson}(\lambda_k)$ as long as the hypergraph is sparse,
as many real-world hypergraphs are.
Denoting by $\rho_k(\alpha_k, \Pos)$ the average probability of hyperedges of size $k$,
we can express the mean $\lambda_k$ of Poisson$(\lambda_k)$ as $\lambda_k = \rho_k(\alpha_k, \Pos)\, {N \choose k}$.
To estimate $\rho_k(\alpha_k, \Pos)$,
we sample $S \ll N$ units and estimate the mean $\rho_k(\alpha_k, \Pos)$ by the corresponding sample mean $\widehat\rho_k(\alpha_k, \Pos)$.
We then generate the number of hyperedges $U_k$ from $\mbox{Poisson}(\widehat\lambda_k)$ with mean $\widehat\lambda_k \coloneqq \widehat\rho_k(\alpha_k, \Pos) \, {N \choose k}$.

\paragraph*{Step 2 Generate $u_k$ hyperedges conditional on $U_k = u_k$}
Conditional on $U_k = u_k$,
we generate a candidate $e \in \mathscr{E}_k$ from proposal distribution $q(e) \coloneqq 1/{N\choose k}$ and accept it with probability $\pi(\alpha_{k}, \Pos_{e}) /(C\, q(e)) = \pi(1, \Pos_{e})$,
where $C \coloneqq \alpha_{k}\, {N \choose k}$ is a convenient upper bound on $\max_{e \in \mathcal{E}_k} \pi(\alpha_{k}, \Pos_{e}) / q(e)$.
We continue sampling until $u_k$ hyperedges 
are accepted.

\section{Theoretical Guarantees}
\label{sec:theory}

We state non-asymptotic and asymptotic theoretical guarantees for minimizers of the sample loss $\widehat\ell(\cdot)$ based on a sample of hyperedges $\mathscr{S} \coloneqq \bigcup_{k=2}^K\, \mathscr{S}_k$, 
consisting of samples $\mathscr{S}_k \coloneqq \mathscr{S}_k^{(0)} \cup \mathscr{S}_k^{(1)}$ of hyperedges of size $k$ ($k = 2, \ldots, K$).
We denote the minimizers of the sample loss $\widehat\ell(\cdot)$ by $\widehat{\bm{\Theta}}$ and $\widehat{\alpha}_2, \ldots, \widehat\alpha_K$,
which are estimators of the true values $\bm{\Theta}^\star$ and $\alpha_2^\star, \ldots, \alpha_K^\star$.
An estimator of the true Gram matrix $\bm{D}^\star \coloneqq \bm\Theta^\star\, \bm{J}\, (\bm\Theta^\star)^\top$ is $\widehat{\bm{D}} \coloneqq \widehat{\bm{\Theta}}\, \bm{J}\,\widehat{\bm{\Theta}}^\top$. 
We provide theoretical guarantees for these estimators assuming that $K \geq 3$,\,
$p \in \mathbb{R} \setminus \{0\}$,\,
and $r \geq 2$ are constants independent of $N$;
recall that $g(\Pos_{e})$ is undefined when $p = 0$.

\s

\begin{assumption}
\label{as:alpha}
{\em  
There exists a sequence of real numbers $\rho_N \in (0,\, 1]$ such that $\alpha_k^\star \in [\rho_N,\, 1]$ ($k = 1, \ldots, K$),
where $\rho_N$ is a non-increasing function of $N$.
}
\end{assumption}

\begin{assumption}
\label{as:theta}
{\em  
There exists a constant $C_1 \in (0,\, \infty)$ such that the elements $\theta_{i,k}^\star$ of\, $\bm\Theta^\star$ satisfy $\max_{1 \leq i \leq N} \max_{2 \leq k \leq r+1} |\theta_{i,k}^\star| \leq C_1$.
}
\end{assumption}

\begin{assumption}
\label{as:gram}
{\em  
There exists a constant $C_2 \in (1,\, \infty)$ such that the elements $D_{i,j}^\star$ of $\bm{D}^\star$ satisfy $\min_{1 \leq i \neq j \leq N} {D}_{i,j}^\star \geq C_2$.
}
\end{assumption}

\begin{assumption}
\label{as:inclusion}
{\em 
There exist constants $L_k \in (0,\, 1)$ such that the sample inclusion probabilities $\mu_e$ satisfy $\mu_e \geq L_k\, |\mathscr{S}_k|\, /\, |\mE_k|$ for all $e \in \mathscr{E}_k$ ($k = 2, \ldots, K$).
}
\end{assumption}

\s

Condition \ref{as:alpha} ensures that there are enough hyperedges to obtain theoretical guarantees,
while allowing the hypergraph to be sparse.
The sparsity of the hypergraph is controlled by $\rho_N$:
If $\rho_N$ is a constant independent of $N$,
the hypergraph is dense,
whereas $\rho_N = o(1)$ implies that the hypergraph is sparse in the sense that the expected number of hyperedges of size $k$ can be of a smaller order of magnitude than the number of possible hyperedges ${N \choose k}$ ($k = 2, \ldots, K$).
Condition \ref{as:theta} is standard and requires that all units reside in a compact set. 
Condition \ref{as:gram} ensures that units are well-separated.
Condition \ref{as:inclusion} guarantees that all hyperedges have a positive probability of being included in the sample.

The main result is Theorem \ref{theorem1},
which considers the non-asymptotic scenario in which the number of units $N$ is fixed,
whereas the sample size $|\mathscr{S}_k|$ of hyperedges of size $k$ can grow until it reaches its finite upper bound ${N \choose k}$ ($k = 2, \ldots, K$). 

\begin{theorem}{\bf Non-asymptotic scenario with fixed $N$}
\label{theorem1}
Consider any $\epsilon \in (0, 1)$ and any $N \in \{3, 4, \ldots\}$.
Then,
under Conditions \ref{as:alpha}--\ref{as:inclusion}, 
the following non-asymptotic error bounds hold with probability at least $1-\epsilon$:
\beno
\dfrac{1}{N(N-1)}\, |\!|\!|\widehat{\bm{D}}-\bm{D}^\star|\!|\!|_F^2
&\leq&  C_3 \left[ \dsum_{k=2}^K \dfrac{L_k\, |\mathscr{S}_k|}{|\mathscr{S}|}\right]^{-1} \dfrac{\Delta_{N,r}}{\epsilon\, \rho^2_N\, \sqrt{|\mathscr{S}|}}  
\ee
and
\beno
|\widehat{\alpha}_k - \alpha_k^\star| 
&\leq & C_4\, \dfrac{\sqrt{|\mathscr{S}|}\; \Delta_{N,r}}{{\epsilon\, \rho_N\, L_k\, |\mathscr{S}_k|}},\;\; k = 2, \ldots, K,
\ee
where 
\beno 
\Delta_{N,r} 
&\coloneqq& \sqrt{(N\, r+K-1)\, (K+4)\, (K+1)} 
\ee
and $C_3 > 0$ and $C_4 > 0$ are constants.
\end{theorem}

A proof of Theorem \ref{theorem1} can be found in Supplement \ref{sec:proof_theo1}. 
The first part of Theorem \ref{theorem1} establishes the rate of convergence of\, $\widehat{\bm{D}}$ as a function of the sample size $|\mathscr{S}_k|$ of hyperedges of size $k$ ($k = 2, \ldots, K$),
the total sample size $|\mathscr{S}| = \sum_{k=2}^K\, |\mathscr{S}_k|$,
and the level of sparsity $\rho_N$.
While there are no existing theoretical guarantees for hyperbolic space models of hypergraphs,
it is instructive to compare Theorem \ref{theorem1} to theoretical results based on graphs.
If $\rho_N$ is a constant independent of $N$,
the rate of convergence is $O(\sqrt{N /\, |\mathscr{S}|})$;
note that the term $\sum_{k=2}^K\, (L_k\, |\mathscr{S}_k|) / |\mathscr{S}|$ is contained in the interval $[\min_{2 \leq k \leq K} L_k ,\,  \max_{2 \leq k \leq K} L_k]$ and $\Delta_{N,r} \asymp \sqrt{N}$ because $K$
and $r$ are constants.
In the special case of graphs, 
the rate $O(\sqrt{N\, /\, |\mathscr{S}|})$ is known to be near-optimal \citep{davenport20141,li2023hyperbolic}. 
The second part of Theorem \ref{theorem1} establishes non-asymptotic error bounds on\, $\widehat\alpha_k$ in terms of\, $|\mathscr{S}_k|$ ($k = 2, \ldots, K$),
$|\mathscr{S}|$,
and $\rho_N$.

Theorem \ref{theorem1} provides the foundation for establishing convergence rates for estimators $\widehat{\bm\Theta}$ of the positions ${\bm\Theta}^\star$.
As discussed in Section \ref{sec:identifiability},
the positions can be estimated up to rotations,
so we assume that the true positions $\bm\Theta^\star$ satisfy $\bm{\Theta}^\star \coloneqq \bm{U}\, |\bm{S}|^{1/2}\bm{J}$.
To state the convergence rate of\, $\widehat{\bm\Theta}$,
we need a condition on the eigenvalues of $\bm{D}^\star$.

\s 

\begin{assumption}
\label{as:eigenvalue}
{\em  The smallest positive eigenvalue of\, $\bm{D}^\star$ satisfies $\lambda_{\min}(\bm{D}^\star) \asymp N$.
}
\end{assumption}

\s

Condition \ref{as:eigenvalue} implies that the positions do not degenerate to a geodesic in the hyperbolic space.  
Under Conditions \ref{as:alpha}--\ref{as:eigenvalue},
we can establish consistency of the estimator $\widehat{\bm{\Theta}}$ for the true positions $\bm\Theta^\star$ in $r$-dimensional hyperbolic space up to rotations.

\begin{theorem}{\bf Asymptotic scenario with $N \to \infty$}
\label{theorem2}
Under Conditions \ref{as:alpha}--\ref{as:eigenvalue}, 
assuming that $N$ and $|\mathscr{S}|$ increase without bound,
\beno
\dfrac{1}{N} \inf_{\bm{R} \in \mathscr{R}}\, |\!|\!|\widehat{\bm{\Theta}}\, \bm{R}  - \bm{\Theta}^\star|\!|\!|_F^2  
&=& O_p\left(\dfrac{\Delta_{N,r}}{\rho_N^2\, \sqrt{|\mathscr{S}|}}\right), 
\ee
where $\mathscr{R} \coloneqq \{\bm{R} \in \mathbb{R}^{(r+1) \times (r+1)}:\; \bm{R}\, \bm{J}\, \bm{R}^\top = \bm{J}\}$ is the set of hyperbolic rotation matrices.
\end{theorem} 

A proof of Theorem \ref{theorem2} can be found in Supplement \ref{sec:proof_theo2}. 
In contrast to Theorem \ref{theorem1}, 
Theorem \ref{theorem2} considers the asymptotic regime in which $N$ and $|\mathscr{S}|$ grow without bound,
assuming that $K \geq 3$,\,
$p \in \mathbb{R} \setminus \{0\}$,\,
and $r \geq 2$ are constants independent of $N$.

{\bf Sample size}
To ensure that,
for a given $\epsilon > 0$,
the error $(1 / N)\, \inf_{\bm{R} \in \mathscr{R}}\, |\!|\!|\widehat{\bm{\Theta}}\, \bm{R} - \bm{\Theta}^\star|\!|\!|_F^2$ is less than $\epsilon$ with high probability,
the sample size $|\mathscr{S}|$ must satisfy $|\mathscr{S}| > \epsilon\, K^2\, N\, r / \rho_N^4$ in light of $\Delta_{N,r} \asymp K \sqrt{N\, r}$. 
The sample size $|\mathscr{S}|$ increases with $K$ and $r$,
because the model complexity increases with $K$ and $r$.
The sparser the hypergraph is,
the larger the sample size $|\mathscr{S}|$ needs to be:
e.g.,
if $K = 4$,\,
$r = 2$,
then $|\mathscr{S}| > 32\, \epsilon\, N$ when $\rho_N = 1$ (dense hypergraph) and $|\mathscr{S}| > 32\, \epsilon\, N^2$ when $\rho_N = 1/N^{1/4}$ (sparse hypergraph).
While sample sizes of order $N$ or $N^2$ may appear to be high,
the number of possible hyperedges is ${N \choose 2} + {N \choose 3} + {N \choose 4} \asymp N^4$.
In other words,
the sample size is of a smaller order than the number of possible hyperedges.

{\bf Comparison with existing results}
While the literature does not provide theoretical guarantees for hyperbolic space models for hypergraphs,
there are some related results on graphs.
For example,
in the special case of graphs---corresponding to hypergraphs with $K = 2$,
the convergence rate of Theorem \ref{theorem2} recovers the rate of \cite{li2023hyperbolic},
who study hyperbolic space models for graphs rather than hypergraphs.
In addition,
the result in Theorem \ref{theorem2} resembles results based on generalized random dot product graphs \citep{rubin2022statistical} (up to logarithmic factors),
although those results focus on graphs rather than hypergraphs and are not based on hyperbolic geometry. 

\section{Simulation Study}
\label{sec:simulation_study}

We demonstrate that the theoretical results in Section \ref{sec:theory} are supported by simulation results.
We consider hyperedges $e$ of sizes $|e| \in \mathscr{K} = \{2, 3, 4\}$,
so that $K = 4$.
We set $\alpha_2 = 5 \times 10^{-1}$, 
$\alpha_3 = 5 \times 10^{-4}$, 
and $\alpha_4 = 5 \times 10^{-6}$.
The positions of $N$ units on the two-dimensional \pd{ }are generated as described in Supplement \ref{sec:positions}.
In each scenario,
we generate 100 population hypergraphs using the method described in Section \ref{sec:simulation}.
For each generated population hypergraph,
we sample hyperedges using the sampling design described in Section \ref{sec:sampling}.
In other words,
we sample all realized hyperedges and,
for each realized hyperedge,
we sample $n$ unrealized hyperedges,
called controls.

\begin{figure}[t!]
    \centering
    \includegraphics[width=\textwidth]{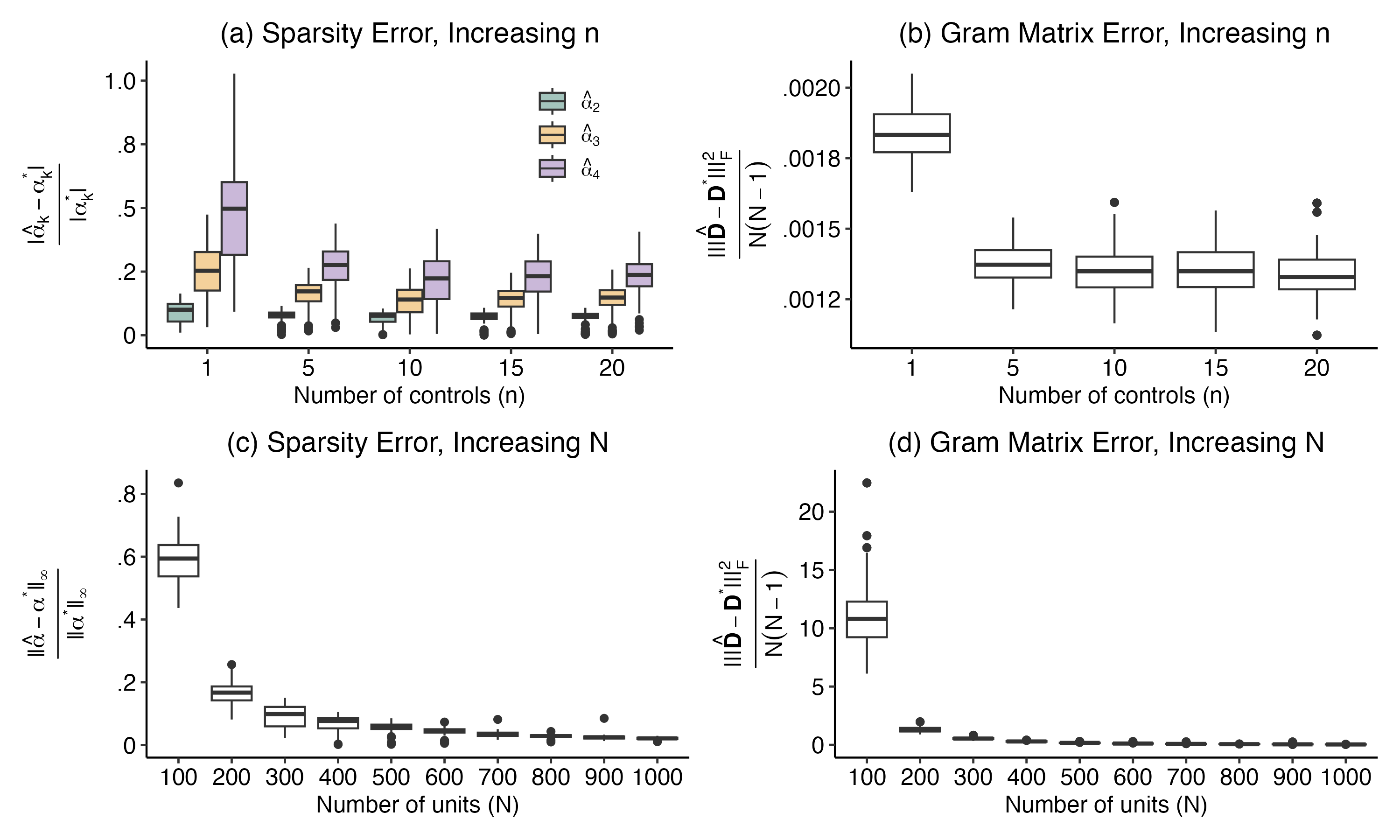} 
    \caption{Simulation results: 
    error of estimating sparsity parameter vector $\bm{\alpha} \coloneqq (\alpha_2, \ldots, \alpha_K)^\top$ and Gram matrix $\bm{D}$ as a function of the number of controls $n$ (the number of unrealized hyperedges sampled for each realized hyperedge) and the number of units $N$.\s
    }
    \label{fig:simulation}
\end{figure}

\paragraph*{Number of controls $n$}
We consider $N = 400$ units and assess the statistical error as a function of the number of controls $n \in \{1, 5, 10, 15, 20\}$.
Figure \ref{fig:simulation} suggests that $n=5$ strikes the best balance between statistical error and computing time when $N = 400$.

\paragraph*{Number of units in hypergraph $N$}
To assess how the statistical error behaves as $N$ increases,
we consider $N \in \{100,\, 200,\, \ldots,\, \mbox{1,000}\}$ and $n = 10$.
The simulation results in Figure \ref{fig:simulation} support the theoretical results in Theorem \ref{theorem1}:
As $N$ increases,
the statistical error of estimating the Gram matrix $\bm{D}$ and the sparsity parameters $\alpha_2, \ldots, \alpha_K$ decreases.

\section{Application to U.S.\ Newswire Articles}
\label{sec:application}

We showcase hyperbolic space models by applying them to U.S.\ Newswire articles between 1900 and 1977 \citep{silcockNewswireLargeScaleStructured2024}.  
Newswire articles by the Associated Press, Reuters, and other news organizations provide news sources for many local newspapers and form a comprehensive archive of publicly available information.  
We focus on the $N = 678$ U.S.\ politicians who were mentioned at least five times in the U.S.\ Newswire data, 
who were born after 1900,
and whose political affiliation is available. 
The data set is available at 
\begin{center}
\url{https://huggingface.co/datasets/dell-research-harvard/newswire}    
\end{center}
The number of realized hyperedges of size $k \in \{2, \ldots, N\}$ is: 
$|\mE_{2}^{(1)}| = \mbox{4,624}$, 
$|\mE_{3}^{(1)}| = \mbox{2,815}$, 
$|\mE_{4}^{(1)}| = \mbox{1,208}$, 
$|\mE_{5}^{(1)}| = \mbox{607}$, 
$|\mE_{6}^{(1)}| = \mbox{261}$, 
$|\mE_{7}^{(1)}| = \mbox{104}$,
$|\mE_{8}^{(1)}| = \mbox{70}$, 
$|\mE_{9}^{(1)}| = \mbox{20}$, 
$|\mE_{10}^{(1)}| = \mbox{3}$,
and $|\mE_{k}^{(1)}| = 0$ for all $k \in \{11, \ldots, N\}$.
Since hyperedges of sizes 2, 3, and 4 comprise $89\%$ of the 9,712 realized hyperedges,
we focus on hyperedges $e$ of size $|e| \in \mathscr{K} = \{2, 3, 4\}$,
so that $K = 4$.
The small number of hyperedges $e$ of size $|e| \geq 5$ implies that the observed hypergraph does not contain much information about hyperedges of size $|e| \geq 5$.
A more detailed description of the data is provided in Supplement \ref{sec:addition_Data}.
While the entire hypergraph is observed,
there are more than $8$ billion possible hyperedges of sizes $2$, $3$, and $4$.
We thus sample hyperedges as described in Section \ref{sec:sampling},
by sampling all realized hyperedges and sampling $n = 40$ unrealized hyperedges for each realized hyperedge.
We consider two-dimensional hyperbolic space to facilitate interpretation,
and estimate the positions of politicians in hyperbolic space by conducting 100 estimation runs with starting values chosen at random,
as described in Supplement \ref{sec:initialization}.
All following results are based on the run minimizing $\widehat{\ell}(\cdot)$.
We demonstrate in Supplement \ref{sec:sensitivity} that the results are not too sensitive to the choice of $n = 40$,\,
$p = -20$,\,
and the starting values used in the 100 estimation runs. 
We do not report comparisons with existing approaches for computational reasons:
e.g.,
we were unable to obtain results using the approaches of \citet{brusa2024,brusa2024a}, \citet{turnbull2023latent}, and \citet{grose2024} within 24 hours.

\subsection{Estimated Positions}

\begin{figure}[t!]
  \centering
  \begin{subfigure}[b]{0.47\textwidth}   
    \centering
    \includegraphics[width=\textwidth]{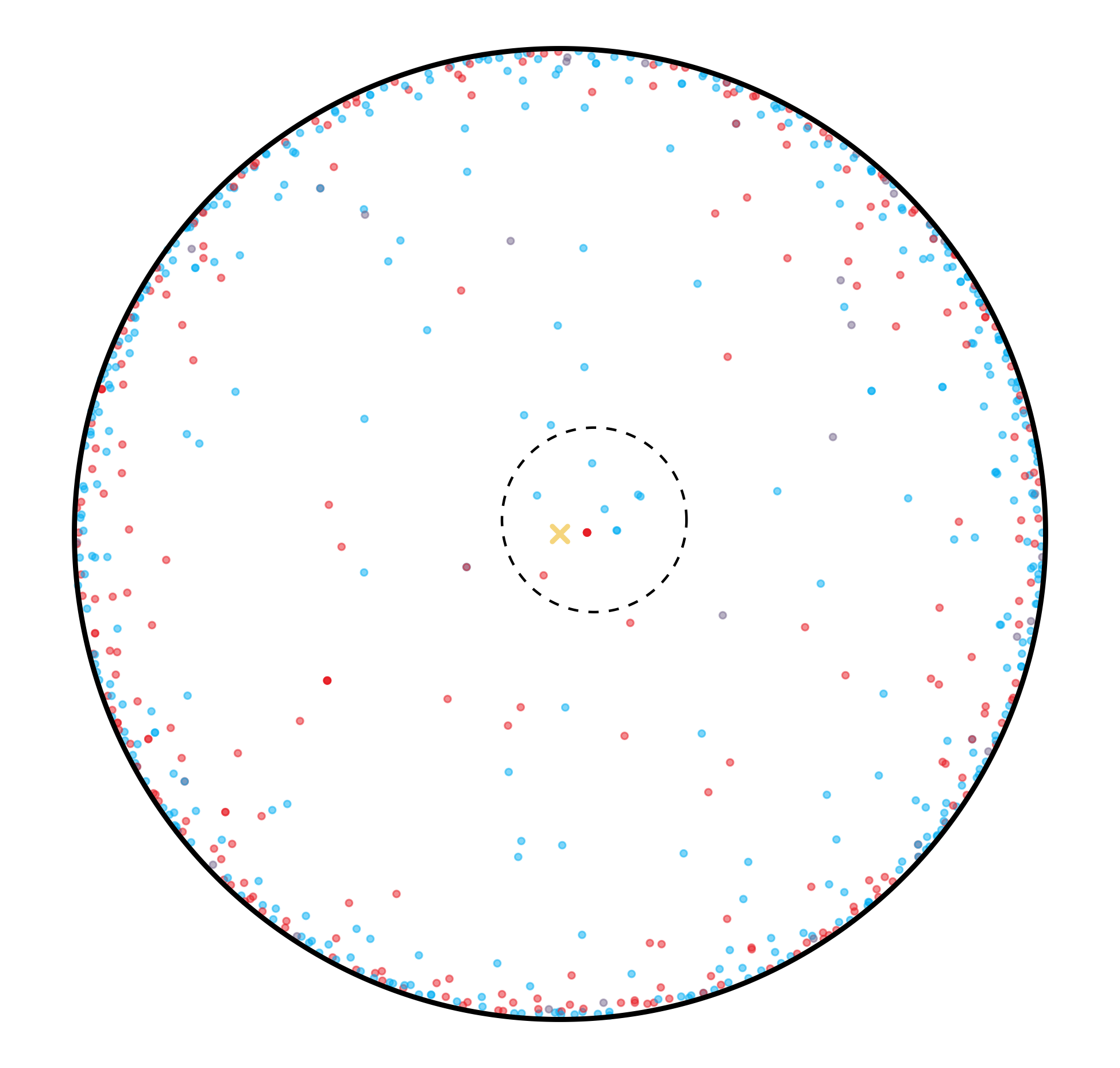}  
    \caption{\hangindent=1.6em Estimated positions of $N = 678$ politicians on the \pd.}
    \label{fig:pos_all}
  \end{subfigure}\;
  \begin{subfigure}[b]{0.47\textwidth}   
    \centering
    \includegraphics[width=\textwidth]{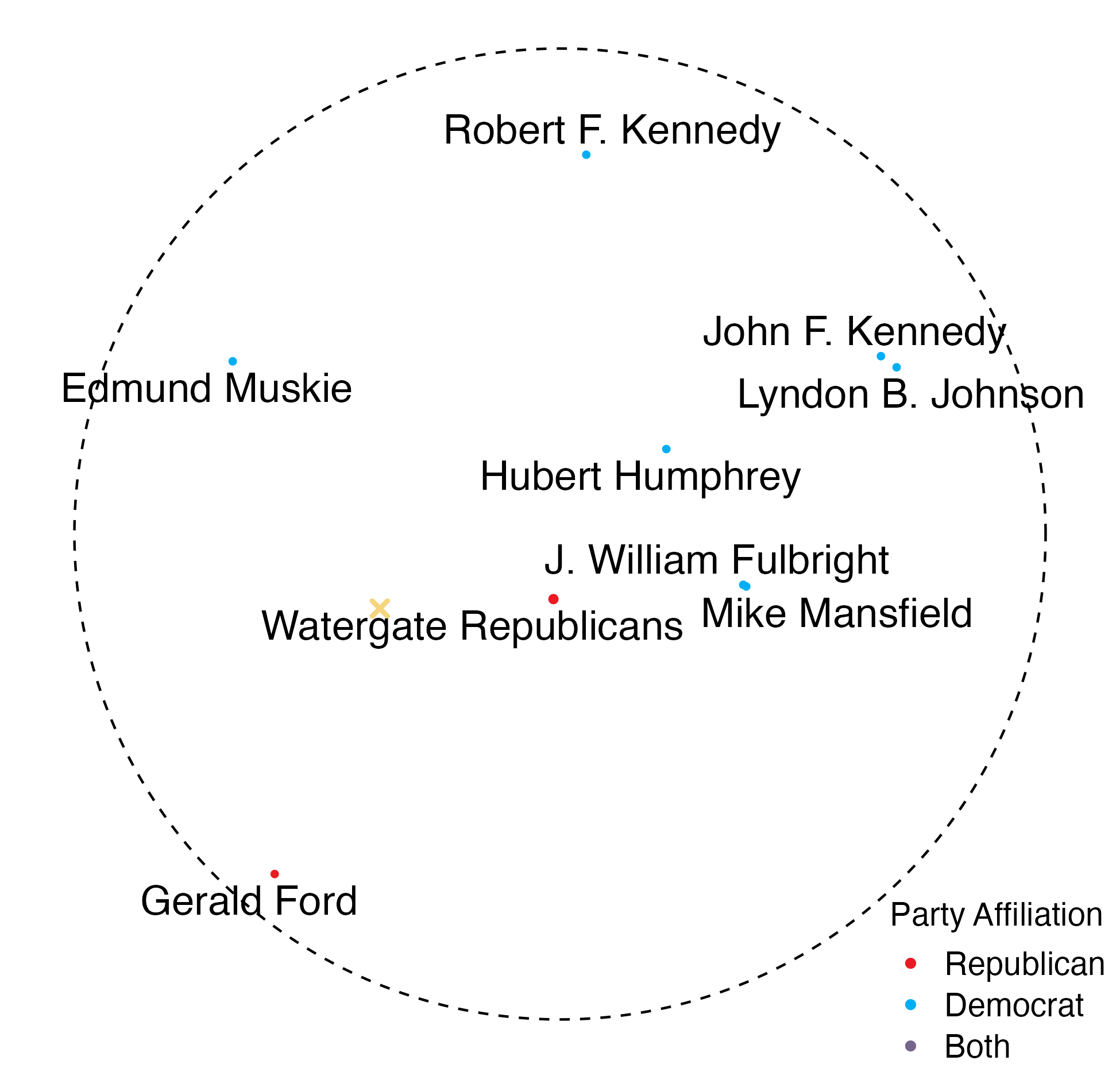} 
    \caption{\hangindent=1.85em Estimated positions of the 13 politicians closest to the center of the \pd.}
    \label{fig:pos_zoom}
  \end{subfigure}
  \caption{Newswire data: 
  Estimated positions of politicians on the \pd, 
  with party affiliation indicated by color;
  politicians who switched parties are labeled ``both.''
  The origin of the \pd\, is represented by $\textcolor[HTML]{F5D57D}{\times}$. 
  ``Watergate Republicans'' refers to Watergate era Republicans Richard Nixon, Henry Kissinger, Melvin Laird, Barry Goldwater, and Nelson Rockefeller, 
  whose positions are almost indistinguishable. 
  }
  \label{fig:application_res_a}
\end{figure}

Figure \ref{fig:application_res_a} (a) shows the estimated positions $\bm\Theta$ of politicians on the \pd.
Figure \ref{fig:application_res_a} (b) reveals that the politicians close to the center of the \pd{ }were all political heavyweights between 1960 and 1977,
the time frame for which we have most data.
Many presidents and their vice presidents are close to each other,
including Richard Nixon and Gerald Ford,
John F.\ Kennedy and Lyndon B.\ Johnson,
and Lyndon B.\ Johnson and Hubert Humphrey;
note that Lyndon B.\ Johnson first served as vice president under President John F.\ Kennedy and then went on to serve as president after the assassination of President John F. Kennedy.
So are the two Kennedy brothers (John F.\ and Robert F.\ Kennedy).

\subsection{Comparison of Hyperbolic and Euclidean Geometry}
\label{sec:assessment}

While the specification of $g(\cdot)$ in \eqref{eq:conc} is based on hyperbolic distance,
it can be extended to Euclidean distance as follows:
\beno
g_{\mathscr{E}}(\Pos_{e}) 
\;\coloneqq\; \left\{\dfrac{1}{|e|}\, \dsum_{i \in e} \left[d^{(e)}_{\mathscr{E},i}(\Pos_e)\right]^p \right\}^{1/p},
\ee 
where
\beno
d^{(e)}_{\mathscr{E},i}(\Pos_e)
&\coloneqq& \dsum_{j \in e \setminus \{i\}}\, |\!|\pos_i - \pos_j|\!|_2.
\ee
We compare results based on hyperbolic and Euclidean space models in terms of expressive power (hierarchical structure and embeddedness) and predictive power.

\paragraph*{Expressive power}

\begin{figure}[t!]
  \centering
    \includegraphics[width=0.9\textwidth]{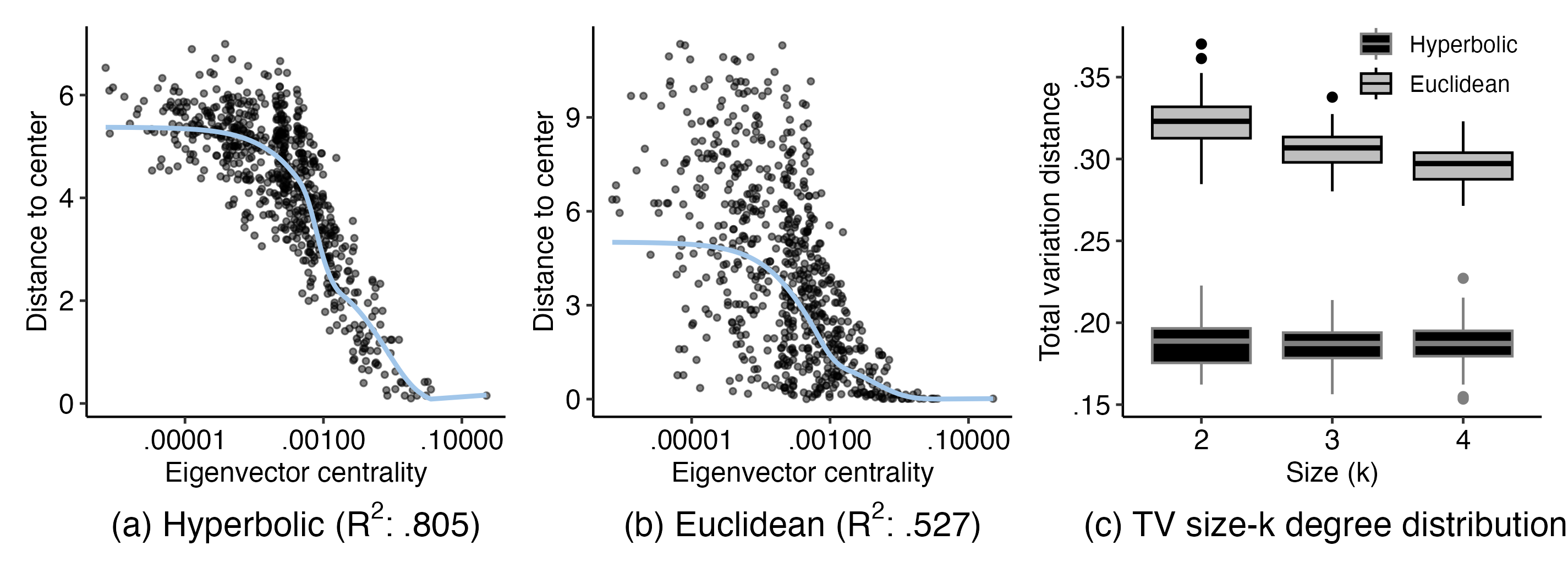}  
    \caption{
    Newswire data: distances of politicians to the center of (a) hyperbolic space or (b) Euclidean space plotted against eigenvector centrality scores.
    The blue lines in (a) and (b) represent the smoothed averages of the distance to the center by eigenvector centrality score. 
    The $R^2$ value is calculated from the accuracy of using the blue line for predicting the distance to the center based on the eigenvector centrality. 
    (c) Total variation distance between the size-$k$ degree distributions of observed and 100 simulated hypergraphs.
    }
    \label{fig:hyperbolic_euclidean}
\end{figure}

We assess the expressive power of hyperbolic and Euclidean space models by comparing how these geometric models capture the hierarchical structure and the embeddedness of politicians.

A measure of unit-specific hierarchy in hypergraphs is the eigenvector centrality score of \citet{bonacich2004}.
The eigenvector centrality scores are the coordinates of the normalized leading eigenvector of $\bm{A}^\top \bm{A}$, 
where $\bm{A} = (A_{m,i})\in \{0, 1\}^{M \times N}$ is the incidence matrix of the hypergraph,
$M$ is the number of realized hyperedges,
and $A_{m,i} \coloneqq 1$ indicates that hyperedge $m$ includes unit $i$ and $A_{m,i} \coloneqq 0$ otherwise.
To compute the distances of politicians to the center of the space,
we compute the hyperbolic distance to the origin $(0, 0) \in \mathbb{R}^2$ of the \pd{ }and the Euclidean distance to the centroid of the positions of politicians.
Figure \ref{fig:hyperbolic_euclidean} reveals how well each model can represent the hierarchy as quantified by eigenvector centrality scores.
Under hyperbolic space models,
politicians with high eigenvector centrality scores are close to the center of the \pd,
whereas under Euclidean space models some politicians with high eigenvector centrality scores are far from the center of the Euclidean space.
\begin{figure}[t!]
  \centering
  \includegraphics[width=0.75\textwidth]{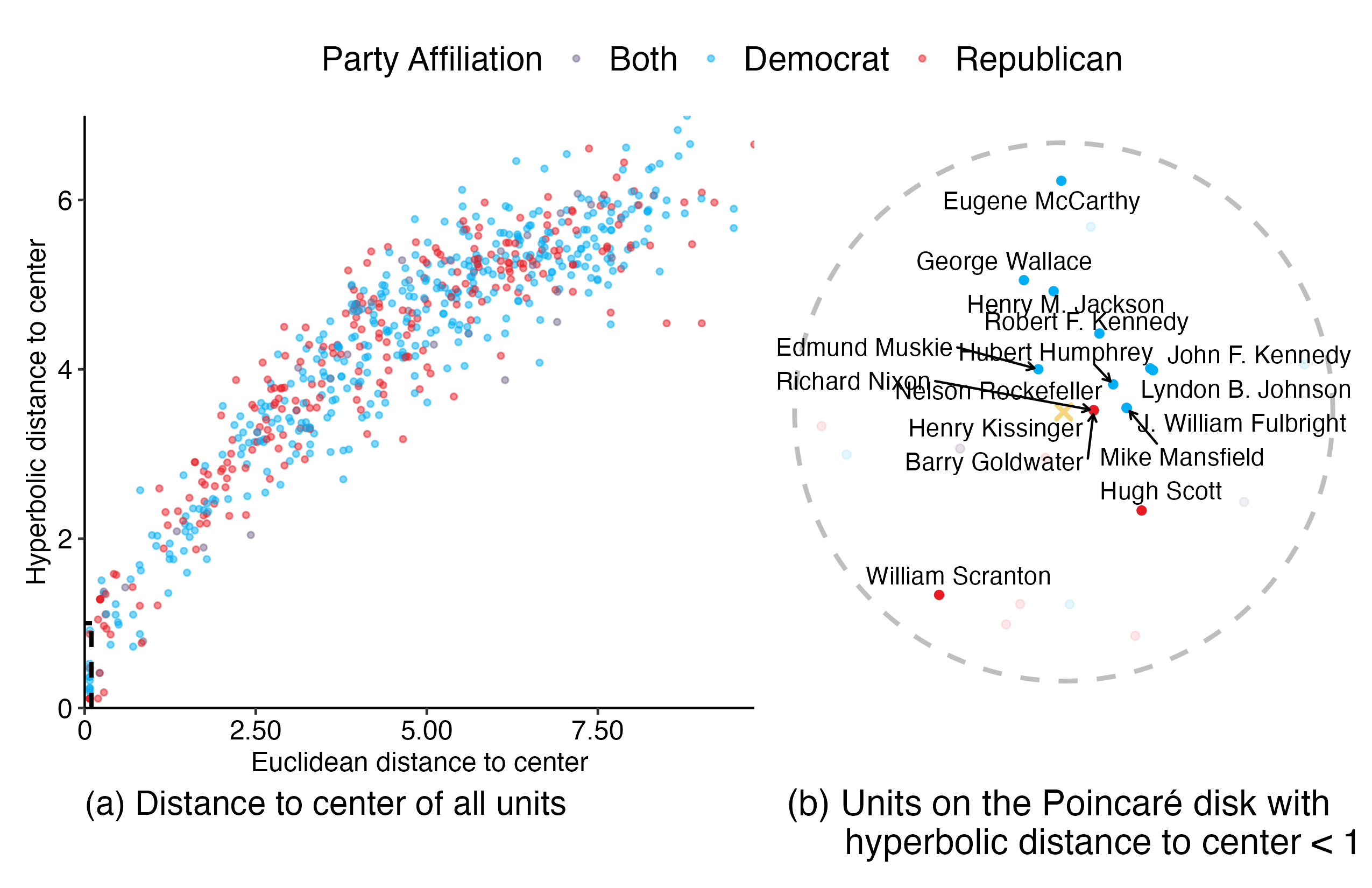}  
  \caption{
  (a) Hyperbolic and Euclidean distances of politicians to the center of the respective space.
  The dotted rectangle in (a) contains politicians with Euclidean distance of less than 1/10 to the center.
  (b) Positions of units with estimated hyperbolic distance less than $1$ on the \pd.  
  The politicians contained in the dotted rectangle in (a) are singled out and named in (b). 
  The party affiliation of politicians is indicated by color;
  politicians who switched parties are labeled ``both.''
  }
  \label{fig:positions_distance_center}
\end{figure}
Figure \ref{fig:positions_distance_center} (a) reveals that there is a positive but nonlinear relationship between the hyperbolic and Euclidean distances of units to the center of the respective space.
The dashed rectangle in Figure \ref{fig:positions_distance_center} (a) highlights a small subset of politicians whose distances to the center according to the Euclidean geometry are less than 1/10.
Figure \ref{fig:positions_distance_center} (b) displays the positions of those politicians on the \pd{ }and provides the names of the politicians involved.
These politicians,
all political heavyweights of the twentieth century,
are approximately equally close to the center of Euclidean space,
but hyperbolic space differentiates among them in terms of political hierarchy (distance to the center) and political orientation (angle).
For example,
the \pd{ }reveals that politicians are separated by party lines:
Democrats tend to be located in the upper semi-circle, 
while most Republicans are located in the lower semi-circle.
At the center of the \pd{ }sits President Nixon, 
surrounded by Democratic figures such as Humphrey and Muskie --- the presidential and vice presidential candidates defeated by President Nixon and his Vice President Agnew in the presidential elections of 1968. 

To assess how well the models capture the embeddedness of politicians in the hypergraph,
we compare the size-$k$ degree distribution of the observed hypergraph to the size-$k$ degree distributions of 100 hypergraphs simulated from hyperbolic and Euclidean space models.
The size-$k$ degree distribution is defined as the empirical distribution of unit-specific size-$k$-degrees,
where the size-$k$ degree of a politician is the number of hyperedges of size $k$ in which the politician is involved.
The total variation distance between the observed and simulated size-$k$ degree distributions in Figure \ref{fig:hyperbolic_euclidean} (c) suggests that hyperbolic space models capture the embeddedness of politicians in the hypergraph considerably better than Euclidean space models.

\begin{figure}[t!]
    \centering
   \includegraphics[width=0.9\textwidth]{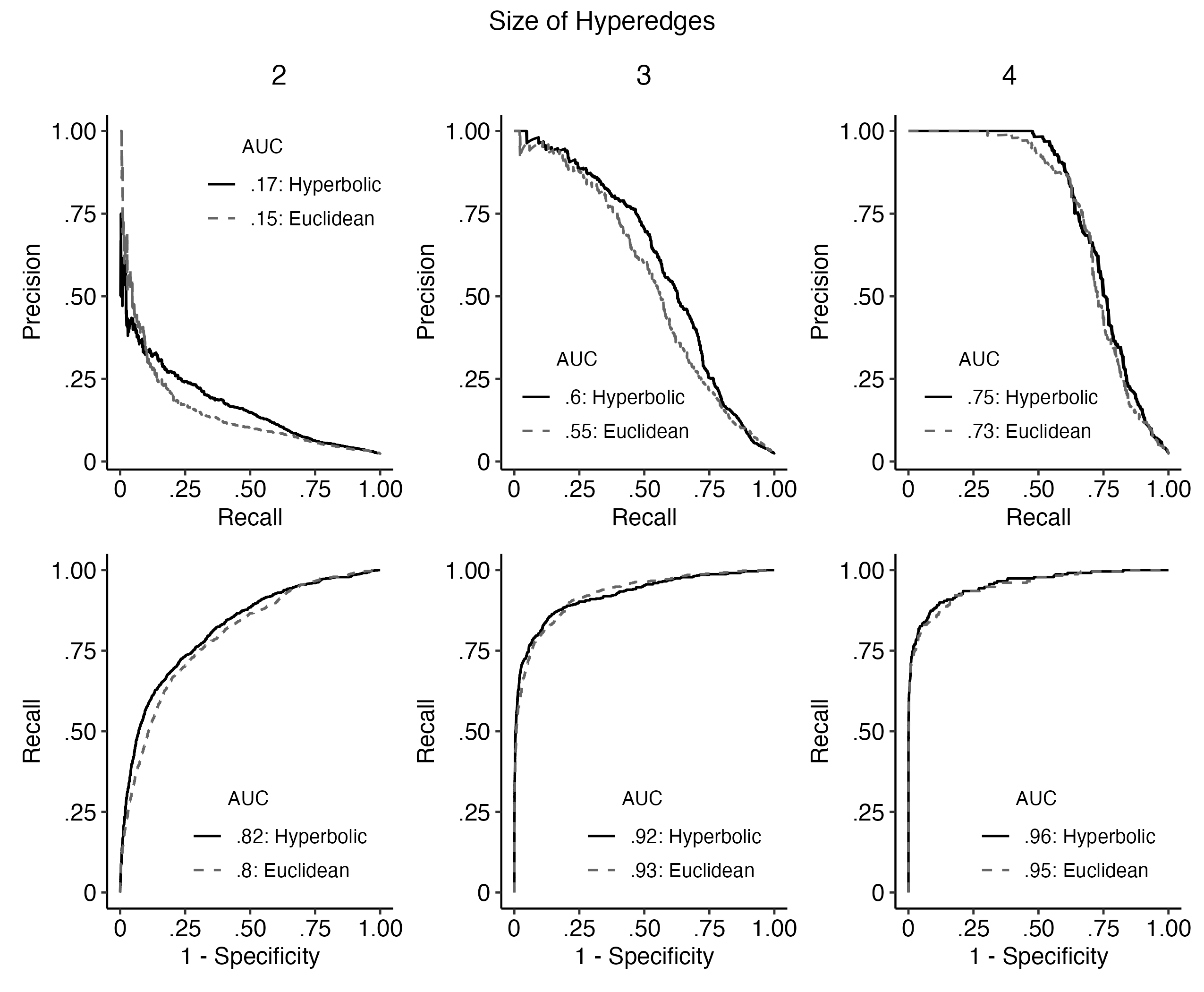}  
    \caption{Newswire data: Receiver-Operator Characteristic (ROC) and Precision-Recall (PR) curve of out-of-sample predictions of hyperedges,
    based on embedding U.S.\ politicians in hyperbolic and Euclidean space.}\s
    \label{fig:pr}
\end{figure}

\paragraph*{Predictive power}
To assess the out-of-sample performance of the model, 
we divide the realized hyperedges of sizes 2, 3, and 4 at random into training and test hyperedges encompassing $80\%$ and $20\%$ of the data, 
respectively.
We use Precision-Recall (PR) and Receiver Operating Characteristic (ROC) curves to compare the performance of hyperbolic and Euclidean space models in Figure \ref{fig:pr}.
The PR curves in the first row plot precision, 
the proportion of predicted positive hyperedges that are true positives, 
against recall, 
the proportion of true positive hyperedges correctly identified.
The ROC curves in the second row plot recall (sensitivity) against specificity, 
the proportion of negatives incorrectly classified as positives.
Figure \ref{fig:pr} suggests that hyperbolic space models have a slight edge over Euclidean space models in terms of predictive power.

\section{Discussion}
\label{sec:disc}

The Newswire data demonstrate that the choice of geometry matters:
Hyperbolic space is superior to Euclidean space in its ability to represent core-periphery and local structure in the Newswire data.
That said,
the choice of geometry is less important for the purpose of predicting hyperedges in the Newswire data.
An interesting direction for future research is incorporating attributes (e.g., party affiliation) to improve predictive power.

\begin{center}
{\large\bf Supplementary Materials}
\end{center}

\noindent
The supplementary materials provide proofs of all theoretical results along with additional details on the simulations and application in Sections \ref{sec:simulation_study} and \ref{sec:application}.

\begin{center}
{\large\bf Data Availability Statement}
\end{center}

\noindent
The data set used in Section \ref{sec:application} is available at 
\begin{center}
\url{https://huggingface.co/datasets/dell-research-harvard/newswire}    
\end{center}

\begin{center}
{\large\bf Disclosure Statement}
\end{center}

\noindent
The authors report there are no competing interests to declare.

\if0\blind
{
\begin{center}
{\large\bf Funding}
\end{center}

\noindent
The authors acknowledge support in the form of U.S.\ Army Research Office award ARO W911NF-21-1-0335 and U.S.\ National Science Foundation award NSF DMS-2515763.

}

\fi
\bibliographystyle{chicago}

\bibliography{base}

\newpage 

\setcounter{page}{1}

\setcounter{footnote}{0}
\renewcommand{\thefootnote}{\fnsymbol{footnote}} 

\begin{center}
{\LARGE\textbf{Supplementary Materials\\}}
\end{center}

\appendix
\numberwithin{equation}{section}
\renewcommand{\cftsecfont}{\mdseries}
\renewcommand{\cftsecpagefont}{\mdseries}
\setlength{\cftbeforesecskip}{0 pt} 
\renewcommand{\cftsecleader}{\cftdotfill{\cftdotsep}} 
\startcontents
\printcontents{ }{1}{}
\newpage 

\renewcommand{\thefootnote}{\arabic{footnote}} 
\setcounter{equation}{0}

\section{Proofs}
\setcounter{prop}{0}

\subsection{Proof of Proposition 1}
\label{sec:proof_prop1}

We first establish identifiability of the Gram matrix $\bm{D}$ and the sparsity parameters $\alpha_2, \ldots, \alpha_K$.

\begin{prop}
The Gram matrix $\bm{D} \coloneqq \bm{\Theta}\, \bm{J}\, \bm{\Theta}^\top$ and the sparsity parameters $\alpha_2, \ldots, \alpha_K$ are identifiable provided that $N>r+2$.
\end{prop}

Consider any $k \in \{3, \ldots, K\}$.
Let $f: \mathbb{R} \mapsto \mathbb{R}$ be defined by $f(y) \coloneqq \alpha\, \frac{2\exp(x)}{1+\exp(x)} - \frac{2\exp(y\,x)}{1+\exp(y\,x)}$,
where $x \in \mathbb{R}$ is fixed. 
For any $\alpha \in (0,\, 1]$ and $x<0$, 
we have $\frac{\partial f(y)}{\partial y }>0$,
$f(0)<0$,
and $f(+\infty)>0$. 
Therefore, 
for a given $\alpha_k$, 
we can find a corresponding $\widetilde{\alpha}_{e}$ such that  
\beno 
\pi(\alpha_{|e|}, \Pos_{e}) 
&=& \alpha_k\, \sigma(- g(e)) 
&=& \sigma(-\widetilde{\alpha}_{e}\, g(e)),
\ee
where $\sigma(x) \coloneqq 2\, \exp(x)/(1+\exp(x))$ and $g(e)$ is defined by 
\be
\label{eq:g}
g(\Pos_{e}) 
\;\coloneqq\; \left(\dfrac{1}{|e|}\, \dsum_{i \in e} \left(\dsum_{j \in e,\; j \neq i}\, d_{\mathscr{L}}(\pos_i,\, \pos_j) \right)^p \right)^{1/p}.
\ee  
If there exists another set of sparsity parameters $\beta_k$ and Gram matrix $\widetilde{\bm{D}}$ such that
\beno 
\alpha_k\, \sigma(- g(e)) 
&=& \beta_k\, \sigma(- \widetilde{g}(e)),
\ee
then we obtain, 
for some $\widetilde{\beta}_{e}$,
\beno 
\sigma(\widetilde{\alpha}_{e}\, g(e)) 
&=& \sigma(\widetilde{\beta}_{e}\, \widetilde{g}(e)),
\ee
where $\widetilde{g}(e)$ 
is defined by 
\beno 
\widetilde{g}(e)
&\coloneqq& \left(\dfrac{1}{|e|}\dsum_{i\in e}\text{arcosh}(-\widetilde{D}_{i,j})^p\right)^{1/p}.
\ee
In other words,
we have an overdetermined system with $O(N^k)$ equations involving $O(N^2)$ variables $\text{arcosh}(-\widetilde{\bm{D}})$. 
The equations hold only when $\alpha_k = \beta_k$ and $\text{arcosh}(-\bm{D}) = \text{arcosh}(-\widetilde{\bm{D}})$. 
Since the $\text{arcosh}(\cdot)$ is a monotone function, 
we obtain $\bm{D} = \widetilde{\bm{D}}$. 

Last,
but not least,
consider the case $K=2$.
For another set of sparsity parameters $\beta_2$ and Gram matrix $\widetilde{\bm{D}}= \widetilde{\bm{\Theta}}\bm{J}\widetilde{\bm{\Theta}}^\top$, 
we obtain
\beno 
\alpha_2\, \sigma\left(- \text{arcosh}(-\bm{\Theta}\bm{J}\bm{\Theta}^\top)\right)  
&=&  \beta_2\, \sigma\left(-\text{arcosh}(-\widetilde{\bm{\Theta}}\bm{J}\widetilde{\bm{\Theta}}^\top)\right).
\ee
Since $\bm\Theta,\, \widetilde{\bm\Theta} \in \mathbb{R}^{N\times (r+1)}$ and the function $\text{arcosh}(\cdot)$ is monotonic, 
we obtain $N\,(N-1)$ equations involving $N\, (r+1)$ variables. 
If $N > r+2$, 
then $\alpha_2 = \beta_2$ and $\bm{D} = \widetilde{\bm{D}}$, 
completing the proof. 

\subsection{Proof of Theorem 1}
\label{sec:proof_theo1}

\setcounter{theorem}{0}

\begin{theorem}
\label{theorem1}
Consider any $\epsilon \in (0, 1)$ and any $N \in \{3, 4, \ldots\}$.
Then,
under Conditions \ref{as:alpha}--\ref{as:inclusion}, 
the following non-asymptotic error bounds hold with probability at least $1-\epsilon$:
\beno
\dfrac{1}{N(N-1)}\, |\!|\!|\widehat{\bm{D}}-\bm{D}^\star|\!|\!|_F^2
&\leq&  C \left[ \dsum_{k=2}^K \dfrac{|\mathscr{S}_k|\, L_k}{|\mathscr{S}|}\right]^{-1} \dfrac{\Delta_{N,r}}{\epsilon\, \rho^2_N\, \sqrt{|\mathscr{S}|}}  
\ee
and
\beno
|\widehat{\alpha}_k - \alpha_k^\star| 
&\leq & C'\, \dfrac{\sqrt{|\mathscr{S}|}\; \Delta_{N,r}}{{\epsilon\, \rho_N\, L_k\, |\mathscr{S}_k|}},\;\; k = 2, \ldots, K,
\ee
where 
\beno 
\Delta_{N,r} 
&\coloneqq& \sqrt{(N\, r+K-1)\, (K+4)\, (K+1)} 
\ee
and $C > 0$ and $C' > 0$ are constants independent of $N$.
\end{theorem}

In the following, 
we denote $C(k) = {N \choose k}$. 
Denote possible latent positions and sparsity parameters as $\bm{\Theta}$ and $\bm{\alpha}$, and the corresponding Gram matrix $\bm{D}\in \mathbb{R}^{N\,\times\, N}$. 
The true values are denoted by $\bm{\Theta}^\star$,
$\bm{\alpha}^\star$, 
and $\bm{D}^\star = \bm{\Theta}^\star\bm{J}\, (\bm{\Theta}^{\star})^\top$, 
while the minimizers of the sample loss $\widehat\ell(\cdot)$ are $\widehat{\bm{\Theta}}$, 
$\widehat{\bm{\alpha}}$, 
and $\widehat{\bm{D}}$. 
We define the multivariate function $f_k(\bm{D}) = \{f^{(e)}_k(\bm{D})\}_{e \in \mE_k}: \mathbb{R}^{N\times N} \mapsto [0,\, 1]^{C(k)}$, 
where, 
for all $e \in \mE_k$,
\be
\label{eq:D}
f^{(e)}_k(\bm{D}) &
\coloneqq&  \sigma(-g(\Pos_e)) 
\ee
is the vector of probabilities corresponding to all possible hyperedges of size $k\in \{2, ..., K\}$ if the sparsity parameters $\alpha_2, ..., \alpha_K$ are set to 1. 
The function $g$ is given in \eqref{eq:g}, which can be perceived as a function of $\bm{D}$ in \eqref{eq:D}.
To better illustrate the proof and emphasize the node composition of hyperedge $e$, we introduce the following notation for the  probability to observed $e = (i_1,i_2,\ldots, i_k)\in \mE_k$ with $k$ nodes where $k = 2, \ldots, K$:
\be
\label{A_eq_1}
P_{i_1,\ldots,i_k} 
&\coloneqq& \alpha_k\, f_k^{(e)}(\bm{D}) = \pi(\alpha_k, \Pos_e),
\ee
where $\pi(\alpha_k, \Pos_e)$ is the hyperedge probability defined in Section 3. Accordingly, hyperedge probability tensors  are defined as follows 
\beno 
\bm{P}_k^\star &\coloneqq& \{ P^\star_{i_1,\ldots, i_k}\}_{1\leq i_1,\ldots,i_k\leq N} \text{, ~~~~~~~} \bm{\widehat{P}}_k &\coloneqq& \{ \widehat{P}_{i_1,\ldots, i_k}\}_{1\leq i_1,\ldots,i_k\leq N},
\ee
where $P^\star_{i_1,\ldots, i_k}$ and $\widehat{P}_{i_1,\ldots, i_k}$ result from plugging in $(\bm{D}^\star,\bm{\alpha}^\star)$ and $(\bm{\widehat{D}},\bm{\widehat{\alpha}})$ into the probabilities defined in \eqref{A_eq_1}. 

Consider the loglikelihood function based on $\mathscr{S} = \bigcup_{k=2}^K \mathscr{S}_k$ of all sampled hyperedges of different sizes where $\mathscr{S}_k = \mathscr{S}^{(1)}_k\cup\mathscr{S}^{(0)}_k$, we introduce the following expansion for the sample loss $\hat{\ell}(\bm\Lambda)$ defined in Section 4.2 as 
\begin{align*}
   \hat{\ell}(\bm\Lambda) &=  \sum_{e\in \mathscr{S}} \frac{1}{\mu_e}\ell(Z_e\mid \bm{\widehat{D}},\bm{\widehat{\alpha}}) = \sum_{k=2}^K \sum_{e \in \mathscr{S}_k} \frac{1}{\mu_e}\ell_k(Z_e\mid \bm{\widehat{D}},\bm{\widehat{\alpha}}),\\
   \ell_k(Z_e\mid \bm{\widehat{D}},\bm{\widehat{\alpha}}): &=  
   Z_{e}\log \widehat{\alpha}_k f_k(\bm{\widehat{D}}) + (1-Z_{e})\log (1-\widehat{\alpha}_k f_k(\bm{\widehat{D}})),
\end{align*}
where $Z_e = 1$ if $e\in \mathscr{S}^{(1)}_k$ and $Z_e = 0$ if $e\in \mathscr{S}^{(0)}_k$.
We can write 
\be
\label{A_eq_2}
    &~&\dsum_{e\in \mathscr{S}} \frac{1}{\mu_e}\left[\ell(Z_e\mid \bm{\widehat{D}},\bm{\widehat{\alpha}}) 
     - \ell(Z_e\mid \bm{{D^\star}},\bm{{\alpha^\star}})\right]\s 
     \\
    &= & \dsum_{k=2}^K 
    \dsum_{e\in \mathscr{S}_k} \frac{1}{\mu_e}\bigg[\ell_k(Z_e\mid \bm{\widehat{D}},{\widehat{\alpha}_k}) - \mathbb{E}\, \ell_k(Z_e\mid\bm{\widehat{D}},{\widehat{\alpha}_k})\s
    \\
    &-& (\ell_k(Z_e\mid \bm{{D^\star}},{{\alpha^\star_k}}) - \mathbb{E}\, \ell_k(Z_e\mid \bm{{D^\star}},{{\alpha^\star_k}}))\s
    \\ 
    &+& \mathbb{E}\, \ell_k(Z_e\mid \bm{\widehat{D}},{\widehat{\alpha}_k}) - \mathbb{E}\, \ell_k(Z_e\mid \bm{{D^\star}},{{\alpha^\star_k}})\bigg]\s 
    \\
    &\leq & \dsum_{k=2}^K  \dsum_{e\in \mathscr{S}_k} \frac{1}{\mu_e}\left[\mathbb{E}\, \ell_k(Z_e\mid \bm{\widehat{D}},{\widehat{\alpha}_k}) - \mathbb{E}\, \ell_k(Z_e\mid \bm{{D^\star}},{{\alpha^\star_k}})\right]\s
    \\
    &+& \sup\limits_{\bm{D},\, \bm{\alpha} \,}\left| \displaystyle\sum_{k=2}^K\sum_{e\in \mathscr{S}_k} \frac{1}{\mu_e}\ell(Z_e\mid \bm{{D}},\bm{{\alpha}}) - \mathbb{E}\, \ell(Z_e\mid \bm{{D}},\bm{{\alpha}})\right|
\ee

We use the Kullback–Leibler divergence to measure the discrepancy between hyperedge probability distribution, and introduce $\bm{Q}_k$ as the hyperedge probability tensor encompassing the probabilities defined in \eqref{A_eq_1} with $(\bm{\widehat{D}},\bm{\alpha}^\star)$. 
\hide{
i.e., 
\be
(\bm{Q}_k^{(1)})_{i_1,\ldots,i_k} &\coloneqq& \alpha^{\star}_k f_k^{(i_1,\ldots,i_k)}(\widehat{\bm{D}}),\;  (\bm{Q}_k^{(2)})_{i_1,\ldots,i_k} &\coloneqq& \widehat{\alpha}_k f_k^{(i_1,\ldots,i_k)}({\bm{D}}^{\star}).
\ee
}
The average KL-divergence and Hellinger distance for the $m$-size hypergraph are
\beno 
    D_k(\bm{P}_k^\star |\!| \widehat{\bm{P}}_k) &\coloneqq& \dfrac{1}{C(k)}\dsum_{1\leq i_1,\ldots, i_k \leq N} P^\star_{i_1,\ldots,i_k}\log \dfrac{P^\star_{i_1,\ldots,i_k}}{\widehat{P}_{i_1,\ldots,i_k}} + (1-P^\star_{i_1,\ldots,i_k}) \log \dfrac{(1-P^\star_{i_1,\ldots,i_k})}{(1-\widehat{P}_{i_1,\ldots,i_k})}\s\\
    D^H_k(\bm{P}_k^\star |\!| \widehat{\bm{P}}_k) &\coloneqq& \dfrac{1}{C(k)}\dsum_{1\leq i_1,\ldots, i_k \leq N} \left(\sqrt{P^\star_{i_1,\ldots,i_k}} -   \sqrt{\widehat{P}_{i_1,\ldots,i_k}}\right)^2 \s\\ &+& \left(\sqrt{1- P^\star_{i_1,\ldots,i_k}} -   \sqrt{1- \widehat{P}_{i_1,\ldots,i_k}}\right)^2,
\ee
respectively. 
Note that $D_k(\bm{P}_k^\star |\!| \widehat{\bm{P}}_k)\geq  D^H_k(\bm{P}_k^\star |\!| \widehat{\bm{P}}_k)$. Given that the sampling probability of a hyperedge $e = (i_1,\ldots,i_k)$ to be sampled in $ \mathscr{S}_k$ is $ \mu_e \geq L_k  |\mathscr{S}_k|/C(k)$, we have
\beno 
&-&\mathbb{E}\, (\frac{1}{\mu_e}\ell_k(Z_e\mid \bm{\widehat{D}},\bm{\widehat{\alpha}}) - \frac{1}{\mu_e}\ell_k(Z_e\mid \bm{{D}^\star},\bm{{\alpha^\star}}) ) = \mathbb{E}_{e\in \mathscr{S}_k}\mathbb{E}_{Z_e}  \frac{1}{\mu_e}\big\{ \ell_k(Z_e\mid \bm{{D}^\star},\bm{{\alpha^\star}}) -  \ell_k(Z_e\mid \bm{\widehat{D}},\bm{\widehat{\alpha}}) \big\} \\
&=&  \dsum_{1\leq i_1,\ldots, i_k \leq N} \mu_e \big\{\frac{1}{\mu_e} P^\star_{i_1,\ldots,i_k}\log \dfrac{P^\star_{i_1,\ldots,i_k}}{\widehat{P}_{i_1,\ldots,i_k}} + \frac{1}{\mu_e}(1-P^\star_{i_1,\ldots,i_k}) \log \dfrac{(1-P^\star_{i_1,\ldots,i_k})}{(1-\widehat{P}_{i_1,\ldots,i_k})}\s \big\}\\
&\geq& C(k)D_k(\bm{P}_k^\star |\!| \widehat{\bm{P}}_k).
\ee

\noindent Then we have 
\be
\label{A_eq_2_1}
D_k(\bm{P}_k^\star |\!| \widehat{\bm{P}}_k) &=& D_k(\bm{P}_k^\star |\!| \bm{Q}_k) + \bm{D}_k^{(2)}(\bm{P}_k^\star,\widehat{\bm{P}}_k,\bm{Q}_k), 
\ee
where
\beno 
\everymath{\displaystyle} 
  &~&\bm{D}_k^{(2)}(\bm{P}_k^\star,\widehat{\bm{P}}_k,\bm{Q}_k) \, = \\ 
  &~&\dfrac{1}{C(m)}\dsum_{1\leq i_1,\ldots, i_k \leq N} P^\star_{i_1,\ldots,i_k}\log \dfrac{Q_{i_1,\ldots,i_k}}{\widehat{P}_{i_1,\ldots,i_k}} + (1-P^\star_{i_1,\ldots,i_k}) \log \dfrac{(1-Q_{i_1,\ldots,i_k})}{(1-\widehat{P}_{i_1,\ldots,i_k})}. 
\ee

Based on \eqref{as:theta}, 
there exist constants $0< C_3 < C_4 < 1$ such that $f^{(i_1,\cdots, i_k)}_k(\bm{D}) \in [C_3,\, C_4]$ for $k \in \{2,\ldots, K\}$. 
Define $h(x) \coloneqq x_0 \log \frac{x_0}{x} + (1-x_0)\log \frac{1-x_0}{1-x}$, 
where $0<x<1$ and $x_0 = \alpha^{\star}_k\, {f}^{(e)}_k({\bm{D}^{\star}})$,
and denote $x_1 = \widehat{\alpha}_k {f}^{(e)}_k(\widehat{\bm{D}})$ and $x_2 = \alpha^{\star}_k {f}^{(e)}_k(\widehat{\bm{D}})$. Each term in $D_k(\bm{P}_k^\star |\!| \widehat{\bm{P}}_k) $ and $D_k(\bm{P}_k^\star |\!| \bm{Q}_k^{(1)})$ can then be represented as $h(x_1)$ and $h(x_2)$. We compare  $h(x_1)$ and $h(x_2)$ in the following.
Note that 
\beno 
h'(x) &=& \dfrac{x - x_0}{x\, (1-x)} \text{   and   } h''(x) &=& \dfrac{x^2 + x_0 - 2\, x_0\, x}{x^2\, (1-x)^2}.
\ee
We first perform Taylor expansion of $h(x_1)$ from $x_0$ and use the fact that $h(x_0) = 0, h'(x_0) = 0$, then $h(x_1) = \frac{x_1^2 + x_0 - 2x_0x_1}{x_1^2(1-x_1)^2}(x_1 - x_0)^2 + o_N(x_1 - x_0)^2 = \bm{\omega}_N(\widehat{\alpha}_k - \alpha^{\star}_k)^2$, where $a = \bm{\omega}_N(b)$ means that $|a(N) / b(N)| \geq C$ for large enough $N$ for some constant $C>0$. 
In the above claim, 
we use condition \ref{as:alpha}, 
where $\alpha_k \in [\rho_N,\, 1]$ can decreases as $N$ increases. 
Similarly, 
$h(x_2) = h(x_1) + \frac{x_1 - x_0}{x_1(1-x_1)}(x_2 - x_1) + O_N(x_2 - x_1)^2 = h(x_1) +  {O}_N (\frac{ ( \widehat{\alpha}_k - \alpha^{\star}_k)^2}{\widehat{\alpha}_k}) + {O}_N{( \widehat{\alpha}_k - \alpha^{\star}_k)^2}$. 
Then,
by combining $h(x_1)$ and $h(x_2)$, 
we have 
\beno
h(x_2)/h(x_1) \leq  1 +  O_N\left( \frac{ (\widehat{\alpha}_k - \alpha^{\star}_k)^2   }{  \widehat{\alpha}_k(\widehat{\alpha}_k - \alpha^{\star}_k)^2} \right) = O_N\left(\frac{1}{\widehat{\alpha}_k}\right) \leq \rho^{-1}_N. 
\ee
Notice that $D_k(\bm{P}_k^\star |\!| \widehat{\bm{P}}_k) = \frac{1}{C(k)} \dsum_{1\leq i_1,\ldots, i_k \leq N} \; h( x_2 )$ and $D_k(\bm{P}_k^\star |\!| \bm{Q}_k) = \frac{1}{C(k)} \dsum_{1\leq i_1,\ldots, i_k \leq N} \; h( x_1 )$, 

\noindent Then we can apply the above arguments for different $k = 1, \ldots, K$ and have 
\be
\label{A_eq_2_2}
    D_k(\bm{P}_k^\star |\!| \bm{Q}_k) &\leq& \rho_N^{-1} \, D_k(\bm{P}_k^\star |\!| \widehat{\bm{P}}_k),\; k = 1,\cdots, K 
\ee  
Combining (\ref{A_eq_2_1}) and (\ref{A_eq_2_2}), we have 
\be
\label{A_eq_3}
 (1-\rho_N^{-1}) D_k(\bm{P}_k^\star |\!| \widehat{\bm{P}}_k) 
    &\leq& \bm{D}_k^{(2)}(\bm{P}_k^\star,\, \widehat{\bm{P}}_k,\, \bm{Q}_k)    
    &\leq& D_k(\bm{P}_k^\star |\!| \widehat{\bm{P}}_k).
\ee

We define the multivariate function $F_k(\bm{D}): \mathbb{R}^{N\times N} \mapsto [0,\, 1]^{C(k)}$ as 
\beno 
F_k(\bm{D}) 
&\coloneqq& \left\{\sqrt{ \frac{1}{C(k)} f_k^{(e)}(\bm{D})}\right\}_{ e \in \mE_k}.
\ee
Notice that $F_k(\bm{D})$ is an overdetermined system and $F_k(\bm{D}) = 0$ only when $|\!|\bm{D}|\!|_F \rightarrow \infty$ due that $\bm{D}_{i,j}>0$ given Condition 3.
Therefore, $F_k(\bm{D})$ is an injective and continuous mapping. Define a new norm on $\bm{D}$ as $|\!| \bm{D} |\!|: = |\!| \bm{D} |\!|_F /\sqrt{C(2)}$ and then $|\!| \bm{D} |\!| = 1$ define a sphere $|\!|\bm{D}|\!|_F = \sqrt{C(2)}$. Given that compactness of  $|\!| \bm{D} |\!| = 1$ and continuity of $F_k(\bm{D})$ over $|\!| \bm{D} |\!| = 1$, then there exists $c^{(1)}>0$ such that $c^{(1)} = \inf_{|\!|\bm{D}|\!| = 1} |\!|F_k(\bm{D})|\!|_F$. Given Condition 3 that $\bm{D}_{i,j}$ is lower bounded, there exists constant $C_3'$ such that $f_k^{(e)}(\bm{D})\geq C'_3$ and $F_k(\bm{D})_{i_1,\cdots,i_k}\geq \sqrt{C'_3/C(k)}$ then $|\!| F_k(\bm{D})|\!|_F \geq \sqrt{ C_3'}$ for all $\bm{D}$, then we have
\beno
\left\|F_k\left(\frac{\bm{D}}{|\!|\bm{D}|\!|}\right)\right\|_F \geq c^{(1)} \geq \sqrt{ C_3'} \implies |\!|F_k(\bm{D})|\!|^2_F \geq  \frac{C_3'}{C(2)}|\!| \bm{D} |\!|^2_F. 
\ee
Consider ${\bm{D}^{\star}}$ which also satisfies $|\!|F_k({\bm{D}^{\star}})|\!|^2_F \geq  \frac{C_3'}{C(2)}|\!| {\bm{D}^{\star}} |\!|^2_F$ and notice that 
\beno
\frac{2}{C(k)} \sum_{e \in \mE_k} \sqrt{f_k^{e}(\bm{D})f_k^{e}({\bm{D}^{\star}})} \leq 2 \leq \frac{2}{C(2)} \sum_{1\leq i,j\leq N}\bm{D}_{i,j}\bm{D}^{\star}_{i,j},
\ee
where we use the fact that $f_k^{e}(\bm{D}), f_k^{e}(\bm{D}^{\star})\leq 1$ and $\bm{D}_{i,j}, \bm{D}^{\star}_{i,j}>1$ by Condition 3. Then we have 
\beno
|\!| F_k(\bm{D}) - F_k(\bm{D}^\star) |\!|^2_F &\geq& \frac{C_3'}{C(2)} |\!|\bm{D} -  \bm{D}^\star|\!|^2_F,
\ee
and,
upon replacing $\bm{D}$ by $\widehat{\bm{D}}$,
we obtain
\be
\label{A_eq_3_5}
    &~&\dfrac{1}{C(k)}\dsum_{1\leq i_1,\ldots, i_k \leq N} \left( \sqrt{P^\star_{i_1,\ldots,i_k}} -   \sqrt{{Q}_{i_1,\ldots,i_k}}\right)^2 
    \dfrac{1}{C(k)}\dsum_{1\leq i_1,\ldots, i_k \leq N} \left( \sqrt{P^\star_{i_1,\ldots,i_k}} -   \sqrt{{Q}_{i_1,\ldots,i_k}}\right)^2 \s
   \\ &=&  \alpha^{\star}_k |\!| F_k(\widehat{\bm{D}}) - F_k(\bm{D}^\star) |\!|^2_F 
    ~ \geq ~\dfrac{C_3'\alpha^{\star}_k}{C(2)} |\!|\widehat{\bm{D}} -  \bm{D}^\star|\!|^2_F \geq \dfrac{C_3'\rho_N}{C(2)} |\!|\widehat{\bm{D}} -  \bm{D}^\star|\!|^2_F.
\ee

Based on Taylor expansion on $h(x) = \sqrt{1-x_0} - \sqrt{1-x}$, we have for every $e = \{i_1,\cdots,i_k\}$
\beno
\left(\sqrt{1 - P^\star_{i_1,\ldots,i_k}} -   \sqrt{1 - Q^{(1)}_{i_1,\ldots,i_k}}\right)^2
 = \bm{\omega}_N( (\alpha^{\star}_k)^2( f_k^{(e)}(\bm{D}^{\star}) - f_k^{(e)}(\widehat{\bm{D}}))^2)
\ee
Then by defining a multivariate function $\widetilde{F}_k(\bm{D}):\mathbb{R}^{N\times N} \mapsto [0,1]^{C(k)}$ as
\beno 
\widetilde{F}_k(\bm{D}) &\coloneqq& \left\{ \sqrt{\frac{1}{C(k)}} f_k^{(e)}(\bm{D})\right\}_{e\in \mE_k},
\ee
and following above argument above, we can again bound $\widetilde{F}_k(\bm{D})$ from below with some constant $C_3''>0$ 
\be
\label{A_eq_4}
 &~&\dfrac{1}{C(k)}\dsum_{1\leq i_1,\ldots, i_k \leq N} \left(\sqrt{1 - P^\star_{i_1,\ldots,i_k}} -   \sqrt{1 - \widehat{P}_{i_1,\ldots,i_k}}\right)^2 
  \geq \frac{C_3''\rho^2_N}{C(2)} |\!|\widehat{\bm{D}} -  \bm{D}^\star|\!|^2_F.
\ee

Combining \eqref{A_eq_3_5} and \eqref{A_eq_4}, 
we obtain,
for some constant $C_4>0$,
\beno
D^H_k(\bm{P}_k^\star |\!| \bm{Q}_k) 
&\geq& \displaystyle\frac{C_4\, \rho_N}{C(2)}\, |\!|\widehat{\bm{D}} -  \bm{D}^\star|\!|^2_F.
\ee
Noting that $D^H_k(\bm{P}_k^\star |\!| \bm{Q}_k) \leq D_k(\bm{P}_k^\star |\!| \bm{Q}_k) \leq \rho_N^{-1} D_k(\bm{P}_k^\star |\!| \widehat{\bm{P}}_k)$, 
we have for $k = 2,\ldots, K$,
\be
\label{A_eq_5}
D_k(\bm{P}_k^\star |\!| \widehat{\bm{P}}_k) 
&\geq& \dfrac{C_4\, \rho^2_N}{N(N-1)} |\!| \widehat{\bm{D}} -  \bm{D}^\star|\!|^2_F. 
\ee


\noindent Given $\sum_{e\in \mathscr{S}_k} \mathbb{E}\, \frac{1}{\mu_e}\ell_k(Z_e\mid \bm{\widehat{D}},\bm{\widehat{\alpha}}) - \mathbb{E}\, \frac{1}{\mu_e}\ell_k(Z_e\mid \bm{{D}^\star},\bm{{\alpha^\star}}) \leq - C(k)D_k(\bm{P}_k^\star |\!| \widehat{\bm{P}}_k)$ and (\ref{A_eq_5}), we have 
\beno 
     &-&\dfrac{1}{|\mathscr{S}|}\,\mathbb{E}\dsum_{e\in \mathscr{S}} \frac{1}{\mu_e}\big( \ell(Z_e\mid \widehat{\bm{{D}}}, \widehat{\bm{{\alpha}}})- \ell(Z_e\mid \bm{{D}^\star},\bm{{\alpha}^\star}) \big) \\
    &=& -\dsum_{k=2}^K \dfrac{1}{|\mathscr{S}|}\,\mathbb{E}\frac{1}{\mu_e}\big(\dsum_{e\in \mathscr{S}_k} \ell_k(Z_e\mid \widehat{\bm{{D}}},\widehat{\alpha}_k) - \ell_k(Z_e\mid {\bm{{D}}^\star},\alpha^\star_k)\big)\s\\
    &\geq &   \dsum_{k=2}^K \frac{C(k)}{|\mathscr{S}|} D_k(\bm{P}_k^\star |\!| \widehat{\bm{P}}_k) \s\\
    &\geq&  \dsum_{k=2}^K \dfrac{C(k)}{|\mathscr{S}|} \dfrac{C_4 \rho_N^2}{N(N-1)} |\!| \widehat{\bm{D}} -  \bm{D}^\star|\!|^2_F. 
\ee
Since $(\widehat{\bm{D}},\widehat{\bm{\alpha}})$ maximizes the loglikelihood, 
i.e., 
$\sum_{e\in \mathscr{S}} \big(\ell(Z_e\mid \bm{\widehat{D}},\bm{\widehat{\alpha}}) - \ell(Z_e\mid\bm{{D^\star}},\bm{{\alpha^\star}})\big) \geq 0$, we get for \eqref{A_eq_2}
\be
\label{A_eq_7}
&~& \sup\limits_{\bm{D},\bm{\alpha}}\left|\dfrac{1}{|\mathscr{S}|}\dsum_{e\in \mathscr{S}} \frac{1}{\mu_e}\ell(Z_e\mid\bm{{D}},\bm{{\alpha}}) - \mathbb{E}\, \frac{1}{\mu_e}\ell(Z_e\mid \bm{{D}},\bm{{\alpha}})\right| \\ 
&\geq& - \dsum_{k=2}^K \frac{1}{|\mathscr{S}|} \dsum_{e\in \mathscr{S}_k}\mathbb{E}\, \frac{1}{\mu_e}\ell_k(Z_e\mid\bm{\widehat{D}},{\widehat{\alpha}_k}) - \mathbb{E}\, \frac{1}{\mu_e}\ell_k(Z_e\mid\bm{{D^\star}},{{\alpha^\star_k}}) \s\\
&\geq &  \dsum_{k=2}^K  \dfrac{C_4 C(k)\, \rho_N^2}{|\mathscr{S}|\, N \,(N-1)} |\!|\bm{D} -  \bm{D}^\star|\!|^2_F. 
\ee
Also notice that given $\mu_e \geq L_k\, |\mathscr{S}_k|\, /\, |\mE_k|$ for all $e \in \mathscr{E}_k$ and $k = 2, \ldots, K$, then
\begin{align}\label{A_eq_7_2}
  &\sup\limits_{\bm{D},\bm{\alpha}}\left|\dfrac{1}{|\mathscr{S}|}\dsum_{e\in \mathscr{S}} \frac{1}{\mu_e}\ell(Z_e\mid\bm{{D}},\bm{{\alpha}}) - \mathbb{E}\, \frac{1}{\mu_e}\ell(Z_e\mid \bm{{D}},\bm{{\alpha}})\right| \\ \leq &\max_{k}\frac{C(k)}{L_k\, |\mathscr{S}_k|} \times \sup\limits_{\bm{D},\bm{\alpha}} \dfrac{1}{|\mathscr{S}|} \sum_{k=2}^K \left|\dsum_{e\in \mathscr{S}_k} \ell_k(Z_e\mid\bm{{D}},\bm{{\alpha}}) - \mathbb{E}\, \ell_k(Z_e\mid \bm{{D}},\bm{{\alpha}})\right|  
\end{align}

Next, we bound the empirical process $\sup_{\bm{D},\bm{\alpha}}|1/|\mathscr{S}| \, \sum_{k=2}^K\sum_{e\in \mathscr{S}_k} \ell_k(Z_e\mid\bm{{D}},\bm{{\alpha}}) - \mathbb{E}\, \ell_k(Z_e\mid \bm{{D}},\bm{{\alpha}})|$ with high probability. 
Based on Condition 2 and 3, we define the parameter space for latent positions by 
\beno 
\bm{\Phi} &\coloneqq& \{ \bm{\Theta}\in \mathbb{R}^+ \times \mathbb{R}^r \mid  -C_2 \leq  \bm{\Theta}_{i,j} \leq C_2, j=2,\ldots, r+1\},
\ee
and the parameter space for hyperbolic Gram matrix $\bm{D}$ by 
\beno 
\bm{\Omega} &\coloneqq& \{\bm{\Theta}\bm{J}\bm{\Theta}^\top \mid \bm{\Theta}\in \mathbb{R}^+ \times \mathbb{R}^r, \;  -C_2 \leq  \bm{\Theta}_{i,j} \leq C_2, j=2,\ldots, r+1,\; D_{i,j} \geq C_3\}.
\ee
We introduce the $\delta$-covering number of a metric space $(\bm{X},\, d)$ as 
\beno 
N(\delta,\, \bm{X},\, d) \coloneqq \inf\{N \in \mathcal{N}: \exists \; \text{a $\delta$-covering $x_1,\ldots, x_N$ of $\bm{X}$}\}.
\ee
In this context, the set $\{a_1, ..., a_k\}$ for some $M\in\{1, 2, ... \}$ is a $\delta$-covering of $X$ if there exists some $k = 1, ..., M$ for any $x \in X$ such that $d(x,a_k) \leq \epsilon$.
Then we have $N(\delta,\, \bm{\Phi},\, |\!| \cdot |\!|_{\infty}) \leq (1+C_2/\delta)^{N r}$,
where $|\!|\cdot|\!|_{\infty}$ denotes the $\ell_{\infty}$-norm. 
Then, 
for any $\bm{D} = \bm{\Theta}\, \bm{J}\, \bm{\Theta}^\top \in \bm{\Omega}$, 
we can find $\bm{\widetilde{\Theta}}\in \bm{\Phi}$ such that $|\!|\bm{\Theta} - \bm{\widetilde{\Theta}}|\!|_{\infty} \leq \delta$,
and 
\beno 
    |\!|\bm{D} - \bm{\widetilde{D}}|\!|_{\infty} 
    &=& |\!| \bm{\Theta}\, \bm{J}\, \bm{\Theta}^\top - \bm{\widetilde{\Theta}}\, \bm{J}\, \bm{\widetilde{\Theta}}^\top |\!|_{\infty} 
    &\leq& 2\, |\!| \bm{\widetilde{\Theta}}\, \bm{J}\, (\bm{\Theta} - \bm{\widetilde{\Theta}})|\!|_{\infty} 
    &\leq& 6\, C_2\, \delta.
\ee
Therefore, we obtain 
\beno 
N(6\, C_2\, \delta,\, \bm{\Omega},\, |\!|\cdot |\!|_{\infty})&\leq& N(\delta,\, \bm{\Phi},\, |\!|\cdot |\!|_{\infty}) \s\\
N(\delta,\, \bm{\Omega},\, |\!|\cdot |\!|_{\infty}) &\leq& \left(1 + \dfrac{6\, C_2^2}{\delta}\right)^{N\, r}.
\ee 

Now we introduce the element-wise loss function for a size-$k$ hyperedge $e = (i_1,\ldots,i_k)$ as
\beno 
\ell_k(e) &=& \mathbf{1}_{\{e \in \mathscr{S}_k\}}\big(Z_e\log \alpha_k f^{(e)}_k(\bm{D}) + (1-Z_e) \log (1- \alpha_k f^{(e)}_k(\bm{D})) \big),
\ee
where $\mathbf{1}_{\{\cdot\}}$ is the indicator random variable, and is 1 for $e\in \mathscr{S}_{k}$. 
With the function space $\bm{L}_k := \{ \ell_k(Z_e \mid \alpha_k,\bm{D}),\; \alpha_k\in [C_1,1],\; \bm{D}\in \bm{\Omega}\}$, we get for any $\bm{D},\widetilde{\bm{D}}\in \bm{\Omega}$
and any $\alpha_k, \widetilde{\alpha}_k \in [C_1,1]$:   \beno 
&&| \ell_k(Z_e \mid  \alpha_k,\bm{D}) -  \ell_k(Z_e \mid \widetilde{\alpha}_k,\widetilde{\bm{D}})| \\
&=& \left|\mathbf{1}_{\{(i_1,\ldots,i_k)\}} \left( Z_e \log \dfrac{\alpha_k f^{(e)}_k(\bm{D})}{\widetilde{\alpha}_k f^{(e)}_k(\widetilde{\bm{D}})} + (1-Z_e) \log \dfrac{1 - \alpha_k f^{(e)}_k(\bm{D})}{1 - \widetilde{\alpha}_k f^{(e)}_k(\widetilde{\bm{D}})} \right) \right| \s\\
&\leq & |\mathbf{1}_{\{(i_1,\ldots,i_k)\}}| \\
&\times &\left|\dfrac{A^{e}_k }{f_k^{(e)}(\bm{\epsilon})(1 - f_k^{(e)}(\bm{\epsilon}))} \, \dsum_{(i,j)\subset e}\dfrac{\partial f^{(e)}_k(\bm{\epsilon})}{\partial \bm{D}_{i,j}} (\bm{D}_{i,j} -  \widetilde{D}_{i,j})  +    \dfrac{A^{e}_k}{\alpha(1-\alpha f_k^{(e)}(\bm{\epsilon}))} \, (\alpha_k - \widetilde{\alpha}_k)\right|\s\\
&\leq & |\mathbf{1}_{\{(i_1,\ldots,i_k)\}}|\, \bigg(\dsum_{(i,j)\subset e}\dfrac{1}{f^{(e)}_k(\bm{\epsilon})(1 - f^{(e)}_k(\bm{\epsilon}))}\left|\dfrac{\partial f_k^{(e)}(\bm{\epsilon})}{\partial \bm{D}_{i,j}}\right| \,  
 |\bm{D}_{i,j} - \widetilde{D}_{i,j}| \\ 
 &+&  \dfrac{1}{\alpha (1-\alpha f_k^{(e)}){(\bm{\epsilon}})}|\alpha_k - \widetilde{\alpha}_k|\bigg)\s\\
&\leq & \max\left( \left|\dfrac{\partial f_k^{(e)}(\bm{\epsilon})}{f^{(e)}_k(\bm{\epsilon})(1 - f^{(e)}_k(\bm{\epsilon}))\, \partial \bm{D}_{i,j}}\right|,\dfrac{1}{\alpha(1-\alpha f_k^{(e)}(\bm{\epsilon}))}\right)\, (k+1) \s\\
&\times&\max(|\!|\bm{D} - \widetilde{\bm{D}}|\!|_{\infty}, |\alpha_k - \widetilde{\alpha}_k |),
\ee
where $A^{e}_k = Z_e(1 - \alpha f_k^{(e)}(\bm{\epsilon})) + (1-Z_e)(- \alpha f_k^{(e)}(\bm{\epsilon}))$, and $\alpha \in (\alpha_k,\widetilde{\alpha}_k)$ and $\bm{\epsilon}= \{\epsilon_{i,j}\},\; \epsilon_{i,j}\in (\bm{D}_{i,j}, \widetilde{D}_{i,j})$. 
Based on Conditions 1, 2 and 3, 
there exist constants $C(C_1,C_2,C_3)>0$ and $C'(C_1,C_2,C_3)>0$ where $C$ and $C'$ are functions of $C_1,C_2,C_3$ such that 
\beno 
\left|\dfrac{1}{f^{(e)}_k(\bm{\epsilon})(1 - f^{(e)}_k(\bm{\epsilon}))}
\frac{\partial f_k^{(e)}(\bm{\epsilon})}{\partial \bm{D}_{i,j}}\right| &\leq& \max\limits_{i\in e}\left( \dsum_{j\in e} \bm{\epsilon}_{i,j} \right) \, \left( \dfrac{1}{\dsum_{j\in e} \bm{\epsilon}_{i,j}} + \dfrac{1}{\sum_{k\in e} \bm{\epsilon}_{jk}}\right) \, \dfrac{1}{\sqrt{\bm{\epsilon}^2_{i,j} -1 }} \\ &\leq& C \s\\ 
\dfrac{1}{\alpha(1-\alpha f_k^{(e)}(\bm{\epsilon}))} &\leq& C'.
\ee
Therefore,
we have with $C_0 \coloneq \max(C,\, C')$:
\be
\label{A_eq_6}
| \ell_k(Z_e \mid  \alpha_k,\bm{D}) -  \ell_k(Z_e \mid \widetilde{\alpha}_k,\widetilde{\bm{D}})| &\leq& C_0(k+1) \, \max(|\!|\bm{D} - \widetilde{\bm{D}}|\!|_{\infty}, |\alpha_k - \widetilde{\alpha}_k |).
\ee
In the following, 
we consider the function space $(\bm{L}_k, d_{\bm{L}_k} )$ defined above with the metric as $d_{\bm{L}_k}((\bm{D},\alpha_k), (\widetilde{\bm{D}},\widetilde{\alpha}_k)): = \max(|\!| \bm{D} - \widetilde{\bm{D}}|\!|_{\infty}, |\alpha_k - \widetilde{\alpha}_k |)$. 
Based on Lemma 2.14 in \citetsupp{sen2018gentle}, the bracketing number $N_{[\;]}(2\delta|\!|l_{\max}|\!|_F, \bm{L}_k, |\!|\cdot|\!|_F) \leq N(\delta, [C_1,1]\otimes \bm{\Omega}, d_{\bm{L}_k})$ where $l_{\max} = C_0(k+1)$, and $\otimes$ denotes the Cartesian product between two spaces.  
\beno 
  && N(\delta,\, [C_1,\, 1] \otimes \bm{\Omega},\, d_{\bm{L}_k}) \s\\
   &\leq& N(\delta,\, \bm{\Omega},\, |\!|\cdot|\!|_{\infty}) \, N(\delta,\, [C_1,\, 1], |\cdot|) \s\\
   &\leq& \left(1+\dfrac{6\, C_2^2}{\delta}\right)^{N\, r}\, \left(1+\dfrac{1-C_1}{\delta}\right)\s\\ 
   &\leq& \left( 1 + \dfrac{\max(6\, C_2^2, 1-C_1)}{\delta}\right)^{N\, r+1},
\ee
which yields
\beno 
N_{[\;]}(\delta,\, \bm{L}_k,\, |\!|\cdot|\!|_F) &\leq& \left( 1 + \dfrac{2\, C_0\, (k+1)\max(6\, C_2^2,\, 1-C_1)}{\delta}\right)^{N\, r+1}.
\ee

Notice that the above argument holds for $k \in \{2, \ldots, K\}$, 
which implies the following result for the loglikelihood function:
\beno 
\left|  \ell(Z_e \mid \bm{D}, \bm{\alpha}_k) - \ell(Z_e \mid \widetilde{\bm{D}}, \widetilde{\bm{\alpha}}_k)   \right| &\leq& \dsum_{k=2}^K \left| \ell_k(Z_e \mid \bm{D},\bm{\alpha}_k) -   \ell_k(Z_e \mid \widetilde{\bm{D}},\widetilde{\bm{\alpha}}_k \right|\s\\
    &\leq& C_0\dsum_{k=2}^K (k+1)\,  \max(|\!|\bm{D} - \widetilde{\bm{D}}|\!|_{\infty}, |\!|\bm{\alpha} - \widetilde{\bm{\alpha}}|\!|_{\infty})
\ee
Similarly, we consider the function space $(\bm{L}, d_{\bm{L}} )$ with 
\beno 
\bm{L} &\coloneqq& \left\{ \ell(\cdot \mid \bm{D},\bm{\alpha}),\; \bm{\alpha} \in [C_1,1]^{K-1},\; \bm{D}\in \bm{\Omega}\right\}
\ee
and metric as $d_{\bm{L}}(( \widetilde{\bm{D}},\widetilde{\bm{\alpha})}, (\bm{D},\bm{\alpha})): = \max(|\!|\bm{D} - \widetilde{\bm{D}}|\!|_{\infty}, |\!|\bm{\alpha} - \widetilde{\bm{\alpha}}|\!|_{\infty})$. In that context, we get 
\beno 
  N(\delta, [C_1,1]^{K-1} \otimes \bm{\Omega}, d_{\bm{L}}) &\leq& N(\delta, \bm{\Omega}, |\!|\cdot|\!|_{\infty})\, N(\delta, [C_1,1]^{K-1}, |\!|\cdot|\!|_{\infty})
\s\\    &\leq& \left( 1 + \dfrac{\max(6\,C_2^2, 1-C_1)}{\delta}\right)^{Nr+K-1},
\ee
and $N_{[\;]}(2\delta|\!|l_{\max}|\!|_F, \bm{L}, |\!|\cdot|\!|_F) \leq N(\delta, [C_1,1]^{K-1} \otimes \bm{\Omega}, d_{\bm{L}})$ where $l_{\max} = C_0\sum_{k=2}^K(k+1)$. Then we have 
\beno 
N_{[\;]}(\delta, \bm{L}, |\!|\cdot|\!|_F) &\leq& \left( 1 + \dfrac{C_0(K+4)(K-1)\max(6C_2^2, 1-C_1)}{\delta}\right)^{Nr+K-1}.
\ee
Then, 
from Theorem 4.12 in \citetsupp{sen2018gentle},
\be
\label{A_eq_8}
&~& \mathbb{E}\sup\limits_{\bm{D} \in \bm{\Omega},\, \bm{\alpha} \in [C_1,\, 1]^{K-1}} \left|\dfrac{1}{|\mathscr{S}|}\dsum_{e\in \mathscr{S}_k} \ell(Z_e\mid\bm{{D}},\bm{{\alpha}}) - \mathbb{E}\, \ell(Z_e\mid \bm{{D}},\bm{{\alpha}})\right|\s
\\
&\leq& \dfrac{\dint_0^{C_{\max}} \sqrt{\log N_{[\;]}(\delta, \bm{L}, |\!|\cdot|\!|_F)} \bm{d}\delta }{\sqrt{|\mathscr{S}|}}\s
\\  
&\leq& \dfrac{C_0\, \sqrt{(N\, r + K - 1)\, (K+4)\, (K+1)}}{\sqrt{|\mathscr{S}|}},
\ee
where $C_{\max} \coloneqq |\!|\sup_{l\in \bm{L}}\ell(Z_e)|\!|_{F}$ is independent of $N$ and $|\mathscr{S}|$, and\break 
$C \coloneqq 2\, \sqrt{2\; C_{\max}\, C_0\, \max(6\, C_2^2,\, 1-C_1)}$. 
Then,
by combining (\ref{A_eq_7}), (\ref{A_eq_7_2}) and (\ref{A_eq_8}), 
and using Markov inequality we have with probability at least $1 - \epsilon$:
\beno 
\dfrac{1}{N(N-1)} |\!|\!|\widehat{\bm{D}}-\bm{D}|\!|\!|_F^2 
&\leq& C\, \left(\dsum_{k=2}^K \dfrac{C_4\, |\mathscr{S}_k|\, L_k}{|\mathscr{S}|}\right)^{-1}
\dfrac{\sqrt{(Nr + K-1)\, (K+4)\, (K+1)}}{\epsilon\, \rho_N^2\, \sqrt{|\mathscr{S}|}},
\ee
where $C > 0$ is a constant.

Last,
but not least,
we establish non-asymptotic error bounds on estimators of $\bm{\alpha}$. 
By \eqref{A_eq_7} and (\ref{A_eq_7_2}), 
we have  
\beno 
\sup_{\bm{D},\bm{\alpha}}\left|\dfrac{1}{|\mathscr{S}|}\dsum_{e\in \mathscr{S}_k} \ell(Z_e\mid\bm{{D}},\bm{{\alpha}}) - \mathbb{E}\, \ell(Z_e\mid \bm{{D}},\bm{{\alpha}})\right| 
&\geq&  \dfrac{L_k |\mathscr{S}_k|}{|\mathscr{S}|} D_k(\bm{P}_k^\star |\!| \widehat{\bm{P}}_k). 
\ee
Also from \eqref{A_eq_3}, 
we have $\bm{D}_k^{(2)}(\bm{P}_k^\star,\widehat{\bm{P}}_k,\bm{Q}_k) \leq \max(1,|1-\rho_N^{-1}|)D_k(\bm{P}_k^\star |\!| \widehat{\bm{P}}_k)$. 
Then,
by similarly using (\ref{A_eq_8}) and the Markov inequality, 
we have with probability at least $1 - \epsilon$:
\beno 
\bm{D}_k^{(2)}(\bm{P}_k^\star,\widehat{\bm{P}}_k,\bm{Q}_k) 
&\leq& C\, \dfrac{|\mathscr{S}|\, \sqrt{(N\, r + K - 1)\, (K+4)\, (K+1)}}{\epsilon\, \rho_N\, L_k\, \sqrt{|\mathscr{S}|}\, |\mathscr{S}_k|}. 
\ee
Finally,
by defining $h(x): =  a\log \frac{b}{x} + (1-a) \log \frac{1-b}{x}$ with $a = \alpha^{\star}_k\, f^{(e)}_k({\bm{D}^{\star}})$ and $b = \alpha^{\star}_k\, f^{(e)}_k(\widehat{\bm{D}})$, we use a first-order Taylor expansion for $x = \widehat{\alpha}_k\, f^{(e)}_k(\widehat{\bm{D}})$ at $b$, 
we obtain 
\beno
\bm{D}_k^{(2)}(\bm{P}_k^\star,\widehat{\bm{P}}_k,\bm{Q}_k) 
&\geq& C_5\, |\widehat{\alpha}_k - {\alpha}_k |, 
\ee 
where $C_5$ is independent of $N$ and $|\mathscr{S}|$ based on Conditions 1 and 3. 

\subsection{Proof of Theorem 2}
\label{sec:proof_theo2}

\begin{theorem}
Under Conditions \ref{as:alpha}--\ref{as:eigenvalue}, 
assuming that $N$ and $|\mathscr{S}|$ increase without bound,
\beno
\dfrac{1}{N} \inf_{\bm{R} \in \mathscr{R}}\, |\!|\!|\widehat{\bm{\Theta}}\, \bm{R}  - \bm{\Theta}^\star|\!|\!|_F^2  
&=& O_p\left(\dfrac{\Delta_{N,r}}{\rho_N^2\, \sqrt{|\mathscr{S}|}}\right), 
\ee
where $\mathscr{R} \coloneqq \{\bm{R} \in \mathbb{R}^{(r+1) \times (r+1)}:\; \bm{R}\, \bm{J}\, \bm{R}^\top = \bm{J}\}$ is the set of hyperbolic rotation matrices.
\end{theorem} 

According to Theorem 1,
we know that $|\!|\widehat{\bm{D}} - \bm{D}^\star|\!|^2_F = O_p(N^2/(\rho_N^2 M))$, 
where $M = \sqrt{|\mathscr{S}|}/\Delta_{N,r}$. 
We perform a singular value decomposition (SVD) of $\bm{D} = \bm{U} \bm{S}\, \bm{U}^\top$, 
with a matrix consisting of the left singular vectors $\bm{U} \in \mathbb{R}^{N\times (r+1)}$ and the diagonal matrix of all singular values $\bm{S} = \text{diag}(s_1, s_2, \cdots, s_{r+1}) \in \mathbb{R}^{(r+1)\,\times\,(r+1)}$, 
and denote its estimator by $\widehat{\bm{D}} = \widehat{\bm{U}}\widehat{\bm{S}}\, \widehat{\bm{U}}^\top$. 
Based on \citetsupp{tabaghi2020hyperbolic}, we can arrange eigenvalues as $s_1< 0 < s_2 < \cdots < s_{r+1}$. 
Define the operator norm induced by a vector norm $|\!|\cdot|\!|$ by $|\!| \bm{A} |\!|_{op} = \sup_{|\!|\mathbf{x}|\!| = 1} |\!| \bm{A} \mathbf{x} |\!|$, where the supremum is taken over all non-zero vectors $\mathbf{x}  \in \mathbb{R}^n$, and $|\!| \cdot |\!|$ denotes the vector $l_2$ norm.   
Given that $|\!| \bm{D} |\!|_{op} \leq |\!|\bm{D}|\!|_F \asymp N$ and Condition 5 hold, we have $|s_i| \asymp N,\; i=1,\cdots,r+1$. 
Therefore, 
$|\widehat{s}_i| \asymp N$ ($i=1,\cdots, r+1$) by \citetsupp{tang2018limit}.

In the following, we establish the result for $\widetilde{\bm{\Theta}}:= \bm{U}\, |\bm{S}|^{1/2}\bm{J}$ relying on Theorem 3 in \citetsupp{rubin2022statistical}. We follow the proof 
of Theorem 3 in \citetsupp{rubin2022statistical} by \( {\bm{D}} \) and \( \widehat{\bm{D}} \). We split the columns of $\bm{U} = \big[ \bm{U}_{(-)}\mid \bm{U}_{(+)} \big]$ where $\bm{U}_{(-)}$ and $\bm{U}_{(+)}$ are eigenvectors corresponding to negative and positive eigenvalues, respectively. We use same notations for $\widehat{\bm{U}}$. We check and decompose the matrix $\bm{U}^\top\widehat{\bm{U}}$ as
\beno 
    \bm{U}^{\top} \hat{\bm{U}} &=&  \left[ \begin{array}{l|l}
\bm{U}_{(-)}^{\top} \hat{\bm{U}}_{(-)} & \bm{U}_{(-)}^{\top} \hat{\bm{U}}_{(+)} \\
\hline \bm{U}_{(+)}^{+} \bm{U}_{(-)} & \bm{U}_{(+)}^{\top} \hat{\bm{U}}_{(+)}
\end{array} \right] \in \mathbb{R}^{(r+1)\times (r+1)}
\ee
We perform a singular value decomposition on $\bm{U}_{(+)}^{\top} \hat{\bm{U}}_{(+)} = \mathbf{W}_{(+), 1} \boldsymbol{\Sigma}_{(+)} \mathbf{W}_{(+), 2}^{\top}$ and $\bm{U}_{(-)}^{\top} \hat{\bm{U}}_{(-)} = \mathbf{W}_{(-), 1} \boldsymbol{\Sigma}_{(-)} \mathbf{W}_{(-), 2}^{\top}$. And we define matrix $\bm{W}$ as
\beno 
    \bm{W}&=&  \left[ \begin{array}{l|l}
\mathbf{\mathbf { w } _ { ( - ) } ^ { * }} & \mathbf{0} \\
\hline \mathbf{0} & \mathbf{w}_{(+)}^*
\end{array}  \right] \in \mathbb{R}^{(r+1)\times (r+1)},
\ee
where $\bm{ W } _ { ( + ) } ^ { * } = \mathbf{W}_{(+), 1} \mathbf{W}_{(+), 2}^{\top}$ and $\bm{ W } _ { ( - ) } ^ { * } = \mathbf{W}_{(-), 1} \mathbf{W}_{(-), 2}^{\top}$. Notice that $\mathbf{W}$ is an orthogonal matrix and $\bm{W}\bm{J}\bm{W}^\top = \bm{J}$. 
In the following, 
we show that $\bm{U}^\top\widehat{\bm{U}}$ is close to $\bm{ W }$. 
First, 
we have 
\beno 
|\!|\bm{U}_{(+)}^{\top} \hat{\bm{U}}_{(+)}-\mathbf{W}_{(+)}^{\star}|\!|_F
&=& |\!|\boldsymbol{\Sigma}_{(+)}-\mathbf{I}|\!|_F 
&\leq& |\!|\bm{U}_{(+)} \bm{U}_{(+)}^{\top}-\hat{\bm{U}}_{(+)} \hat{\bm{U}}_{(+)}^{\top}|\!|_F^2 .
\ee
Upon invoking the Davis-Kahan theorem, 
we obtain
\beno
|\!|\bm{U}_{(+)}^{\top} \hat{\bm{U}}_{(+)}-\mathbf{W}_{(+)}^{\star}|\!|_F 
&\leq& |\!|\bm{U}_{(+)} \bm{U}_{(+)}^{\top}-\hat{\bm{U}}_{(+)} \hat{\bm{U}}_{(+)}^{\top}|\!|_F^2 
&=& O_p\left(\dfrac{|\!|\bm{D}-\widehat{\bm{D}}|\!|^2}{s^2_{\min}(\bm{D})}\right),
\ee
where $s_{\min}(\bm{D})$ is the smallest eigenvalue of $\bm{D}$ in magnitude. 
Based on Condition 5, 
we have $s_{\min}(\bm{D}) \asymp N$,
hence
\beno 
|\!|\bm{U}_{(+)}^\top \widehat{\bm{U}}_{(+)} - \bm{W}^\star_{(+)}|\!|_F 
&=& O_p\left(\dfrac{1}{\rho_N^2\, M}\right).
\ee
We have same result for $|\!|\bm{U}_{(-)}^\top \widehat{\bm{U}}_{(-)} - \bm{W}^\star_{(-)} |\!|_F = O_p\left(1/(\rho_N^2\, M)\right)$. 
Next we bound $|\!|\bm{U}_{(+)}^\top \widehat{\bm{U}}_{(-)} |\!|_F$, 
i.e., 
we bound each $(i,j)$ element in $\bm{U}_{(+)}^\top \widehat{\bm{U}}_{(-)}$ for $i = 1,\cdots, r$ and $j = 1$. We denote the $i$-th column of $\bm{U}_{(+)}$ as $\bm{u}_{i}$ and the corresponding eigenvalues are $s_{i+1}$ for $i =1, \cdots, r$. Then 
\beno 
(\bm{U}_{(+)}^\top \widehat{\bm{U}}_{(-)})_{i,1}  &=&\left(\hat{s}_{1}-s_{i+1}\right)^{-1}\left(\boldsymbol{u}_{i}\right)^{\top}(\widehat{\bm{D}}-\bm{D}) \widehat{\bm{U}}_{(-)} \\
&=&\left(\hat{s}_{1}-s_{i+1}\right)^{-1}\left(\boldsymbol{u}_{i}\right)^{\top}(\widehat{\bm{D}}-\bm{D}) \bm{U}_{(-)} \bm{U}_{(-)}^{\top} \widehat{\bm{U}}_{(-)}  \\
& +&\left(\hat{s}_{1}-s_{i+1}\right)^{-1}\left(\boldsymbol{u}_{i}\right)^{\top}(\widehat{\bm{D}}-\bm{D})\left(\mathbf{I}-\bm{U}_{(-)} \bm{U}_{(-)}^{\top}\right) \widehat{\bm{U}}_{(-)}.
\ee
We first consider term \( \bm{u}_{i}^\top (\widehat{\bm{D}} - \bm{D})\, \bm{U}_{(-)} \in \mathbb{R}\)
\beno 
|\!| \bm{u}_{i}^\top (\widehat{\bm{D}} - \bm{D})\, \bm{U}_{(-)} |\!|_F
&=& \text{trace}\left((\widehat{\bm{D}} - \bm{D})\, \bm{U}_{(-)}\, \bm{u}^\top_{i} \right) 
\\   
&\leq& |\!|\widehat{\bm{D}} - \bm{D}|\!|_{op} \, |\!|\bm{U}_{(-)}\, \bm{u}^\top_{i}|\!|_{op}\s
\\
&=& O_p\left(\dfrac{N}{\sqrt{\rho_N^2 M}}\right),
\ee
where we use $|\!|\widehat{\bm{D}} - \bm{D}|\!|_{op} \leq |\!|\widehat{\bm{D}} - \bm{D}|\!|_{F} \asymp N/(\rho_N^2 M)^{1/2}$ based on the von Neumann trace inequality along with $|\!| \bm{U}_{(-)}\, \bm{u}^\top_{i}  |\!|_{\text{op}} = 1$. 
In addition, 
using the fact that $\hat{s}_i, s_i \asymp N$ ($i = 1, \cdots, r$), 
we obtain
\beno 
|\!|\left(\hat{s}_{1}-s_{i+1}\right)^{-1}\left(\boldsymbol{u}_{i}\right)^{\top}(\widehat{\bm{D}}-\bm{D}) \bm{U}_{(-)} \bm{U}_{(-)}^{\top} \widehat{\bm{U}}_{(-)} |\!|_F &=& O_p\left(\dfrac{1}{\sqrt{\rho_N^2\, M}}\right)
\ee
We notice that
\beno 
 |\!|\bm{u}_{i}^\top (\widehat{\bm{D}} - \bm{D})|\!|_F
&\leq& |\!|\widehat{\bm{D}} - \bm{D}|\!|_{\text{op}} &=& O_p\left(\dfrac{N}{\sqrt{\rho_N^2\, M}}\right),
\ee
and,
according to the Davis-Kahan theorem, 
\beno 
|\!|\left(\mathbf{I}-\bm{U}_{(-)} \bm{U}_{(-)}^{\top}\right) \widehat{\bm{U}}_{(-)}|\!|_F 
&=& |\!|(\mathbf{I}-\bm{U}_{(-)} \bm{U}_{(-)}^{\top}) \widehat{\bm{U}}_{(-)}\widehat{\bm{U}}_{(-)}^{\top}|\!|_F 
&\leq& |\!|\bm{U}_{(-)} \bm{U}_{(-)}^{\top} - \widehat{\bm{U}}_{(-)}\widehat{\bm{U}}_{(-)}^{\top}|\!|_F \s
\\
&&&=& O_p\left(\dfrac{|\!|\bm{D}-\widehat{\bm{D}}|\!|^2}{s^2_{\min}(\bm{D})}\right)\s
\\
&&&=& O_p\left(\dfrac{1}{\rho_N^2\, M}\right).
\ee
Then we have 
\beno 
|\!|(\hat{s}_{1}-s_{i+1})^{-1}\left(\boldsymbol{u}_{i}\right)^{\top}(\widehat{\bm{D}}-\bm{D})\left(\mathbf{I}-\bm{U}_{(-)} \bm{U}_{(-)}^{\top}\right) \widehat{\bm{U}}_{(-)}|\!|_F &=& O_p\left(\dfrac{1}{(\rho_N^2\, M)^{3/2}}\right).
\ee
Thus we have $|\!|\bm{U}_{(+)}^\top \widehat{\bm{U}}_{(-)})|\!|_F = O_p(1/\sqrt{\rho_N^2\, M})$. 
Upon combining the results on the diagnoal block, 
we obtain
\beno
|\!| \bm{U}^{\top}\widehat{\bm{U}} - \bm{W}|\!|_F 
&=& O_p\left(\dfrac{1}{\sqrt{\rho_N^2\, M}}\right). 
\ee


Given the matrix equation $\hat{\bm{U}} \hat{\bm{S}}-(\widehat{\bm{D}}-\bm{D}) \hat{\bm{U}}=\bm{D} \hat{\bm{U}}$, we have the matrix series expansion following proof of Theorem 3 in  Supplement C.2 of  \citetsupp{rubin2022statistical}. 
Specifically, we place $A - P$ as $\widehat{\mathbf{D}}-\mathbf{D}$, and obtain 
\beno
\hat{\mathbf{U}}|\hat{\mathbf{S}}|^{1 / 2}  &=&  \dsum_{k=0}^{\infty}(\widehat{\mathbf{D}}-\mathbf{D})^k \mathbf{U S} \mathbf{U}^{\top} \hat{\mathbf{U}} \mathbf{J}^{k+1}|\hat{\mathbf{S}}|^{-k-1 / 2} \\
&= &  \mathbf{U}|\mathbf{S}|^{1/2}\bm{W} + (\widehat{\mathbf{D}}-\mathbf{D})\mathbf{U} |\mathbf{S}|^{-1 / 2} \mathbf{W} \mathbf{J} + 
\underbrace{\dsum_{k=2}^{\infty}(\widehat{\mathbf{D}}-\mathbf{D})^k \mathbf{U} |\mathbf{S}|^{-k+1 / 2} \mathbf{W} \mathbf{J}^{k+2}}_{\bm{R}_{V_1}}\\
& +&\underbrace{\dsum_{k=0}^{\infty}(\widehat{\mathbf{D}}-\mathbf{D})^k \mathbf{U} \mathbf{J}|\mathbf{S}|^{-k+1 / 2}\left(\mathbf{U}^{\top} \hat{\mathbf{U}}-\mathbf{W}\right) \mathbf{J}^{k+1}}_{V_2} \\
& + &\underbrace{\dsum_{k=0}^{\infty}(\widehat{\mathbf{D}}-\mathbf{D})^k \mathbf{U} \mathbf{S}\left(\mathbf{U}^{\top} \hat{\mathbf{U}} \mathbf{J}^{k+1}|\hat{\mathbf{S}}|^{-k-1 / 2}-|\mathbf{S}|^{-k-1 / 2} \mathbf{U}^{\top} \hat{\mathbf{U}} \mathbf{J}^{k+1}\right)}_{V_3}
\ee
Following Lemma 12 in \citetsupp{rubin2022statistical}, 
we have the following expansion:
\be\label{ap_1}
\hat{\mathbf{U}}|\hat{\mathbf{S}}|^{1 / 2} &=& {\mathbf{U}}|{\mathbf{S}}|^{1 / 2} \, \bm{W} + (\widehat{\bm{D}} - \bm{D})\, \bm{U}\, |\bm{S}|^{-1/2}\, \bm{W}\, \bm{J} + \bm{R}_{V_1} + \bm{V}_2 + \bm{V}_3.
\ee
We estimate the order of all terms on the left hand side except ${\mathbf{U}}|{\mathbf{S}}|^{1 / 2}\bm{W}$ following the proof in C.2 of Theorem 3. 
Specifically, 
we replace $|\!|\cdot |\!|_{2\rightarrow \infty}$ by $|\!|\cdot|\!|_{F}$ using the fact that for any matrix $\bm{A}\in \mathbb{R}^{n\times n}$, $|\!|\bm{A}|\!|_F \leq \sqrt{ \text{rank}(\bm{A})}\, |\!|\bm{A}|\!|_{op} \leq \sqrt{ \text{rank}(\bm{A})\, n}|\!|\bm{A}|\!|_{2\rightarrow \infty}$. 

We obtain the asymptotic order of four above residual terms similar to the argument in sections (\text{C.2.1}), (\text{C.2.2}), and (\text{C.2.3}) of the proof for  Theorem 3 from \citetsupp{rubin2022statistical}. 
First, 
given $\text{rank}(\widehat{\bm{D}} - \bm{D}) = O(1)$ and the submultiplication of $|\!|\cdot|\!|_F$,
we have
\beno 
&|\!|(\widehat{\bm{D}} - \bm{D})\, \bm{U}|\bm{S}|^{-1/2}\, \bm{W}\, \bm{J}|\!|_F 
    = O_p((N/(\rho_N^2 M)^{1/2}) \, N^{-1/2}) 
    = O_p((N/(\rho_N^2 M))^{1/2}).
\ee
With regard to $\bm{R}_{V_1}$, 
notice that $|\!|\bm{U}|\!|_F,\, |\!|\bm{W}|\!|_F,\, |\!|\bm{J}|\!|_F = O(1)$ and $M>1$, 
we have
\beno 
|\!|\bm{R}_{V_1}|\!|_F 
&\leq& \dsum_{k=2}^\infty O_p(|\!| \widehat{\bm{D}} - \bm{D} |\!|^k_F |\!| |\bm{S}||\!|^{-k+1/2}_F) 
&\leq& \dsum_{k=2}^\infty O_p( (N/(\rho_N^2 M)^{1/2})^k N^{-k+1/2})\s
\\
&&&=& O_p\left(\dfrac{N^{1/2}}{\rho_N^2\, M}\right). 
\ee
With regard to $\bm{V}_2$, following the argument in \text{C.2.2}, we have 
\beno 
|\!|\bm{V}_2|\!|_F & \leq&  |\!| |\mathbf{S}||\!|_F^{1 / 2} |\!|\mathbf{U}^{\top} \hat{\mathbf{U}}-\mathbf{W}|\!|_F 
  +  |\!|\widehat{\mathbf{D}}-\mathbf{D}|\!|_F  |\!|\mathbf{S}||\!|_F^{-1 / 2}  |\!|\mathbf{U}^{\top} \hat{\mathbf{U}}-\mathbf{W}|\!|_F + o_p(|\!|  \bm{R}_{V_1}  |\!|_F) \s \\
  & =& O_p\left(N^{1/2}\times \dfrac{1}{(\rho_N^2 M)^{1/2}}\right) + O_p\left(\dfrac{N}{(\rho_N^2 M)^{1/2}} \times \dfrac{1}{N^{1/2}} \times \dfrac{1}{(\rho_N^2 M)^{1/2}}\right)\s
  \\
  &=& O_p\left( (\dfrac{N}{\rho_N^2 M})^{1/2} + \dfrac{N^{1/2}}{\rho_N^2 M}\right)
\ee
With regard to $\bm{V}_3$, 
we introduce the matrix $\bm{M}_k$ defined in (\text{C.2.3}) as\break 
$\bm{M}_k : = \mathbf{U}^{\top} \hat{\mathbf{U}} \mathbf{J}^{k+1}|\hat{\mathbf{S}}|^{-k-1 / 2}-|\mathbf{S}|^{-k-1 / 2} \mathbf{U}^{\top} \hat{\mathbf{U}} \mathbf{J}^{k+1}$. We decompose $\bm{M}_k$ following (\text{C.2.3}) as
\beno
\bm{M}_k &=& -\mathbf{H}_k \circ\left(\left(\mathbf{U}^{\top} \hat{\mathbf{U}} \hat{\mathbf{S}}-\mathbf{S} \mathbf{U}^{\top} \hat{\mathbf{U}}\right) \mathbf{J}^k+|\mathbf{S}| \left(\mathbf{J} \mathbf{U}^{\top} \hat{\mathbf{U}}-\mathbf{U}^{\top} \hat{\mathbf{U}} \mathbf{J}\right) \mathbf{J}^k\right),
\ee
where $\mathbf{H}_k $ is defined as (\text{C.2.3}). 
Notice that $\mathbf{U}^{\top} \hat{\mathbf{U}} \hat{\mathbf{S}}-\mathbf{S} \mathbf{U}^{\top} \hat{\mathbf{U}}=\mathbf{U}^{\top}(\widehat{\bm{D}}-\bm{D}^{\star}) \hat{\mathbf{U}}$, which gives us $ |\!|  \mathbf{U}^{\top} \hat{\mathbf{U}} \hat{\mathbf{S}}-\mathbf{S} \mathbf{U}^{\top} \hat{\mathbf{U}} |\!|_F = O_p( |\!| \widehat{\bm{D}}-\bm{D}^{\star} |\!|_F) = O_p(\frac{N}{\rho_N M^{1/2}})$. In addition, following equation (13) in (\text{C.2.3}), we have
$ |\!| \mathbf{J} \mathbf{U}^{\top} \hat{\mathbf{U}}-\mathbf{U}^{\top} \hat{\mathbf{U}} \mathbf{J}  |\!|_F = O_p(|\!|  \bm{U}_{(+)}^\top \widehat{\bm{U}}_{(-)})  |\!|_F) = O_p(1/\sqrt{\rho_N^2 M})$. Given that the ranks of $\mathbf{U}^{\top} \hat{\mathbf{U}} \hat{\mathbf{S}}-\mathbf{S} \mathbf{U}^{\top} \hat{\mathbf{U}}$ and $\mathbf{J} \mathbf{U}^{\top} \hat{\mathbf{U}}-\mathbf{U}^{\top} \hat{\mathbf{U}} \mathbf{J}$ are bounded and the fact that $|\!| A  |\!|_F  \asymp  |\!| A  |\!|_{op}$ when rank of $A$ is bounded. 
Then following (\text{C.2.3}), 
we have 
\beno
|\!| \bm{M}_k  |\!|_F  
&\asymp& |\!|\bm{M}_k  |\!|_{op}
&=& O_p\left(N^{-k-3/2} \times \left(\dfrac{N}{\rho_N\, M^{1/2}} + N \times 1/\sqrt{\rho_N^2 M}\right)\right)\s
\\ 
&&&=& O_p\left(\dfrac{N^{-k-1/2}}{\rho_N\, M^{1/2}}\right).
\ee
Then we have 
\beno
|\!|\bm{V}_3 |\!|_F 
&\leq& |\!| \mathbf{U} \mathbf{S} \bm{M}_0|\!|_F + |\!|\widehat{\bm{D}} - \bm{D}|\!|_F\, |\!| \mathbf{U} \mathbf{S} \bm{M}_1|\!|_F  + o_p(|\!|  \bm{R}_{V_1}  |\!|_F)\s 
\\
&\leq& O_p\left(N\, \dfrac{N^{-1/2}}{\rho_N\, M^{1/2}}\right) + O_p\left( \dfrac{N^2}{\rho_N\, M^{1/2}}\, \dfrac{N^{-3/2}}{\rho_N\, M^{1/2}}\right)\s 
\\
&=& O_p\left(\dfrac{N^{1/2}}{\rho_N\, M^{1/2}}\right).
\ee
The transformed latent positions $\widetilde{\bm{\Theta}} = \bm{U}\, |\bm{S}|^{1/2}\bm{J}$ are identifiable and estimated by $\widehat{\bm{\Theta}} = \widehat{\bm{U}}|\widehat{\bm{S}}|^{1/2}\bm{J}$, 
which indicates that $\bm{W} = \bm{I}$. 
According to \eqref{ap_1} and the above results, 
\beno 
|\!|\widehat{\bm{\Theta}} - \widetilde{\bm{\Theta}}|\!|_F 
&=& O_p\left(\left( \dfrac{N}{\rho_N^2\, M} \right)^{1/2}\right).
\ee
Note that ${\bm{\Theta}} = \widetilde{\bm{\Theta}}\bm{R}$,
where $\bm{R}$ is a hyperbolic rotation such that $\bm{R}\, \bm{J}\, \bm{R}^\top = \bm{J}$. Using the fact that $|\!|\bm{A}\, \bm{R}|\!|_F = |\!|\bm{A}|\!|_F$ if $\bm{R}\,\bm{J} \bm{R}^\top = \bm{J}$, we have 
\beno 
\dfrac{1}{N} \inf_{\bm{R}}|\!| \widehat{\bm{\Theta}}\bm{R} - \bm{\Theta} |\!|^2_F &=& O_p\left(\dfrac{1}{\rho_N^2\, M}\right)  
&=&  O_p\left(\dfrac{\Delta_{N,r}}{\rho_N^2\, \sqrt{|\mathscr{S}|}}\right),
\ee
where the infimum is over all hyperbolic rotation matrices $\bm{R}$. 

\section{Computational Details}

\subsection{Manifold Gradient Descent}
\label{sec:gradient}

We first derive the Euclidean gradient of the population loss with respect to the latent position $\pos_h$ with $h\in \mathscr{V}$. 
The population loss is defined by
\beno
\ell(\bm\Lambda)
&\coloneqq& -\dsum_{k = 2}^K\, \dsum_{e \in \mE_k} z_{e}\, \log \pi(\alpha_{|e|}, \Pos_{e}) + (1-z_{e})\,\log(1 - \pi(\alpha_{|e|}, \Pos_{e})),
\ee
and its gradient  with respect to the latent position $\pos_h$ with $h\in \mathscr{V}$ is:
\beno 
\nabla_{\pos_h}\, \pi(\alpha_{|e|}, \Pos_{e}) &=& \dfrac{-2\, \alpha_{|e|}\, \exp(g(\Pos_e))}{(1+\exp(g(\Pos_e))^2} \, \nabla_{\pos_h}\, \, g(\Pos_e)\s\\
&=&  -\pi(\alpha_{|e|}, \Pos_{e})\left(1-\dfrac{\pi(\alpha_{|e|}, \Pos_{e})}{2\, \alpha_{|e|}}\right)\nabla_{\pos_h}\, \, g(\Pos_e)\s\\
\nabla_{\pos_h}\, \, g(\Pos_e)&=& \nabla_{\pos_h} \left(\dfrac{1}{|e|}\, \dsum_{i \in e} \left(d^{(e)}_i(\Pos_e)\right)^p \right)^{1/p} \s\\
&=& \dfrac{1}{p}\left(\dfrac{1}{|e|}\, \dsum_{i \in e} \left(d^{(e)}_i(\Pos_e)\right)^p \right)^{(1-p)/p} \, \nabla_{\pos_h}\,\left(\dfrac{1}{|e|} \dsum_{i \in e} \left(d^{(e)}_i(\Pos_e)\right)^p\right)\s\\
\nabla_{\pos_h} \left( \dsum_{i \in e} \left(d^{(e)}_i(\Pos_e)\right)^p \right)&=& \dsum_{i \in e} p\, \left(d^{(e)}_i(\Pos_e)\right)^{p-1} \nabla_{\pos_h} d_i^{(e)}(\Pos_e).
\ee
For $i \neq h$, $\nabla_{\pos_h}\, d_i^{(e)}(\Pos_e)$ is 
\beno
\nabla_{\pos_h} d_i^{(e)}(\Pos_e) &=& \nabla_{\pos_h} \left(\dsum_{j \in e,\; j \neq i}\, d_{\mathscr{L}}(\pos_i, \pos_j)\right)\s\\
&=&\nabla_{\pos_h} d_{\mathscr{L}}(\pos_i,\pos_h),
\ee
while we get for $i =h$
\beno
\nabla_{\pos_h} d_h^{(e)} (\Pos_e) &=& \dsum_{i \in e,\; i \neq h} \nabla_{\pos_h} d_{\mathscr{L}}(\pos_i,\pos_h).
\ee
With $\bm{J}~=~\text{diag}(-1, \bm{1}_r) \in \mathbb{R}^{r+1}$, $\bm{1}_n ~=~(1, 1, \ldots)\in \mathbb{R}^n$ being a vector of $n$ ones, and $\Delta_{i,h} ~=~1/(\left\langle\pos_i,\, \pos_h\right\rangle_{\mathscr{L}}^2-1)^{0.5}$, we have 
\be
\label{eq:d}
\nabla_{\pos_h}  d_{\mathscr{L}}(\pos_i,\pos_h) &=& -\bm{J} \Delta_{i,h}\, \pos_i.
\ee
Plugging \eqref{eq:d} into the previous equations gives for $i =h$
\beno
\nabla_{\pos_h} d_h^{(e)}(\Pos_e) &=& -\bm{J}  \dsum_{i \in e,\; i \neq h} \Delta_{i,h}\, \pos_i
\ee
and for $i\neq h$
\beno
\nabla_{\pos_h} d_i^{(e)}(\Pos_e) &=& -\bm{J}\,\Delta_{i,h}\, \pos_i.
\ee
This, in turn, allows us to calculate 
\beno 
\nabla_{\pos_h}  \dsum_{i \in e} \left(d^{(e)}_i(\Pos_e)\right)^p &=& -p\,\bm{J}\,\dsum_{i \in e; i\neq h}  \left(\left(d^{(e)}_i(\Pos_e)\right)^{p-1} + \left(d^{(e)}_h(\Pos_e)\right)^{p-1}\right) \,\Delta_{i,h}\, \pos_i,
\ee
\beno 
\nabla_{\pos_h}\, \, g(\Pos_e)&=& \dfrac{-\bm{J}}{|e|} \left(\dfrac{1}{|e|}\, \dsum_{i \in e} \left(d^{(e)}_i(\Pos_e)\right)^p \right)^{(1-p)/p} 
\s\\&\times& 
\left(\dsum_{i \in e; i\neq h}  \left(\left(d^{(e)}_i(\Pos_e)\right)^{p-1} + \left(d^{(e)}_h(\Pos_e)\right)^{p-1}\right) \,\Delta_{i,h}\, \pos_i\right),
\ee
and 
\beno 
\nabla_{\pos_h}\, \pi(\alpha_{|e|}, \Pos_{e}) &=& -\pi(\alpha_{|e|}, \Pos_{e})\left(1-\dfrac{\pi(\alpha_{|e|}, \Pos_{e})}{2\, \alpha_{|e|}}\right)\, \dfrac{-\bm{J}}{|e|} \left(\dfrac{1}{|e|}\, \dsum_{i \in e} \left(d^{(e)}_i(\Pos_e)\right)^p \right)^{(1-p)/p} \s\\&\times& 
\left(\dsum_{i \in e; i\neq h}  \left(\left(d^{(e)}_i(\Pos_e)\right)^{p-1} + \left(d^{(e)}_h\right)^{p-1}\right) \,\Delta_{i,h}\, \pos_i\right).
\ee
Defining 
\beno 
A_1(e,\Lambda) &=& \left(\dfrac{1}{|e|}\, \dsum_{i \in e} \left(d^{(e)}_i(\Pos_e)\right)^p \right)^{(1-p)/p} \in \mathbb{R}\s\\
\bm{A}_2^h(e,\Lambda) &=&\dfrac{\bm{J}}{|e|} \left(\dsum_{i \in e; i\neq h}  \left(\left(d^{(e)}_i(\Pos_e)\right)^{p-1} + \left(d^{(e)}_h\right)^{p-1}\right) \,\Delta_{i,h}\, \pos_i\right) \in \mathbb{R}^{r+1},
\ee
we can write 
\beno 
\nabla_{\pos_h}\, \pi(\alpha_{|e|}, \Pos_{e}) &=& \pi(\alpha_{|e|}, \Pos_{e})\left(1-\dfrac{\pi(\alpha_{|e|}, \Pos_{e})}{2\, \alpha_{|e|}}\right)\, A_1(e,\Lambda)\,\bm{A}_2^h(e,\Lambda)
\ee
Finally, the derivative can be derived as follows 
\beno 
\nabla_{\pos_h}\, \ell(\bm\Lambda)
&=& -\dsum_{k = 2}^K\, \dsum_{e \in \mE_k} \mathbf{1}_{\{h \in e\}} \, 
\dfrac{z_{e} -\pi(\alpha_{|e|}, \Pos_{e})}{\pi(\alpha_{|e|}, \Pos_{e}) \, (1-\pi(\alpha_{|e|}, \Pos_{e}))}
\,\nabla_{\pos_h}\, \pi(\alpha_{|e|}, \Pos_{e}) \s\\
&=&- \dsum_{k = 2}^K\, \dsum_{e \in \mE_k} \mathbf{1}_{\{h \in e\}}\, \dfrac{z_{e} -\pi(\alpha_{|e|}, \Pos_{e})}{1-\pi(\alpha_{|e|}, \Pos_{e})} \, \left(1-\dfrac{\pi(\alpha_{|e|}, \Pos_{e})}{2\, \alpha_{|e|}}\right) \, A_1(e,\Lambda)\,\bm{A}_2^h(e,\Lambda).
\ee
Note that we derived the population gradient, but letting $e \in \mathcal S_k \coloneqq \mathcal S_k^{(0)} \bigcup \mathcal S_k^{(1)}$ and including the reciprocal weight $\mu_e$ as a multiplicative constant yields the following Euclidean gradient of the sample loss under hyperbolic geometry:
\beno
\nabla_{\pos_h}\, \widehat\ell(\bm\Lambda)
&=& - \dsum_{k = 2}^K\, \dsum_{e \in \mathcal S_k} \dfrac{ \mathbf{1}_{\{h \in e\}}}{\mu_e}\,  \dfrac{z_{e} -\pi(\alpha_{|e|}, \Pos_{e})}{1-\pi(\alpha_{|e|}, \Pos_{e})} \, \left(1-\dfrac{\pi(\alpha_{|e|}, \Pos_{e})}{2\, \alpha_{|e|}}\right) \, A_1(e,\Lambda)\,\bm{A}_2^h(e,\Lambda).
\ee

\subsection{Estimation under Euclidean Geometry}
\label{sec:est_euclidean}

We estimate the latent space model for hyperedges under the assumption that the positions $\Pos$ of units are in Euclidean space or hyperbolic space.
In the Euclidean case,
we employ the same minimization algorithm with the only difference that the cyclical updates of $\pos_h$ for all $h\in \mathscr{V}$ are carried out by the Euclidean variant of the manifold gradient descent algorithm, 
which reduces \eqref{eq:updated_pos} to the gradient descent updates
\be
\label{eq:update_euclidean}
\pos_h^{(t+1)} &=& \pos_h^{(t)} - \eta_h^{(t+1)}\, \nabla_{\pos_h}\, \widehat{\ell}(\bm\Lambda),
\ee
where $\eta_h^{(t+1)}$ is a learning rate that is set by Brent's method to minimize the sample loss and $\nabla_{\pos_h}\,\widehat{\ell}(\bm\Lambda)$ denotes the gradient of the sample loss with $\Pos$ being in the Euclidean space. 

Contrasting the definition in \eqref{eq:dd},
the function $d^{(e)}_i(\Pos_e)$ is substituted by its Euclidean analog $d^{(e)}_{\mathscr{E},i}(\Pos_e)$ given by: 
\beno
d^{(e)}_{\mathscr{E},i}(\Pos_e)
&\coloneqq& \dsum_{h \in e \setminus \{i\}}\, d_{\mathscr{E}}(\pos_i,\, \pos_h),
\ee
with 
\beno
d_{\mathscr{E}}(\pos_i, \pos_h)
&\coloneqq& |\!|\pos_i - \pos_h|\!|_2.
\ee
Therefore, we substitute the derivative of the hyperbolic distance between $\pos_i$ and $\pos_h$ with respect to $\pos_h$ with the partial derivative of the Euclidean distance:
\beno
\nabla_{\pos_h}  d_{\mathscr{E}}(\pos_i,\pos_h) &=& - \dfrac{\pos_i -\pos_h}{\vecnorm{\pos_i -\pos_h}}.
\ee
Defining 
\beno 
B_1(e,\Lambda) &=&\left(\dfrac{1}{|e|}\, \dsum_{i \in e} \left(d^{(e)}_{\mathscr{E},i}(\Pos_e)\right)^p \right)^{(1-p)/p} \in \mathbb{R}\s\\
\bm{B}_2^h(e,\Lambda) &=& \dfrac{1}{|e|} \left(\dsum_{i \in e; i\neq h}  \left(\left(d^{(e)}_{\mathscr{E},i}(\Pos_e)\right)^{p-1} + \left(d^{(e)}_{\mathscr{E},h}(\Pos_e)\right)^{p-1}\right) \,\dfrac{\pos_i-\pos_h}{\vecnorm{\pos_i-\pos_h}}\right) \in \mathbb{R}^{r+1},
\ee
 we can write 
\beno 
\nabla_{\pos_h}\, \pi(\alpha_{|e|}, \Pos_{e}) &=& \pi(\alpha_{|e|}, \Pos_{e})\left(1-\dfrac{\pi(\alpha_{|e|}, \Pos_{e})}{2\, \alpha_{|e|}}\right)\, B_1(e,\Lambda)\,\bm{B}_2^h
(e,\Lambda)
\ee
Then the the gradient of the population loss regarding $\pos_h$ is
\beno 
\nabla_{\pos_h}\, \ell_{\mathscr{E}}(\bm\Lambda)
&=&- \dsum_{k = 2}^K\, \dsum_{e \in \mE_k} \mathbf{1}_{\{h \in e\}}\,  \dfrac{z_{e} -\pi(\alpha_{|e|}, \Pos_{e})}{1-\pi(\alpha_{|e|}, \Pos_{e})} \, \left(1-\dfrac{\pi(\alpha_{|e|}, \Pos_{e})}{2\, \alpha_{|e|}}\right) \, B_1(e,\Lambda)\,\bm{B}_2^h(e,\Lambda).
\ee
The respective gradient of the sample loss is then:
\beno 
\nabla_{\pos_h}\, \widehat\ell_{\mathscr{E}}(\bm\Lambda)
&=&- \dsum_{k = 2}^K\, \dsum_{e \in \mathcal S_k} \dfrac{\mathbf{1}_{\{h \in e\}}}{\mu_e}\,  \dfrac{z_{e} -\pi(\alpha_{|e|}, \Pos_{e})}{1-\pi(\alpha_{|e|}, \Pos_{e})} \, \left(1-\dfrac{\pi(\alpha_{|e|}, \Pos_{e})}{2\, \alpha_{|e|}}\right) \, B_1(e,\Lambda)\,\bm{B}_2^h(e,\Lambda).
\ee
Given this partial derivative, we can carry out update rule \eqref{eq:update_euclidean}.

\subsection{Initialization}
\label{sec:initialization}

For all reported experiments, we sample the positions uniformly from a square with length 0.2 centered at the origin $(0,\, 0)$ of the \pd, 
following  \citet{nickel2017poincare,nickel2018learning},.  
These initial positions are transformed to the Lorentz model by applying the following function:
\be
\label{eq:P_to_L}
f^{-1}(\pos) &=& \left(1+ \vecnormsqrt{\pos},\, \dfrac{2\, \pos}{1- \vecnormsqrt{\pos}}\right),
\ee
which is the inverse of the bijective function $f$ in Section 2 of the main manuscript. 
Our implementation first updates the sparsity parameters and then the positions. 
Therefore, 
no initial values for the sparsity parameters are needed. 

\hide{
In the \pd{ }model for hyperbolic space, 
units with positions close to the center, 
are due to the negative curvature of the space generally more active than units further distant from the center. 
Therefore, we chose the initial position $\pos_i^{(0)}$ of unit $i \in \mathscr{A}$ on a circle with a radius proportional to $\text{deg}(i) = \sum_{e\in \mathscr{S}^{(1)}} \mathbb{I}(i \in e)$, being the number of hyperedges in the train data unit $i$ participates in. 
The radius of the circle where $\pos_i^{(0)}$  lies on is
\beno 
r_i &=& d\left((0,0), \left(0,1-\dfrac{\text{deg}(i) }{\sum_h^N \text{deg}(h)}\right)\right),
\ee
where the distance $d$ is chosen according to the space where the latent positions live, i.e., either hyperbolic or Euclidean space. 
The angle of the circle $\beta_i$ is chosen randomly for each initial position between 0 and $2\,\pi$, which yields the following initial position for $i = 1, ..., N$: 
\beno 
\pos_i^{(0)} &=& (r_i\cos(\beta_i), r_i\sin(\beta_i)).
\ee
Finally, these initial positions are transformed to the Lorentz model by the function
\be
\label{eq:P_to_L}
f^{-1}(\pos) &=& \left(1+ \vecnormsqrt{\pos}, \dfrac{2\, \pos}{1- \vecnormsqrt{\pos}}\right),
\ee
which is the inverse of the bijective function $f$ from Section 2 of the manuscript. 
}


\section{Simulating Positions}
\label{sec:positions}

We generate the positions for $N$ units on the \pd{ }by their polar coordinates.
For simplicity, we detail the procedure with $r=2$.
Polar coordinates are a way to represent each point $(x,y) \in \mathcal{P}^2$ in terms of the radial coordinate $x_\rho \in [0, 1)$, which is its Euclidean distance to the point $(0,0)$, and the angular coordinate $y_\theta \in [0, 2\pi)$, which is the angle between the straight line from $(x,y)$ to $(0,0)$ and the horizontal axis.

The positions are controlled by parameters $\gamma > 0 $ and $\rho \in (0,1]$ and generated according to the following steps:
\begin{enumerate}
    \item Generate the radial coordinates $Z_1, \ldots, Z_N \overset{\text{iid}}{\sim} \text{Z}(\gamma, \rho)$ with density at $z\in [0,\rho]$
    \beno 
    f_Z(z \mid \gamma, \rho) &=& \gamma \dfrac{\text{sinh}( \gamma\, z)}{\text{cosh}(\rho\, \gamma) -1},
    \ee
    which closely approximates an exponential distribution with rate $\gamma$, truncated at $\rho$.
    \hide{
    Generate $u_1, \ldots, u_N \overset{\text{iid}}{\sim} \text{Uniform}[0,\, 1]$ and obtain the radial coordinates via inverse sampling by
   \beno 
   z_i = \frac{1}{\gamma} \cosh^{-1} \left( (\cosh(\gamma\, \rho) - 1)\, u_i + 1\right), \quad i = 1, \ldots, N.
   \ee
    }
   \item Generate the angular coordinates $S_1, \ldots, S_N \overset{\text{iid}}{\sim} \text{Uniform}[0,\, 2\, \pi]$.
   \item Transform the polar coordinates to the Cartesian coordinates: for $i$th position we get $( z_i \cos(s_i),\, z_i \sin(s_i))$.
\end{enumerate}
For all experiments, we set $\gamma = 3$ and $\rho = 1/2$.

\section{Application to Newswire Data}
\label{sec:additional_information}

\subsection{Data}
\label{sec:addition_Data}

We apply our proposed method to the Newswire data set \citepsupp{silcockNewswireLargeScaleStructured2024} containing approximately 2.7 million articles in the public domain.
Using a topic classification based on a neural network architecture, these articles were tagged with entities, i.e., people, that are mentioned in them and topic of the article.
 For this application, we focused on articles with one of the following topics:  \q{Civil Rights}, \q{Defense}, \q{Federal Government Operations}, \q{Macroeconomics},
    \q{International Affairs}, \q{Labor, Immigration, and Employment}.
These entities correspond to the units defined in the manuscript.
\citetsupp{silcockNewswireLargeScaleStructured2024} provided the Wikidata ID for all units. 
This information allows us to link each entity to the database \href{https://www.wikidata.org}{Wikidata}. 
Wikidata is an open-source database feeding into Wikipedia articles, from which we derived each unit's birthday and political party affiliation. 
\citetsupp{silcockNewswireLargeScaleStructured2024} supplement these data by providing the profession for each unit. 
As stated in the main manuscript, our analysis focuses on hyperedges of sizes 2, 3, and 4. The inclusion criteria for units are as follows:
\begin{enumerate}
    \item Profession: The profession of the unit is   
\q{politician}, \q{judge}, \q{lawyer}, \q{military officer}, or \q{military personnel}.
    \item Birth: The birthday of the unit is before the 1st of January, 1900. 
    \item Political affiliation: The unit was registered as a Democrat or Republican (or both) in the U.S.
    \item Activity: The unit was mentioned in at least five articles with other units that also fulfill all inclusion criteria for hyperedges of sizes $k \in \{2,3,4\}$. 
\end{enumerate}
Because the final criterion is now evaluated only with respect to hyperedges of sizes $k \in {2,3,4}$, the total counts of realized hyperedges differ slightly from the reported values in Section \ref{sec:assessment}:
$|\mE_{2}^{(1)}| = \mbox{4,537}$, 
$|\mE_{3}^{(1)}| = \mbox{2,761}$, 
$|\mE_{4}^{(1)}| = \mbox{1,148}$. 

\subsection{Additional Results of Euclidean and Hyperbolic Model}
\label{sec:additional_results}
\paragraph*{Estimates of $\Theta$} 

\begin{figure}[t!]
  \centering
 \includegraphics[width=0.5\textwidth]{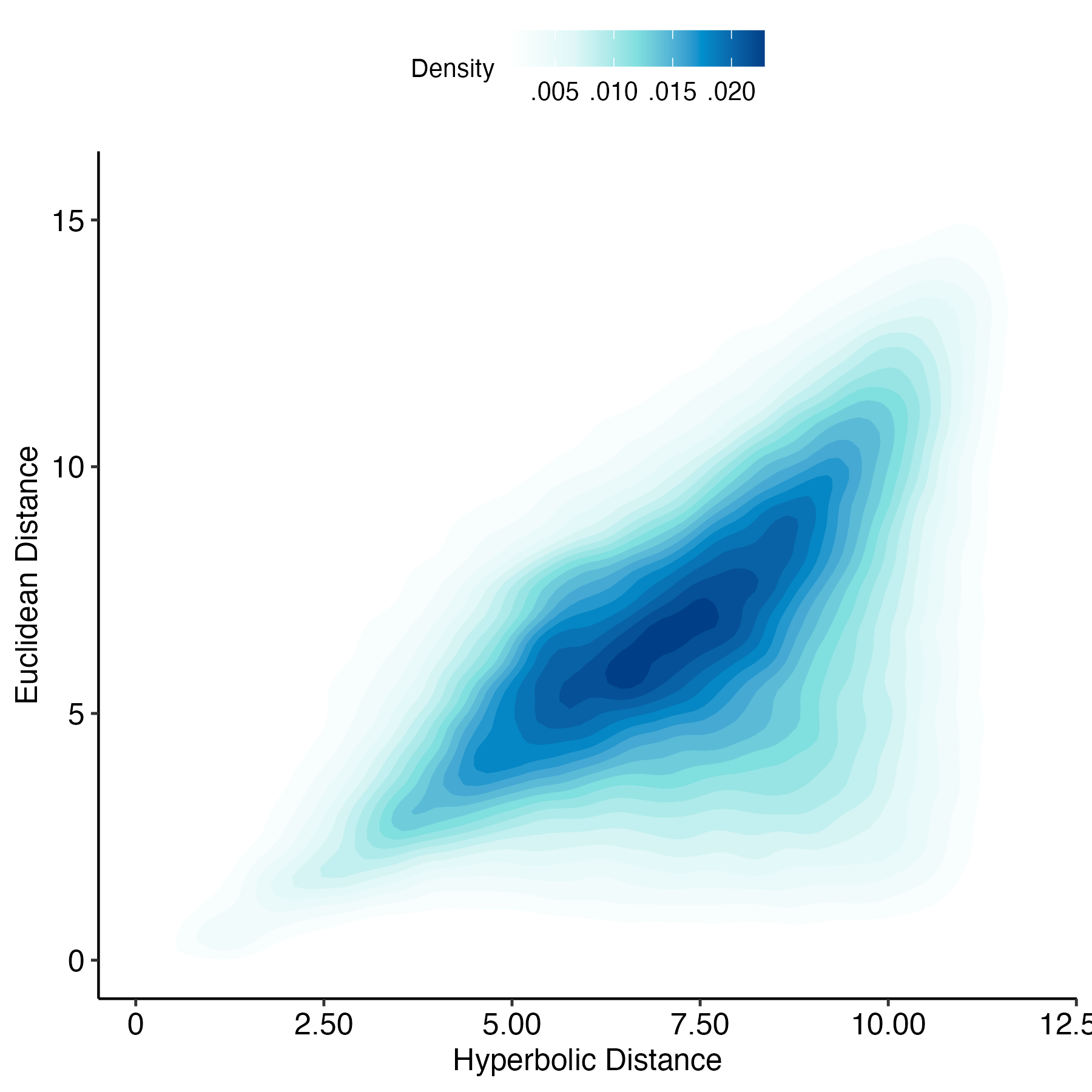}  
  \caption{Two-dimensional kernel density estimator of the differences in pairwise distances according to the model under Euclidean and hyperbolic geometries and the minimal distance to the center within pair.}
  \label{fig:est_distances}
\end{figure}

The correlation between the pairwise distances implied by the hyperbolic and Euclidean latent models is moderate, $.448$, with hyperbolic distances being on average $.738$ longer than the corresponding Euclidean distances. 
In Figure \ref{fig:est_distances}, we present two-dimensional kernel density estimates of the pairwise distances between all units under the hyperbolic and Euclidean latent space models. 
For points $\bm{x}, \bm{y} \in \mathcal{L}^2$, the hyperbolic distance is defined as $\operatorname{arcosh}\left(-\langle \bm{x}, \bm{y} \rangle_{\mathcal{L}} \right)$, where $\langle \cdot, \cdot \rangle_{\mathcal{L}}$ denotes the Lorentzian inner product defined in Section \ref{sec:back}. 
The Euclidean distance between $\bm{x}, \bm{y} \in \mathbb{R}^2$ is $|\!|\bm{x} - \bm{y}|\!|_2$. 
Figure \ref{fig:est_distances} reveals a positive association between the two distance measures, while also highlighting discrepancies that suggest each geometry captures different structural aspects of the hypergraph.  

\begin{figure}[t!]
  \centering
 \includegraphics[width=0.75\textwidth]{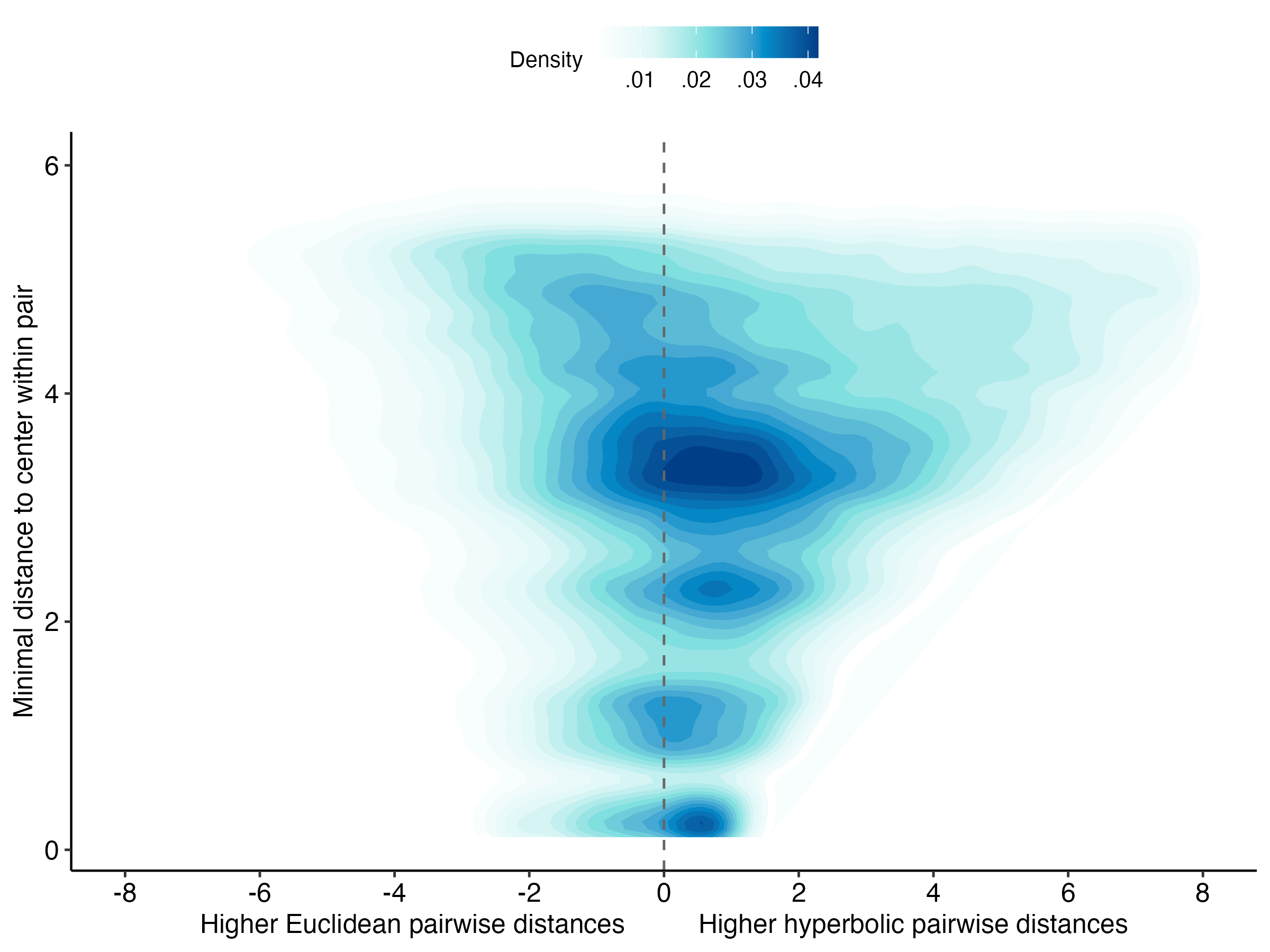}  
  \caption{Two-dimensional kernel density estimator of the pairwise distances between all units according to the model under Euclidean and hyperbolic geometries.}
  \label{fig:comparison_distances}
\end{figure}

Turning to the question of which node pairs exhibit the largest discrepancies in distance under the two geometries, we investigate whether these differences are particularly pronounced for pairs involving units near the center (minimal distance to center within pair is small) or periphery of the \pd (minimal distance to center within pair is large).
Figure \ref{fig:comparison_distances} reveals that pairs for which at least one node lies close to the center tend to have substantially larger distances in the hyperbolic model compared to the Euclidean model.
Euclidean space compresses the center, thus underestimates distances between central units.
In contrast, for pairs where both units are located farther from the center, the distances derived from both geometries  more comparable.

To facilitate a more direct comparison between the estimated positions under hyperbolic and Euclidean geometries, we project the positions obtained in hyperbolic space into Euclidean space with the following procedure:
\begin{enumerate}
    \item Compute all pairwise hyperbolic distances between the positions of all $N$ units.
    \item Apply multidimensional scaling to distance matrix to obtain a configuration in Euclidean space to approximately preserve the hyperbolic distances \citepsupp{Cox2000}.
    \item Use Procrustes analysis to align the resulting configuration with the Euclidean positions, maximizing structural similarity \citepsupp{Young1938a}.
\end{enumerate}
We refer to the hyperbolic positions obtained through this transformation as transformed hyperbolic positions. 

\begin{figure}[t!]
  \centering
 \includegraphics[width=\textwidth]{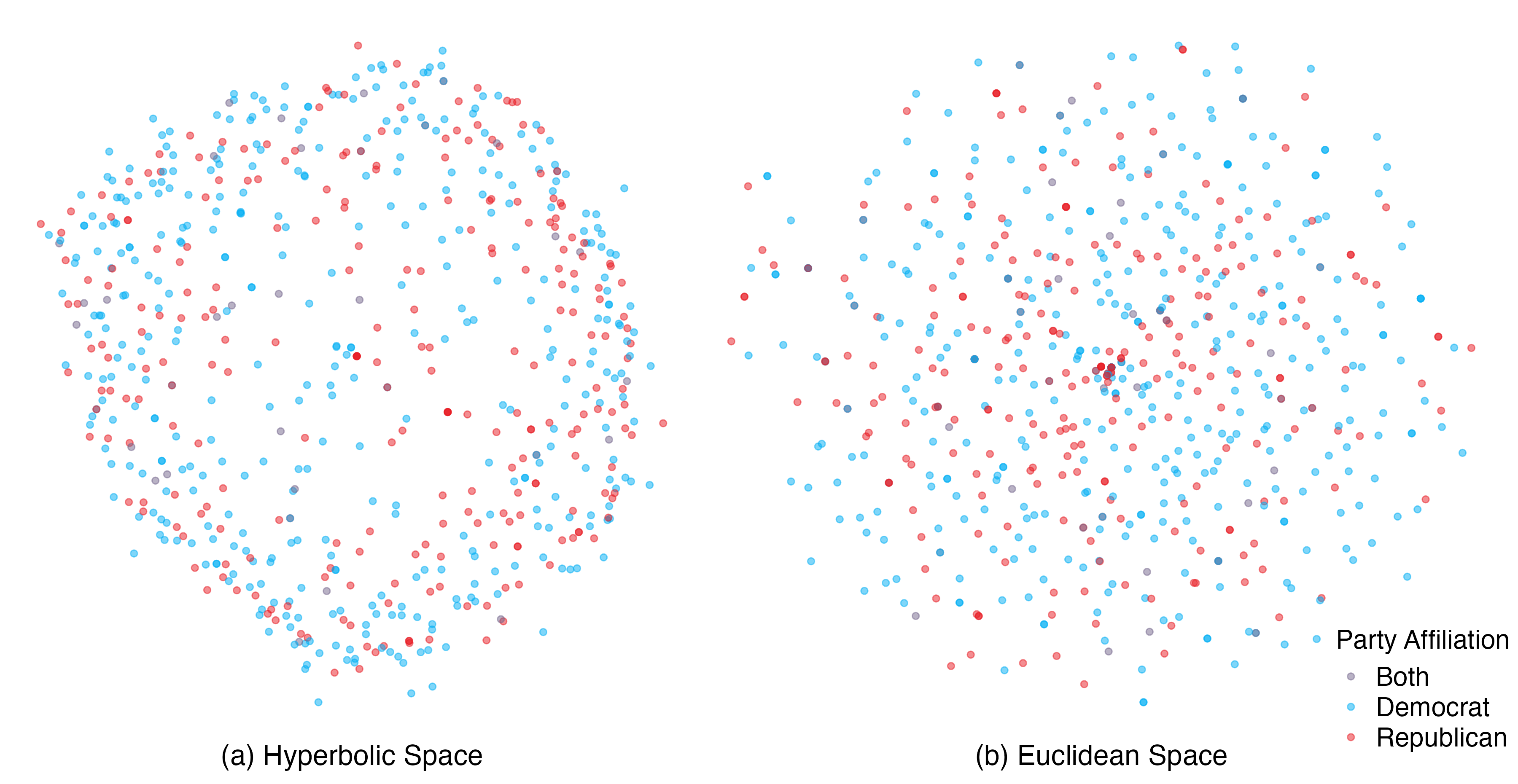}  
  \caption{Estimated latent positions obtained by fitting the model under hyperbolic and Euclidean geometry. The transformed hyperbolic space positions are shown in (a), while the original positions from the model under Euclidean space are shown in (b). }
  \label{fig:est_projected}
\end{figure}

Figure \ref{fig:est_projected} (a) displays these transformed positions, while Figure \ref{fig:est_projected} (b) shows the original Euclidean estimates.
Both position clouds appear visually distinct, demonstrating that positions in hyperbolic space capture different latent similarities than those in Euclidean space. 
The hyperbolic configuration reveals a peripheral clustering pattern. 
In contrast, the Euclidean positions appear more compact and contracted around their centroid.
Compared to Figure \ref{fig:application_res_a} (a) in the main text, the transformed positions appear contracted toward the center of the space, as expected given that hyperbolic distances to the origin are shorter than their Euclidean counterparts.

\paragraph*{Estimates of $\alpha$}

\begin{figure}[t!]
  \centering
         \includegraphics[width=0.9\textwidth]{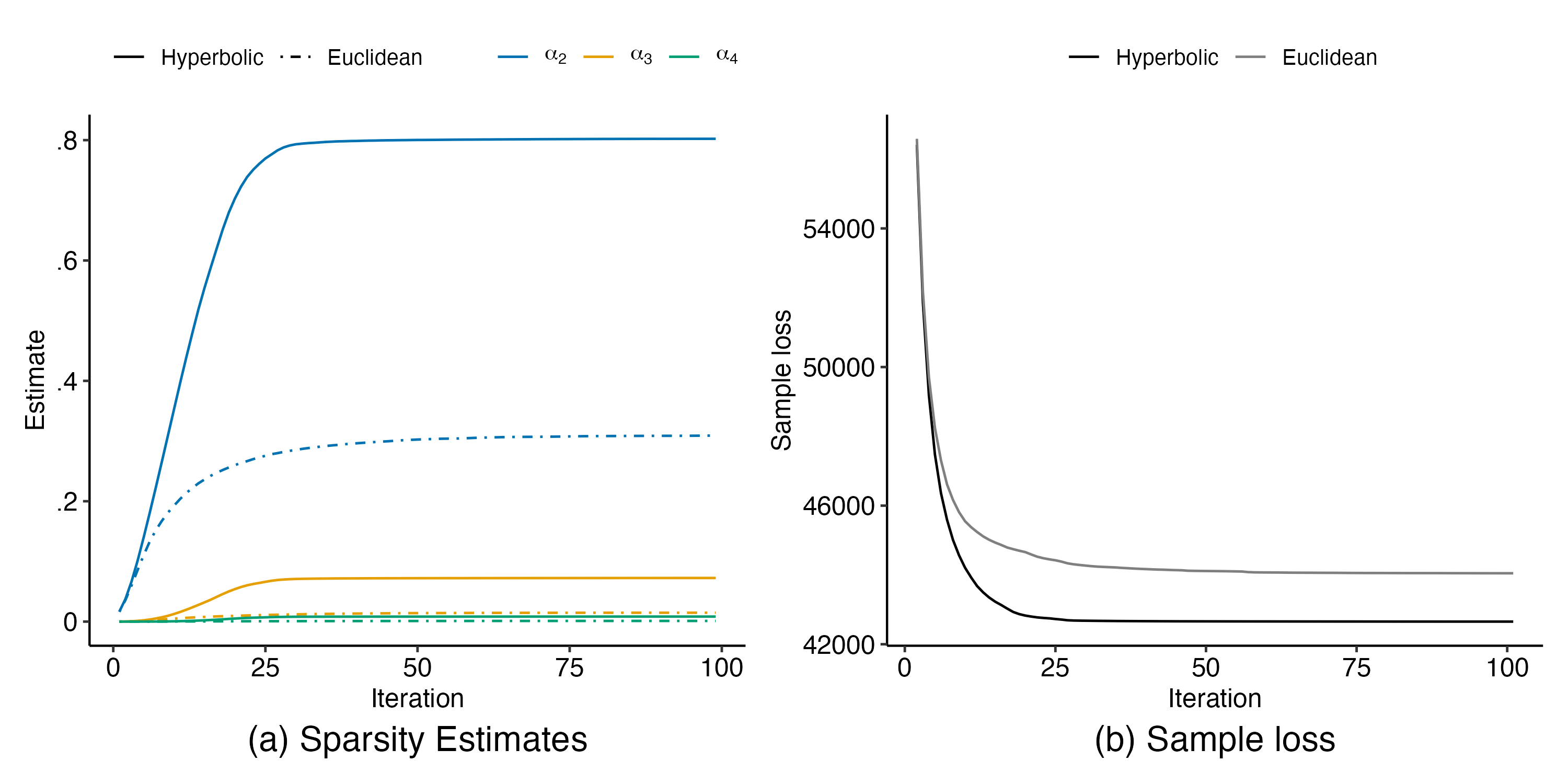}  
  \caption{(a) Trace plot of sparsity parameter estimates of the latent space model based on hyperbolic (solid line: \sampleline{thick}) and Euclidean geometry (dot-dashed line: \sampleline{dash pattern=on 3pt off 2pt on 0.5pt off 2pt, thick}).
  (b) Trace plot of sample loss based on hyperbolic (black) and Euclidean geometry (grey).
  }
  \label{fig:est_sparsity}
\end{figure}

Figure \ref{fig:est_sparsity} (a) shows the estimates of $\alpha_2$, $\alpha_3$, and $\alpha_4$ based on the hyperbolic and Euclidean space models. 
In line with expectations, 
we find that $\widehat\alpha_2 > \widehat\alpha_3 > \widehat\alpha_4$,
suggesting that the density of smaller hyperedges is higher than the density of larger hyperedges. 
This observation holds irrespective of the geometry of the space.
The estimates appear to converge within 100 iterations under both geometries. 

\paragraph*{Sample loss}

Figure \ref{fig:est_sparsity} (b) displays the trace of the sample loss across 100 iterations. 
Convergence appears to be achieved under both geometries after approximately 25 iterations.
That said,
we observe that the hyperbolic geometry is able to achieve a lower value of the sample loss function than the Euclidean geometry.  

\subsection{Sensitivity}
\label{sec:sensitivity}

\subsubsection{Sensitivity to Choice of $p$}
\label{sec:choice_p}
\hide{
\begin{figure}
    \centering
   \includegraphics[width=0.9\textwidth]{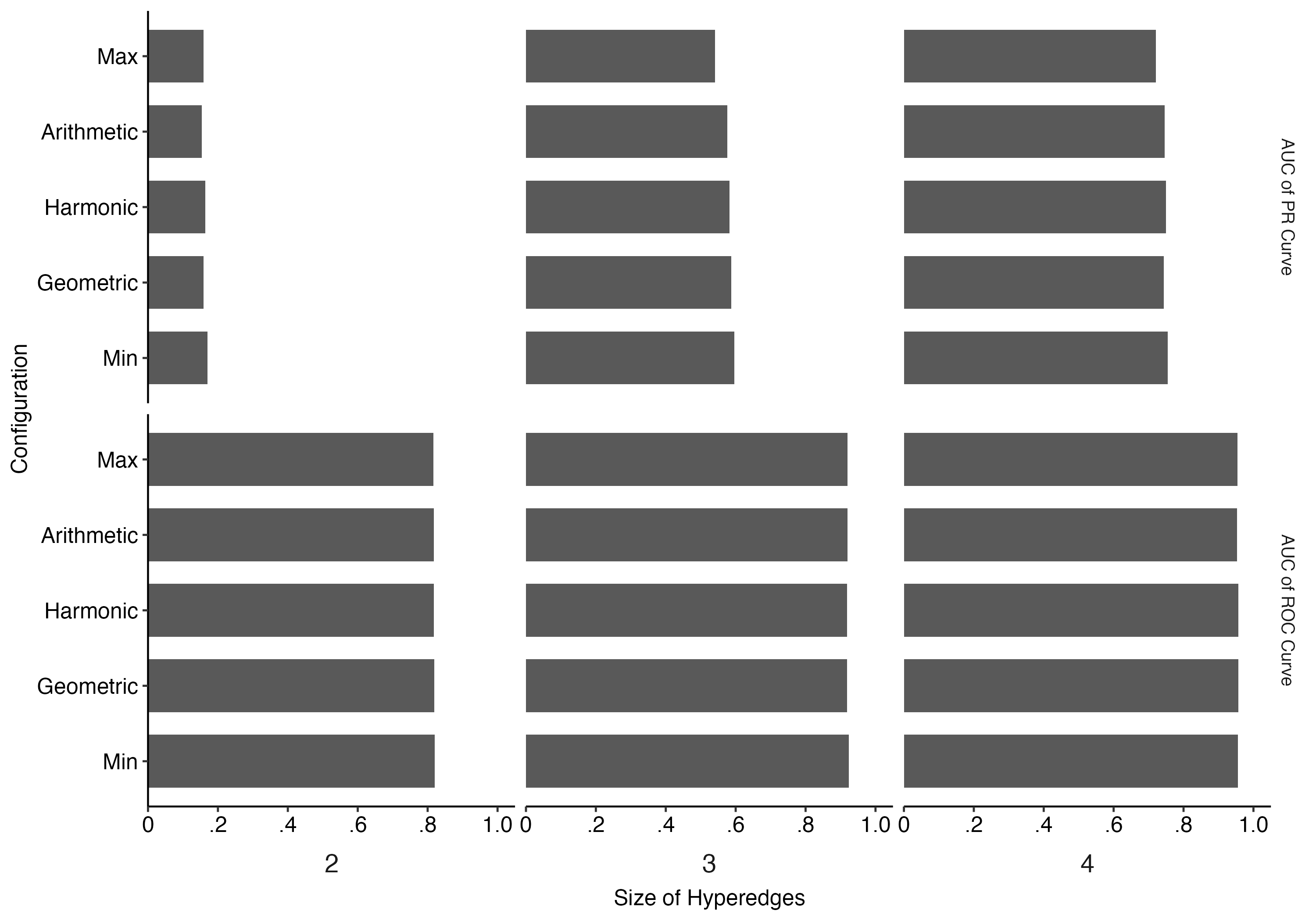}  
    \caption{Newswire data: Area under the Curve (AUC) of the Receiver-Operator Characteristic (ROC) curve of out-of-sample predictions of hyperedges,
    based on hyperbolic geometry with $p$ values approximating the minimum, harmonic mean, geometric mean, arithmetic mean, and maximum via the Hölder mean.}\s
    \label{fig:pr_sens}
\end{figure}
}

\begin{figure}
    \centering
   \includegraphics[width=0.9\textwidth]{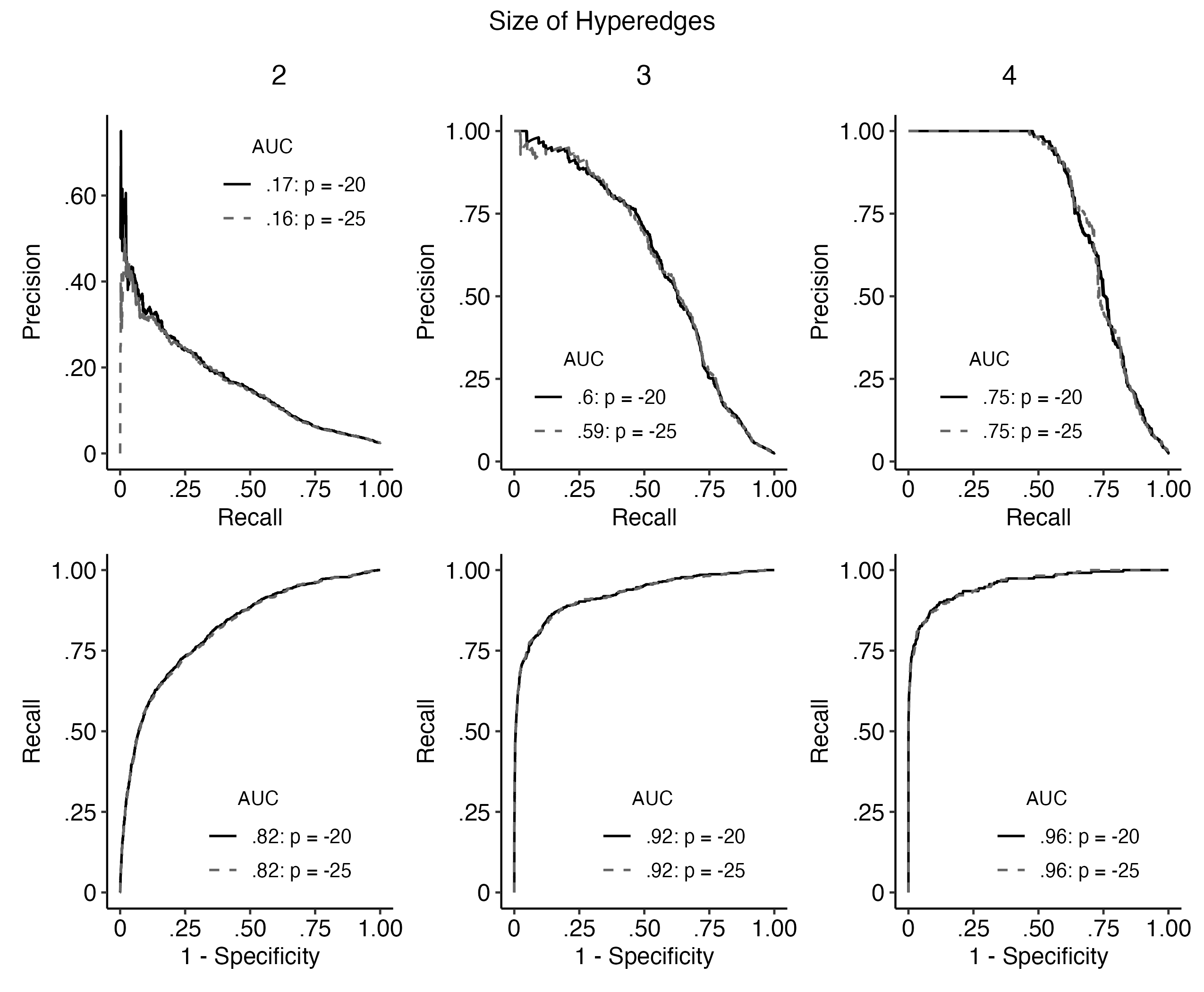}  
    \caption{Newswire data: Receiver-Operator Characteristic (ROC) and Precision-Recall (PR) curve of out-of-sample predictions of hyperedges,
    based on hyperbolic geometry with $p \in \{-25,-20\}$.}\s
    \label{fig:pr_sens}
\end{figure}

\begin{figure}[t!]
  \centering
    
    \includegraphics[width=0.7\textwidth]{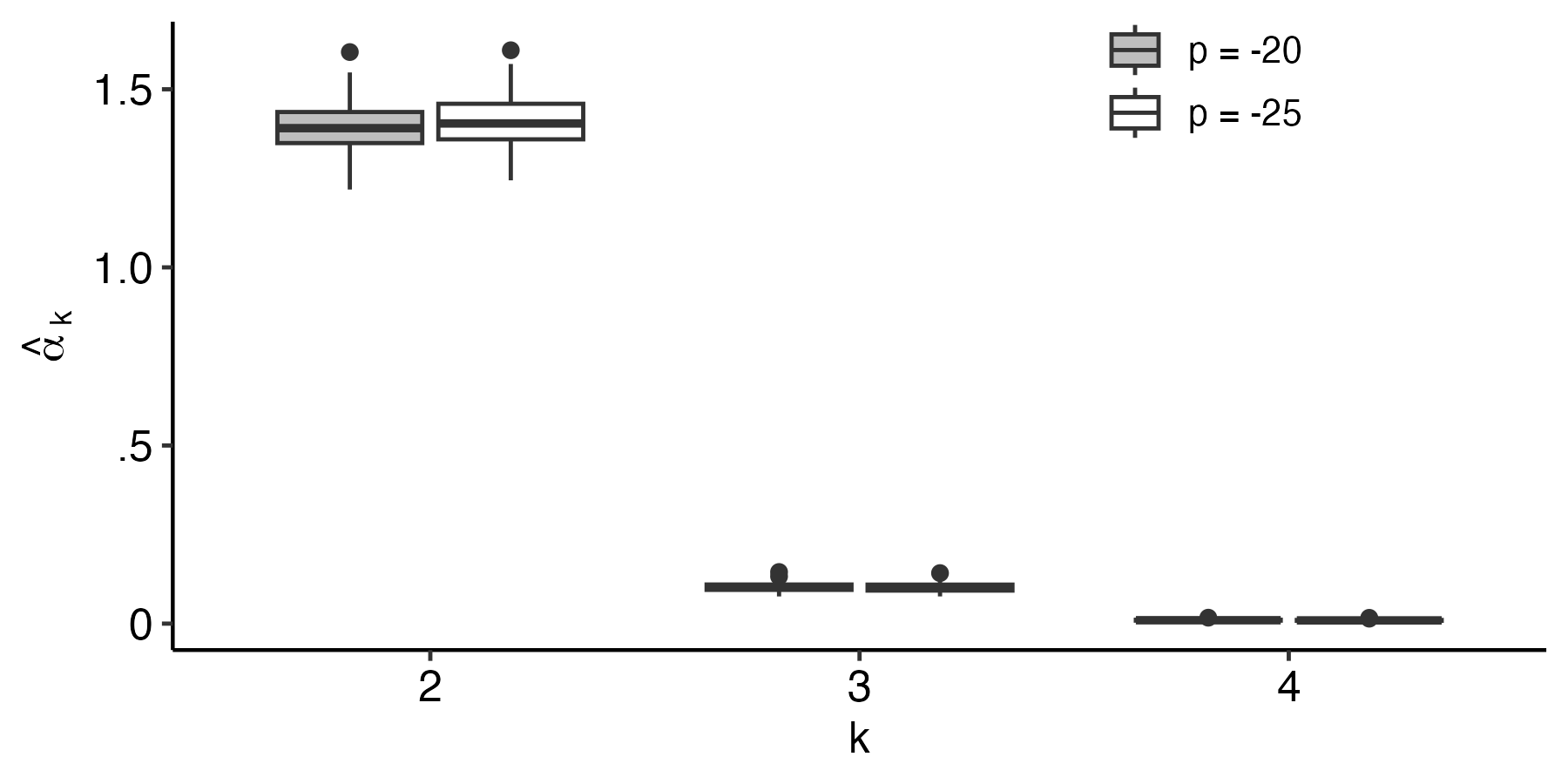}  
    \caption{Boxplots of sparsity parameter estimates $\widehat\alpha_2$, $\widehat\alpha_3$, $\widehat\alpha_4$ based on the latent space model under hyperbolic geometry with $p \in \{-25,-20\}$.}
  \label{fig:robustness_p}
\end{figure}

As mentioned in Section 3.1 of the manuscript,
we choose $p = - 20$ so that $g(\Pos_e)$ is a good approximation of $\min_{i \in  e} \{d^{(e)}_i(\Pos_{e})\}$.
We assess the sensitivity of results to changes of $p$ in two ways.
First,
we conduct out-of-sample predictions as described in Section 6.2 of the manuscript.
Figure \ref{fig:pr_sens} shows that out-of-sample predictions based on $p=-25$ and $p=-20$ are almost identical.
Second, 
we compare sparsity parameter estimates based on $p=-25$ and $p=-20$.
Figure \ref{fig:robustness_p} shows that sparsity parameter estimates based on $p=-25$ and $p=-20$ are similar,
suggesting that the results are not too sensitive to the choice of $p$ as long as $p \leq 20$.

\subsubsection{Sensitivity to Choice of Starting Values and Sample Loss}
\label{sec:initial}

\begin{figure}[t!]
  \centering
  \centering
    \includegraphics[width=0.9\textwidth]{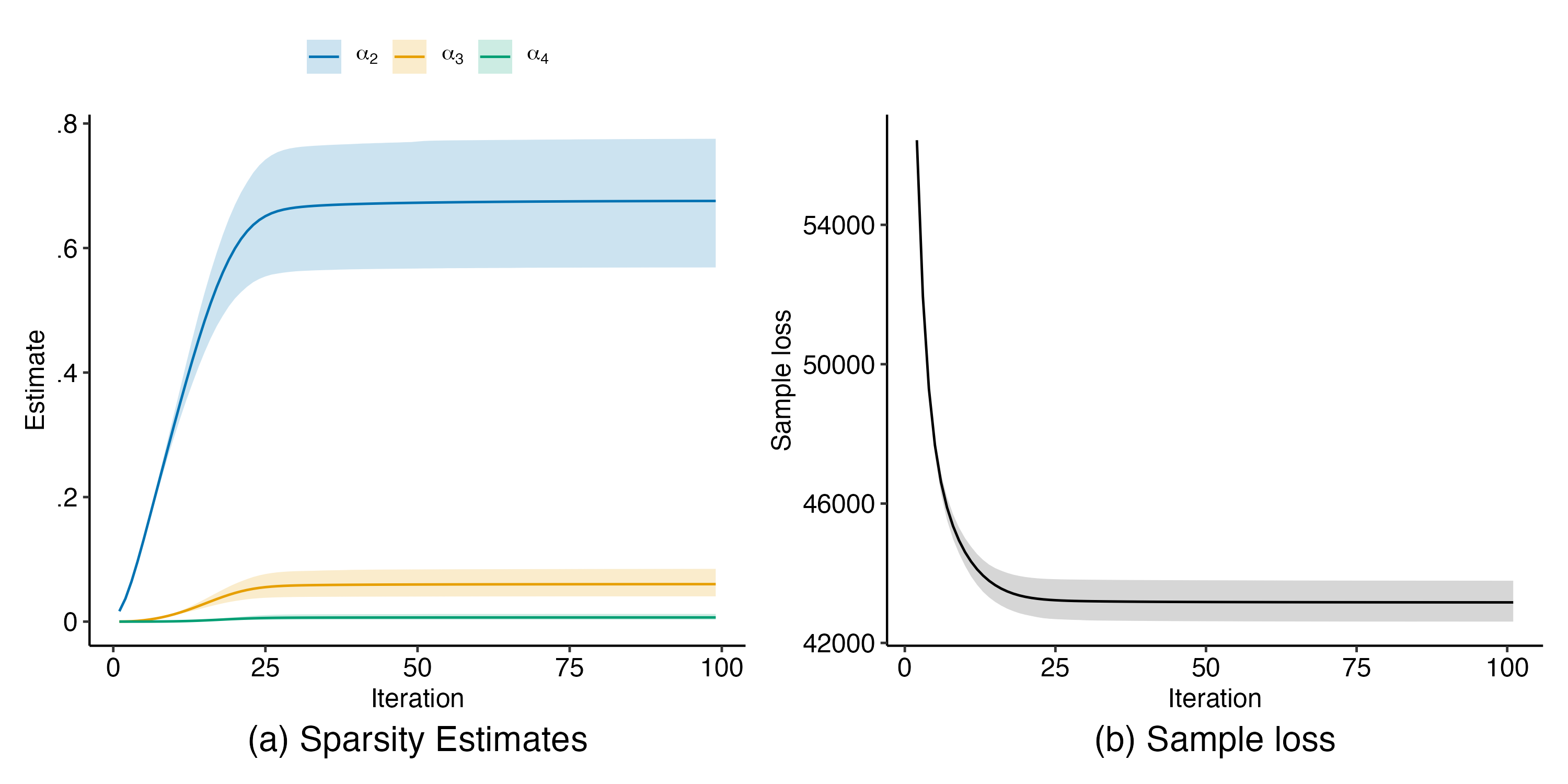}  
  \caption{
   Results based on varying starting values and $\mathscr{S}_k^{(0)} (k \in \mathscr{K})$:
  (a) Trace plots of sparsity parameter estimates based on hyperbolic geometry;
     (b) sample loss per iteration, where the  solid line represents the average sample loss.}
  \label{fig:llh_sensitivity}
\end{figure}

\hide{
\begin{figure}[t!]
  \centering
  \centering
    \includegraphics[width=0.6\textwidth]{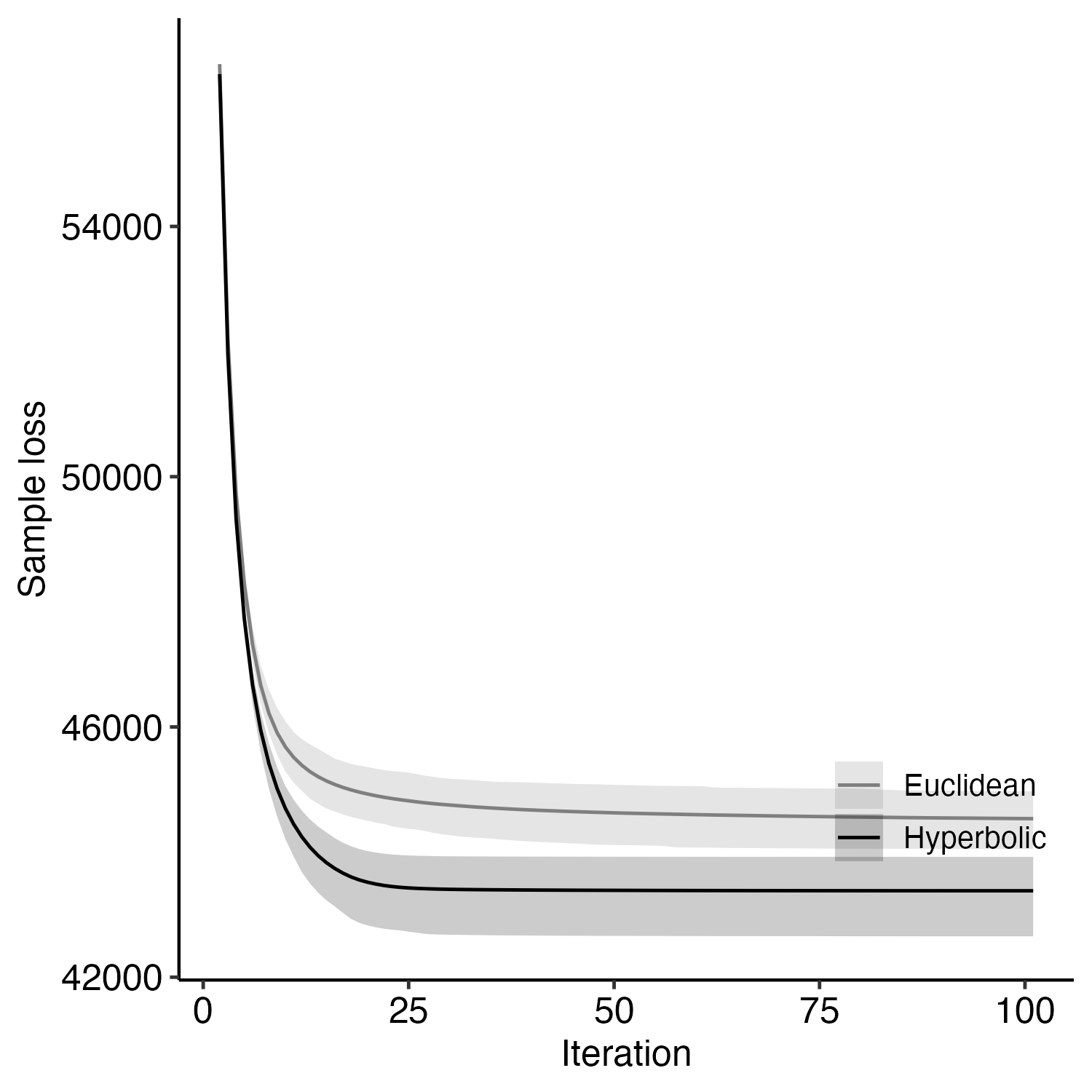}  
  \caption{Range of sample losses per iteration over 100 runs with different initializations and different geometries (Euclidean geometry in grey and hyperbolic in black). The respective solid lines indicate the point-wise average sample loss. }
  \label{fig:llh_comp}
\end{figure}

\begin{figure}[t!]
  \centering
    
    \includegraphics[width=0.6\textwidth]{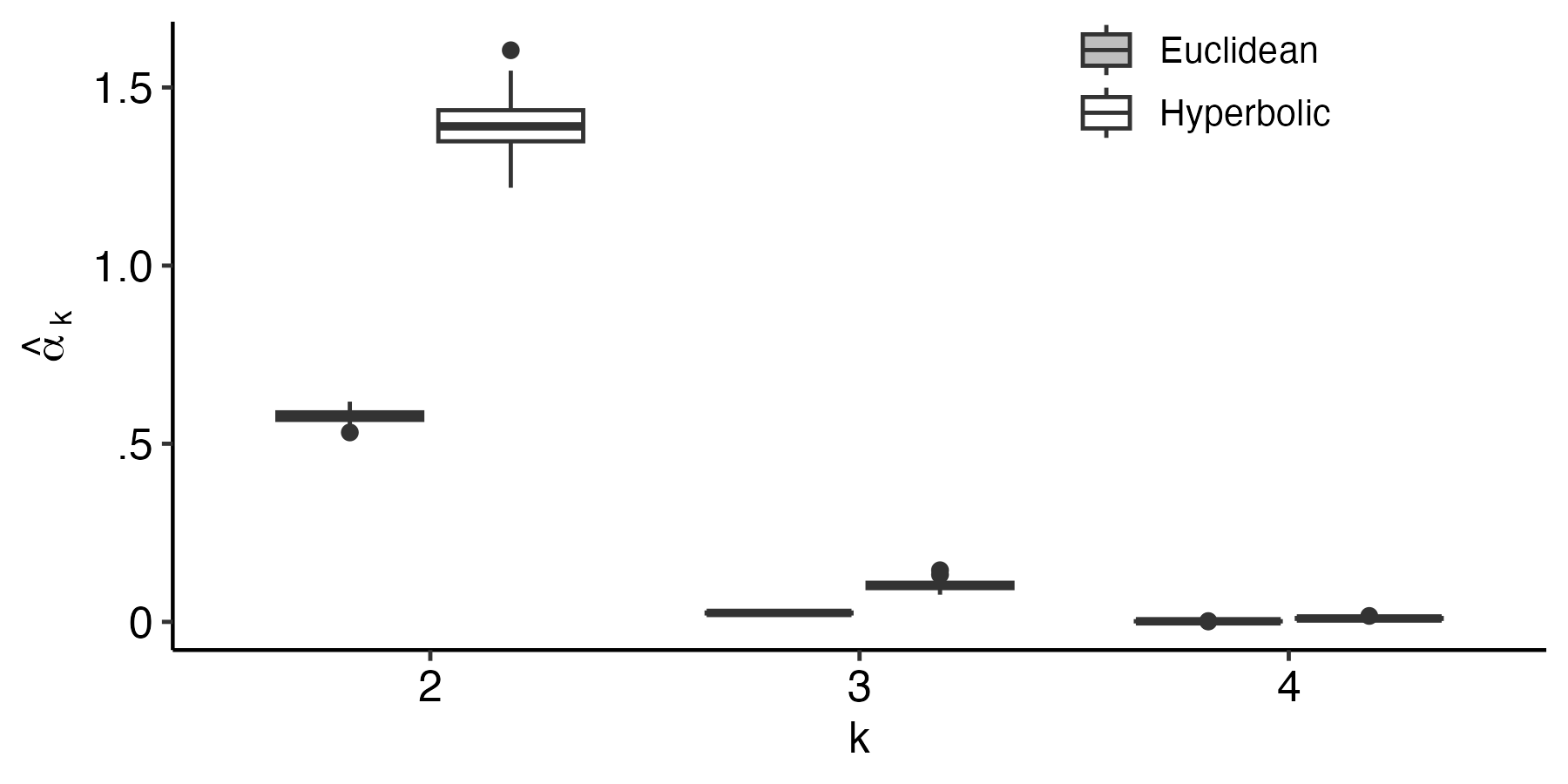}  
    \caption{\hangindent=1.6em  Boxplots of sparsity parameter estimates $\widehat\alpha_2$, $\widehat\alpha_3$, $\widehat\alpha_4$ based on different initializations and different geometries (Euclidean in blue and hyperbolic in yellow).}
  \label{fig:robustness_p}
\end{figure}
}

In the main manuscript, we initialize the positions of units using 100 different initial positions and minimize the same sample loss defined in Section \ref{sec:optimization}, given that the $\mathscr{S}_k^{(0)} (k \in \mathscr{K})$ are unchanged. 
To make sure that the results are not sensitive to this approximation of the population loss, we here resample both the starting positions and  $\mathscr{S}_k^{(0)} (k \in \mathscr{K})$. 
Put differently, we assess how much the results change for different approximations of the population loss.
In line with the sample inclusion probabilities detailed in Section \ref{sec:sampling}, all realized hyperedges are sampled, so $\mathscr{S}_k^{(1)} (k \in \mathscr{K})$ are unchanged for each run.
First, we see in Figure \ref{fig:llh_sensitivity} (a) that the sample loss for each samples converges within approximately 25 iterations, demonstrating robustness to initial positions.
Second, 
the range of trace plots in Figure \ref{fig:llh_sensitivity} (b) suggest that the sparsity parameter estimates are not sensitive to the choice of starting values and the minimization algorithm is not trapped in local minima of the sample loss function $\widehat{\ell}(.)$.

\subsubsection{Sensitivity to Choice of Number of Controls $n$}
\label{sec:control}

\begin{figure}
    \centering
   \includegraphics[width=0.9\textwidth]{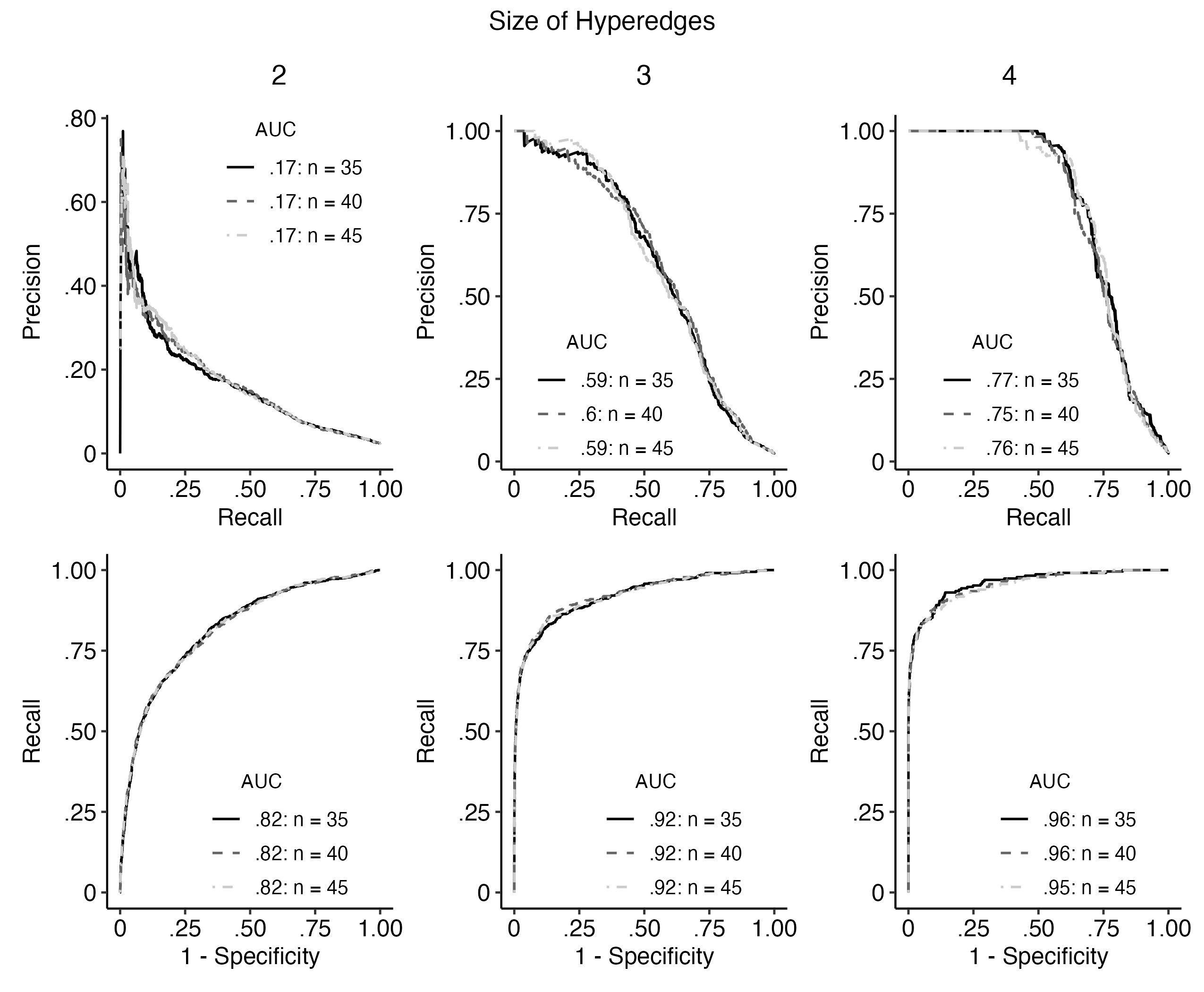}  
    \caption{Newswire data: Receiver-Operator Characteristic (ROC) and Precision-Recall (PR) curve of out-of-sample predictions of hyperedges,
    based on embedding units in hyperbolic space with number of controls $n \in \{35,\, 40,\, 45\}$.}\s
    \label{fig:pr_control}
\end{figure}

\begin{figure}[t!]
  \centering
    
    \includegraphics[width=0.4\textwidth]{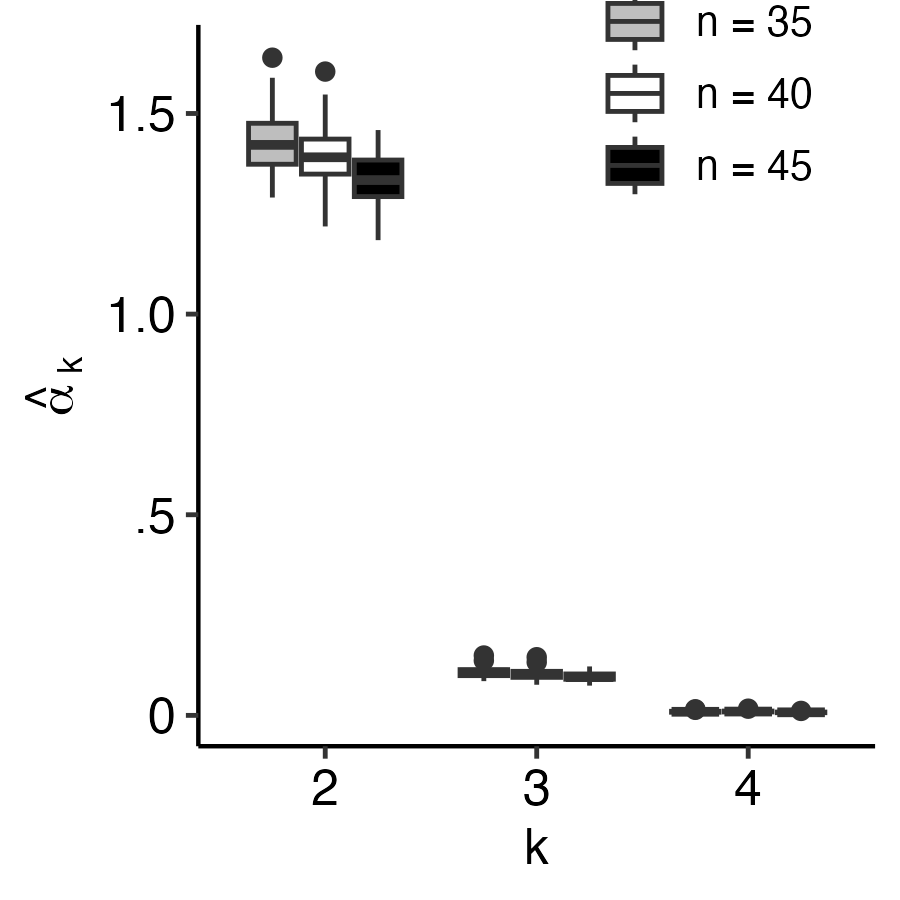}  
    \caption{Boxplots of sparsity parameter estimates $\widehat\alpha_2$, $\widehat\alpha_3$, $\widehat\alpha_4$ based on $n \in \{35,\, 40,\, 45\}$ controls.}
  \label{fig:robustness_sparsity_control}
\end{figure}

In addition,
we assess the sensitivity of the results to the choice of the number of controls $n$;
recall that we sample all realized hyperedges and,
for each realized hyperedge,
we sample $n$ unrealized hyperedges of the same size.
Figure \ref{fig:pr_control} shows the out-of-sample performance when $n \in \{35,\, 40,\, 45\}$. 
Figure~\ref{fig:robustness_sparsity_control} displays the estimated sparsity parameters in all three settings. 
The results indicate that the results are not too sensitive to changes of $n$. 

\section{Computational Settings}

The proposed models and methods are implemented in {\tt R} package $\mathtt{hyperspace}$, 
which is included in the replication package.
Most routines of {\tt R} package $\mathtt{hyperspace}$ are implemented in {\tt C++} and imported to {\tt R} using {\tt R} package {\tt Rcpp} \citepsupp{rcpp}. 
The experiments were carried out on a Red Hat Enterprise Linux 8.10 server running {\tt R} version 4.4.1. 
We used 100 cores and approximately 60 GB of RAM.  

\bibliographystylesupp{chicago}
\bibliographysupp{base}

\end{document}